\documentclass[pra,onecolumn,superscriptaddress,aps,5pt]{revtex4}
\usepackage{amsfonts}
\usepackage{amssymb,latexsym,amsmath}
\usepackage{graphicx}
\usepackage{epstopdf}
\newcommand{\beq}{\begin{equation}}
\newcommand{\eeq}{\end{equation}}

\newcommand{\eps}{\varepsilon}

\def\({\left(\begin{array}{cccccc}}
\def\){\end{array}\right)}

\def\bes{\begin{eqnarray}}
\def\ees{\end{eqnarray}}

\begin{document}
\title{Existence, Stability and Dynamics of Discrete Solitary Waves in a Binary Waveguide Array}
\author{Y. Shen}
\affiliation {Department of Mathematics, Southern Methodist University, Dallas, Texas 75275, USA}

\author{P.G. Kevrekidis}
%\email{kevrekid@math.umass.edu}
\affiliation{Department of Mathematics and Statistics, University of Massachusetts,
Amherst, Massachusetts 01003-4515 USA}

\affiliation{Center for Nonlinear Studies and Theoretical Division, Los Alamos
National Laboratory, Los Alamos, NM 87544}

\author{G. Srinivasan}

\affiliation{Theoretical Division, Los Alamos
National Laboratory, Los Alamos, NM 87544}

\author{A. B. Aceves}
\affiliation{Department of Mathematics, Southern Methodist University, Dallas, Texas 75275, USA}

\begin{abstract}
Recent work has explored binary waveguide arrays in the 
long-wavelength, near-continuum limit, here we examine the \textit {opposite}
limit, namely the vicinity of the so-called anti-continuum limit.  
We provide a systematic discussion of states involving one,
two and three excited waveguides, and provide comparisons that illustrate
how the stability of these states differ from the \textit {monoatomic}
limit of a single type of waveguide. We do so by developing a general theory
which systematically tracks down the key eigenvalues
of the linearized system. When we find the states to be unstable,
we explore their dynamical evolution through direct numerical simulations.
The latter typically illustrate, for the parameter values considered
herein, the persistence of localized dynamics and  the emergence 
for the duration of our simulations 
of robust quasi-periodic states  for two excited sites.  
As the number of excited nodes increase, the unstable dynamics 
feature less regular oscillations of the solution's amplitude.
\end{abstract}

%\date{} % Activate to display a given date or no date (if empty),
         % otherwise the current date is printed 
\maketitle
%$C_1 = -1$
%Dirac devil point at $\delta\beta =0$

\section{Introduction}

Over the past two and a half decades, the study of localized modes in 
nonlinear lattice dynamical systems has been a multi-faceted theme
of research, that has now been summarized in numerous reviews~\cite{review}.
Relevant applications span a wide number of disciplines and themes
including, but not limited to, 
arrays of nonlinear-optical waveguides \cite{moti}, 
Bose-Einstein condensates (BECs) in periodic potentials \cite{ober},
micromechanical cantilever arrays~\cite{sievers}, as well as 
Josephson-junction ladders \cite{alex}, 
halide-bridged transition metal complexes~\cite{swanson},
layered antiferromagnetic crystals~\cite{lars3}, 
dynamical models of the DNA double strand \cite{Peybi}, or
granular crystals of beads interacting through Hertzian contacts~\cite{theo10}.

Arguably, one of the most prototypical settings where such nonlinear
states have emerged is that of optical waveguide arrays, where several
of the relevant ideas and developments were first observed and analyzed,
as has been summarized in for example~\cite{dnc,moti}. 
In fact, in this setting and the related context of photorefractive
crystals, notions such as discrete diffraction~\cite{yaron} and its management~\cite{yaron1}, Talbot revivals~\cite{christo2}, $\mathcal{PT}$-symmetry
and its breaking~\cite{kip}, as well as discrete solitons~\cite{yaron,yaron2} and vortices~\cite{neshev,fleischer} have been experimentally observed.
Another source
of considerable inspiration has been the dynamics of 
Bose-Einstein condensates (BECs) in optical lattices~\cite{ober,konotop1}.
Yet, an additional motivation that has rendered such settings
popular has been the existence of a prototypical (and deceptively simple)
mathematical model that contains the main physical ingredients of
diffraction and nonlinearity. This model is the well-known 
discrete nonlinear Schr{\"o}dinger equation (DNLS)~\cite{book}.

More recently, a number of variants of this theme of optical waveguide
arrays have been studied in detail, notable examples being
multi-component models involving multiple 
polarizations~\cite{dncm,rudy}, waveguides featuring quadratic
(so-called $\chi^2$) nonlinearities~\cite{stege,rudy2}, the
examination of dark-solitonic states~\cite{shandarov,hadi} and
the study of binary waveguide arrays~\cite{akyl,jesuskip,Aceves}.
Here, we focus more specifically on the theme of binary
waveguide arrays and their alternating coupling structure. 

In this study, we adopt the prototypical model of~\cite{akyl,Aceves},
derived on the basis of coupled mode theory and incorporating the
effects of distinct propagation constants on the two \textit {components},
i.e., the even and odd waveguides. In earlier work~\cite{akyl,Aceves},
 emphasis was placed on quasi-continuum, long wavelength limits of
the system. The discussion in those studies also included the
potential for bearing band-gap and gap solitary waves, for supporting
dark-bright solitary waves and strongly localized solutions, 
as well as a modulational instability (discrete and continuous)
of the system. Here, our principal focus is on the highly 
localized solutions in the vicinity of the so-called anti-continuum
(AC) limit~\cite{macaub}. In the neighborhood of the uncoupled limit
between adjacent sites, following the approach of~\cite{Kevrekidis}
but for this considerably more complex problem, we obtain a systematic
set of results about the states that persist for finite coupling
between the adjacent sites. We also explore the stability of the
different configurations, developing a linear stability analysis
and characterizing the dominant eigenvalues therein. We find
that as the coupling varies between the two sets of waveguides,
away from the monoatomic limit (of a single type of waveguides),
the linearization eigenvalues may drastically change (e.g., real
ones may become imaginary etc.). Finally, for the unstable configurations, 
that include populating up to four separate waveguides, 
we examine the dynamics for case examples we think are prototypical and
for which we observe the formation of robust, quasi-periodic waveforms. 

Our presentation is structured as follows. In section II, we briefly 
discuss the relevant physical model and associated
properties/parameters thereof. Section III considers the
existence of solutions near the AC limit. In section IV, the more
complicated issue of the stability of the solutions is analyzed. 
Section V presents systematic computations for configurations of
different types, populating 2-4 waveguides. Finally, in section VI, we
summarize briefly our findings, and present our conclusions, as well
as a number of directions for future work. 

\section{Physical Model and Setup}
Following the earlier formulation of the binary
waveguide model, on the basis of coupled mode theory (CMT) with Kerr nonlinearity~\cite{akyl,Aceves}, the dynamical equations of interest read:
\beq\label{binary}
\begin{array}{c}
i {A_n}'  +{ \Delta\beta\over2 }A_n + \eps(C_1 B_n + B_{n+1}) +\gamma_a |A_n|^2 A_n = 0\\
i {B_n}'  -{ \Delta\beta\over2} B_n + \eps( A_{n-1} + C_1A_n) +\gamma_b |B_n|^2 B_n = 0
\end{array}
\eeq
where $'$ denote $d\over dz$. 
Here $A_n$ and $B_n$ denote the ``even'' and ``odd'' waveguides, and 
$C_1$ denotes the unequal coupling of the n-th waveguide with $n-1$
and $n+1$. 
We also introduce the parameter $\eps$ to control the strength of the coupling. 
%When $C_1=-1$, for $\eps\to \infty$ the difference equation can be viewed as a discretization of differential equation, 
$\Delta \beta$ denotes a detuning parameter, while $\gamma_{a,b}$ the
nonlinear prefactors, are associated with the strength of the Kerr
effect in the system. 
In the limit of $\eps \rightarrow \infty$, for $C_1 =-1$ or $C_1=1$
(in respective vicinities of the Brillouin zone), one can devise
a long wavelength limit of the system; for $C_1=1$ and $\Delta \beta=0$,
we fall back on the DNLS limit of waveguides of the same type.
The key limit that we will utilize herein is 
the AC-limit of $\eps\to0$~\cite{Kevrekidis,Konotop}. 

This system can also be derived from the following Lagrangian
\bes
 L =\sum_n\Big\{{ i\over 2} [A_n^*A_n'-A_n (A_n^*)'] + {\Delta \beta \over 2} |A_n|^2 + {\gamma_a\over 2} |A_n|^4
+{ i\over 2} [B_n^*B_n'-B_n (B_n^*)'] - {\Delta \beta \over 2} |B_n|^2 + {\gamma_b\over 2} |B_n|^4\nonumber
\\+{\eps} {\rm Re} [2 A_nB_{n+1}^*+2C_1 A_nB_n^*]\Big\}
\ees
and conserves the Hamiltonian
\bes
H = -\sum_n\Big\{ {\Delta \beta \over 2} |A_n|^2 + {\gamma_a\over 2} |A_n|^4
 - {\Delta \beta \over 2} |B_n|^2 + {\gamma_b\over 2} |B_n|^4+{\eps} {\rm Re}[
2 A_nB_{n+1}^*+2C_1 A_nB_n^*]\Big\}
\ees
and the total power 
\beq
P= \sum_n(|A_n|^2+|B_n|^2).
\eeq
 With $(A_n, B_n)\propto \exp\{i(nk_x+k_z z)\}$, we obtain the linear dispersion relation  $k_z ^2 = ({\Delta \beta \over 2 })^2 + \eps^2 (C_1^2 +1+2C_1 \cos k_x) $. In the continuum limit, i.e., for $k_x\to0$, we have $k_z^2 = ({\Delta \beta \over 2 })^2 + \eps^2 (C_1+1)^2$. Instead extending work on 
the continuum limit beyond~\cite{akyl,Aceves},
this paper focuses on the formation of discrete solitons bifurcating from the
uncoupled limit of the system, i.e., from the limit of $\eps \rightarrow 0$.

\section{Existence of Discrete Solitons}

Starting from Eqn. (\ref{binary}) without coupling, i.e. $\eps=0$ 
\beq
\begin{array}{c}
i{A_n}'  +{ \Delta\beta\over2 }A_n  +\gamma_a |A_n|^2 A_n = 0\\
i{B_n}'  -{ \Delta\beta\over2} B_n+\gamma_b |B_n|^2 B_n = 0
\end{array}
\eeq
for which, in addition to the trivial solutions, we identify 
nontrivial ones of the form:
%With $A_n(z) = a_n(z)e^{i\phi_n(z)}$ and $B_n(z) = b_n(z)e^{i\varphi_n(z)}$, we get 
%\bes
%i  a_n'- a_n\phi_n'   +{ \Delta\beta\over2 }a_n  +\gamma_a a_n^3=0\\
%-i  a_n' - a_n\phi'   +{ \Delta\beta\over2 }a_n  +\gamma_a a_n^3=0
%i b_n'- b_n\phi_n'   -{ \Delta\beta\over2 }b_n  +\gamma_b b_n^3=0
%\ees
%$a_n' = 0$,  $ \phi_n' = { \Delta\beta\over2 }  +\gamma_a a_n^2$, 
\beq\label{eps0}
\begin{array}{c}
A_n(z) = {a}_ne^{ic_n} e^{i[( { \Delta\beta\over2 }  +\gamma_a {a}_n^2)z ]}\\
B_n(z) = {b}_ne^{id_n} e^{i[(- { \Delta\beta\over2 }  +\gamma_b {b}_n^2)z]}
\end{array}
\eeq
where ${a}_n$, ${b}_n$, ($c_n$, $d_n$) are, without loss of
generality~\cite{Kevrekidis} the real amplitudes (phases). 
%that does not depend on $z$.  
%Set $a_n = b_n = 1$, then $A_n = a_{n}e^{\alpha z}$, $b_n = b_{n}e^{\beta z}$, where $\alpha = { \Delta\beta\over2 }  +\gamma_a$ and $\beta = - { \Delta\beta\over2 }  +\gamma_b$, $a_{n} = e^{ic_n}$, $b_{n} = e^{id_n}$.

%\textbf {Gowri}

As we turn on $\eps$, we consider the solution of Eqn. (\ref{binary}) in
the form of standing waves as $(A_n, B_n) = ({\bf a}_n(z)e^{i\alpha z} ,{\bf
  b}_n(z)e^{i\beta z} )$,  with $\alpha = { \Delta\beta\over2 }  +a^2
\gamma_a$ and $\beta = - { \Delta\beta\over2 }  +b^2 \gamma_b$. Here
$a$, $b$ are amplitudes of $a_n(z)$,
$b_n(z)$ respectively corresponding to the $\epsilon =0$ system. Here 
instead,  ${\bf a}_n(z), {\bf b}_n(z)$ satisfy
 \beq \label{ab}
 \begin{array}{c}
i{\bf a}_n' =  \gamma_a {\bf a}_{n}(a^2-|{\bf a}_{n}|^2) -\eps(C_1 {\bf b}_{n} + {\bf b}_{n+1})e^{-i\rho z},\\ 
i{\bf b}_n' =  \gamma_b {\bf b}_{n}(b^2-|{\bf b}_{n}|^2) -\eps({\bf a}_{n-1} + C_1{\bf a}_{n})e^{i\rho z},
\end{array}
\eeq
%For stationary solution of above equation, we set $(a_n', b_n') = (0,0)$ to get
% \bes
% \gamma_a a_{n}(a^2-|a_{n}|^2) =\eps(C_1 b_{n}e^{-i\rho z} + b_{n+1}e^{-i\rho z})\\
% \gamma_b b_{n}(b^2-|b_{n}|^2) =\eps(a_{n-1}e^{i\rho z} + C_1a_{n}e^{i\rho z})
% \ees
 where $\rho = \alpha-\beta = \Delta\beta+a^2\gamma_a-b^2\gamma_b$.
For a stationary solution to exist, we must 
assume that $\rho = 0$, i.e., that the linear detuning in
the propagation constant balances the corresponding nonlinear one. The
stationary solutions for Eqn. (\ref{ab}) with $({\bf a}_n(z), {\bf
  b}_n(z)) = (a_n, b_n)$, then satisfy 
\beq\label{sab}
\begin{array}{c}
 \gamma_a a_{n}(a^2-|a_{n}|^2) =\eps(C_1 b_{n} + b_{n+1}),\\
 \gamma_b b_{n}(b^2-|b_{n}|^2)  = \eps(a_{n-1} + C_1a_{n}).
 \end{array}
 \eeq
%Without loss of generality in the 1, we can set $a_n$s, $b_n$s to be real numbers.
 When $\eps =0$, as stated before 
%we get either $a_{n} = 0$ or 
%$a_{n} =a e^{ic_n}$, either $b_{n} = 0$ or $b_{n} =b e^{id_n}$, 
%with $c_n,d_n\in{\{0, \pi\}}$. 
 the solutions of Eqn. (\ref{sab})  are in the form:
\beq\label{ab0}
a_n =a_n^{(0)} =  \left\{
\begin{array}{c}
ae^{ic_n},\quad n\in S_1, \\
0, \quad n\in \mathbb{Z} /S_1,
\end{array}
\right.
\qquad
b_n =b_n^{(0)} =  \left\{
\begin{array}{c}
be^{id_n},\quad n\in S_2, \\
0, \quad n\in \mathbb{Z} /S_2,
\end{array}
\right.
\eeq
%\textbf{Gowri}
where $S_i$ denote finite sets of nodes of the lattice and $c_n,
d_n\in \{0, \pi\}$. Only those waveguides that belong to $S_i$ have a
non-zero excitation. For small $\eps\neq0$, the Jacobian matrix of
Eqn. (\ref{sab}) remains invertible, and the solutions of
Eqn. (\ref{sab}) are analytic functions of $\eps$  around $\eps=0$
\cite{Kevrekidis,Konotop}. We can then expand $(a_n, b_n)$ as 
\bes
\begin{array}{cccc} 
a_n = a_n^{(0)} +\sum_{k=1}^{k=\infty} \eps^k a_n^{(k)},\qquad
b_n = b_n^{(0)} +\sum_{k=1}^{k=\infty} \eps^k b_n^{(k)},
\end{array}
\ees
In particular, we can solve Eqn. (\ref{sab}) at O$(\eps)$ to get
\bes\label{ab1}
\begin{array}{cccc} 
a_n^{(1)} = \frac{C_1 b_n^{(0)}+b_{n+1}^{(0)}}{\gamma_a (a^2-3(a_n^{(0)})^2)},\qquad
b_n^{(1)} = \frac{ a_{n-1}^{(0)}+C_1 a_{n}^{(0)}}{\gamma_b (b^2-3(b_n^{(0)})^2)}.
\end{array}
\ees
%\textbf{Gowri} 
While keeping the discussion general, we note here by equating the $O(1)$ terms, that
$(a_n^{(0)})^2 = a^2$ and $(b_n^{(0)})^2 = b^2$ for excited nodes. Following the same
vein, we can extract the solutions order by order in $\eps$. 
We now explore the stability of these solutions, by analyzing their
linearized eigenvalues.

% The linearized equations for Eqn. (\ref{1}-\ref{2}) around $(a_n^{(0)}, b_n^{(0)})$ are
%\bes
%\gamma_a(a^2-2|a_n^{(0)}|^2 )a_n^{(1)} -\gamma_a (a_n^{(0)})^2 (a_{n}^{(1)})^* = \eps(C_1 b_n^{(0)} +b_{n+1}^{(0)})\\
%\gamma_b(b^2-2|b_n^{(0)}|^2 )b_n^{(1)} -\gamma_b (b_n^{(0)})^2 (b_{n}^{(1)})^* = \eps(a_{n-1}^{(0)} +C_1  a_{n}^{(0)})
%\ees
%which are going to be used to get the Jacobian matrix for the newton solver and solve for linear stability. 
%By takeing $a_{n}^*  Eqn. (\ref{1}) - a_{n}  (Eqn. (\ref{1}))^*$, %$a_{n}^*$ times Eqn. (\ref{1}) minus $a_{n}$ times the conjugate of Eqn. (\ref{1}), 
 %and similarly  for Eqn. (\ref{2}), we get

% From Eqn. (\ref{sab}), we have
% \bes
% {\gamma_a\over\eps} |a_{n}|^2(a^2-|a_{n}|^2) =(C_1 b_{n} + b_{n+1})a_{n}^*= (C_1 b_{n}^* + b_{n+1}^*)a_{n},\\
%{ \gamma_b \over\eps}|b_{n}|^2(b^2-|b_{n}|^2)  = (a_{n-1} + C_1a_{n})b_n^*= (a_{n-1}^* + C_1a_{n}^*)b_n.
% \ees
% that means 
% $
% (C_1 b_{n} + b_{n+1})a_{n}^* $ and 
%$ (a_{n-1} + C_1a_{n})b_n^*
% $
% are real and
% \bes
%C_1(a_{n}^* b_{n}-a_{n}b_{n}^*) +(a_{n}^*b_{n+1} - a_{n}b_{n+1}^*)=0\label{3}\\
% C_1(a_{n} b_{n}^*-a_{n}^*b_{n}) +(a_{n-1}b_{n}^* - a_{n-1}^*b_{n})=0\label{4}
%  \ees
%  Adding Eqn. (\ref{3}) and Eqn. (\ref{4}) we have 
% \bes
%a_{n-1}b_{n}^* - a_{n-1}^*b_{n}+a_{n}^*b_{n+1} - a_{n}b_{n+1}^* =0
%\ees
%i.e.
%$
%a_{n-1}b_{n}^*  +a_{n}^*b_{n+1}
%$ 
%is also real. So we can choose
%$(c_{n-1}-d_n)=0$ mod $\pi$ and  $(d_{n+1}-c_n) = 0$ mod $\pi$, $(d_{n}-c_n) = 0$ mod $\pi$. So we can set $a_n$s, $b_n$s to be real numbers. 

\section{Stability of discrete solitons}

To examine the linear stability of the stationary
solutions, we assume small perturbations of the form:
\bes\begin{array}{c}
A_n(z) = e^{i\alpha z}\{a_{n} + \delta p_n  \}\\
B_n(z) = e^{i\beta z}\{b_{n} + \delta q_n\}
\end{array}
\ees
%\textbf {Gowri - Corrected t*delta to delta.}
where $a_n$ and $b_n$ are stationary solutions that satisfy Eqn. (\ref{sab}), and the term proportional to $\delta$ is a small perturbation of order $\delta$. At order $O(\delta)$;
$\delta$ is a formal parameter, denoting the smallness of the perturbation.
%\textbf {Gowri}
We decompose the perturbations, which are complex in nature as
$p_n = (r_n+s_n i)e^{\lambda z}$, $q_n= (f_n  + h_n i)e^{\lambda z} $, to obtain
 the linearization equations at O$(\delta)$:
\bes\label{4es}\begin{array}{cccc} 
\lambda r_n = \gamma_a(a^2-a_n^2) s_n -\eps (C_1 h_n + h_{n+1}),\\
-\lambda s_n =  \gamma_a(a^2-3a_n^2 ) r_n +\eps (C_1 f_n + f_{n+1}),\\
\lambda f_n =  \gamma_b(b^2-b_n^2) h_n -\eps (s_{n-1} +C_1  s_{n}),\\
- \lambda h_n = \gamma_b(b^2-3b_n^2) f_n +\eps (r_{n-1} +C_1  r_{n}).
 \end{array}
\ees
We rewrite this as 
\bes\label{ll}
\begin{array}{cccc} 
\lambda {\bf R} =  \mathcal L_- {\bf S}\\
-\lambda {\bf S} = \mathcal L_+ {\bf R}
\end{array}
\ees
where ${\bf R} =(\cdots,r_{n-1}, f_{n-1}, r_n, f_n, \cdots)^T$, ${\bf S} =(\cdots,s_{n-1}, h_{n-1}, s_n, h_n, \cdots)^T$ and $\mathcal L_\pm$ are  infinite-dimensional symmetric tri-diagonal matrices, which consist of elements  
\bes
(\mathcal L_-)_{n,n} =  
\left( \begin{array}{cccc}
\gamma_a(a^2-a_n^2)  & -\eps C_1\\
-\eps C_1 & \gamma_b(b^2-b_n^2)
\end{array}\right)\nonumber
\\
(\mathcal L_+)_{n,n} =  
\left( \begin{array}{cccc}
\gamma_a(a^2-3a_n^2)  & \eps C_1\\
\eps C_1 & \gamma_b(b^2-3b_n^2)
\end{array}\right)
\\\nonumber
(\mathcal L_\pm)_{n,n+1} =  
\left( \begin{array}{cccc}
 0 & \pm\eps\\
0 & 0
\end{array}\right) = (\mathcal L_\pm)_{n,n-1}^T.
\ees
%\bes
%(\mathcal L_-)_{n,n+1} =  
%\left( \begin{array}{cccc}
% 0 & -\eps\\
%0 & 0
%\end{array}\right)
%\ees
%\bes
%(\mathcal L_-)_{n,n-1} =  
%\left( \begin{array}{cccc}
%0 & 0\\
%-\eps & 0
%\end{array}\right)
%\ees
%\bes
%(\mathcal L_-)_{n,n-1} =  
%\left( \begin{array}{cccc}
%0 & 0\\
%\eps & 0
%\end{array}\right)
%\ees
We can further  write Eqns. (\ref{ll}) in
the  form: 
\bes\label{lambdaJH}
\lambda \mathbf \Psi= \mathcal J \mathcal H \mathbf \Psi,
\ees
where  $\mathbf \Psi = (\cdots, r_n, f_n, s_n, h_n,\cdots)^T$ is the eigenvector, $\mathcal H$ is defined as
\bes
\mathcal H_{n,m} =  
\left( \begin{array}{cc}
{(\mathcal L_{+})_{n,m} }  & {\bf 0}\\
 {\bf 0}& {(\mathcal L_{-})_{n,m} } 
\end{array}\right)
\ees
and the skew-symmetric matrix $\mathcal J$ consists of
$4 \times 4$ blocks of the form:
\bes
%\mathcal J_{n,m} = 
\left( \begin{array}{cccc}
 0&0&1&0\\
 0&0&0&1\\
 -1&0&0&0\\
 0&-1&0&0 
\end{array}\right)
%\delta_{n,m}
\ees
%\textbf {Gowri}
Let's first consider the uncoupled case when $\eps=0$. Since the
coupling term is zero, the analysis is independent of which population
of the waveguides is excited. For a total of $K$ excited waveguides $A_n$ ($B_n$),  the eigenvalues $\mu$ of $\mathcal{H} {\bf \Phi} =\mu {\bf \Phi} $ consist of exactly $K$ eigenvalues $\mu = -2\gamma_a a^2$ ($ \mu = -2\gamma_bb^2$) from $\mathcal L_{+} $ and $K$ eigenvalues of $\mu =0$ from $\mathcal L_{-}$.   Eventually these will map to  $K$ double zeros of $\lambda$. For the remaining infinite unexcited (i.e., vanishing)
$A_n$s ( $B_n$s), we will get an infinite number of $\mu = \gamma_a a^2$ ($\mu = \gamma_b b^2$) from  both $\mathcal L_{\pm} $, which will eventually map to infinite pairs of $\lambda = \pm  i  \gamma_a a^2$ ($\lambda = \pm i \gamma_b b^2$). 

As we turn on $\eps$, the infinite number of $\lambda = \pm  i  \gamma_a a^2$ ($\lambda = \pm i \gamma_b b^2$), will form the branches of 
the continuous spectrum, actually with plane wave eigenfunctions $p_n,q_n \propto e^{\pm i(kn-\omega z)}$. The corresponding eigenvalues $\lambda = i\omega$ 
satisfy 
\beq
(\pm\omega-\gamma_a a^2)(\pm\omega-\gamma_b b^2) = \eps^2(C_1^2+1+2C_1\cos(k)).\label{cs}
\eeq
Now let us focus on how the finite number of zero eigenvalues move
(possibly) away from $\lambda=0$,  as we turn on $\eps$.  We write Eqn. (\ref{ll}) as,
\bes
 \lambda^2  {\bf S} = -\mathcal L_{+}\mathcal L_{-}{\bf S}% :=   \mathcal M {\bf S}
\ees
and notice  that  $\mathcal L_{+} $ is invertible when $\eps=0$, so there exists $\eps_0$, such that when $0\le\eps<\eps_0$,  $\mathcal L_{+} $ is still invertible. Then, we have
\bes\label{ii}
\lambda^2 \mathcal L_{+}^{-1} {\bf S}   =-\mathcal L_{-}{\bf S}.
\ees
Forming the inner product of both sides of Eqn. (\ref{ii}) with $\bf S$, we get
\bes
\lambda^2  = -\frac{<{\bf S}, \mathcal L_{-}{\bf S}> }{<{\bf S}, \mathcal L_{+}^{-1} {\bf S}>},
\ees
where the inner product is defined as $<{\bf u, v}> = \sum_n \bar u_n v_n$. 

As mentioned before we can expand $(a_n, b_n)$ in powers of $\eps$, 
and similarly we can also expand $\mathcal L_\pm $, $\bf S$, etc.
In particular, if $\bf S_a$ is the eigenvector of $ \mathcal L_{-}$ corresponding to $\mu_a$,  such that $\lim_{\eps\to 0} \mu_a =0$, then
%Denote ${\bf S_a^{(0)}} = \lim_{\eps\to 0} {\bf S_a} $,   
%$ \mathcal L_{-}{\bf S_a} = \mu_a {\bf S_a}$
%\bes <{\bf S_a}, \mathcal L_{-}{\bf S_a}>=\mu_a<{\bf S_a}, {\bf S_a}> \ees 
\bes
\lim_{\eps\to 0} <{\bf S_a}, \mathcal L_+^{-1}{\bf S_a}> = <{\bf S_a^{(0)}}, (\mathcal L_+^{-1})^{(0)}{\bf S_a^{(0)}} > \neq0.
\ees
where $(\mathcal L_+^{-1})^{(0)}$ will be a diagonal matrix with
\bes
(\mathcal L_+^{-1})^{(0)}_{n,n} &=& 
\left( \begin{array}{cccc}
\frac{1}{\gamma_a(a^2-3(a_n^{(0)})^2)}  & 0\\
 0 & \frac{1}{\gamma_b(b^2-3(b_n^{(0)})^2)}
\end{array}\right).
\ees
Then the leading order in $\epsilon$ of $\lambda^2 $ will be
\beq\label{lambda}
 \lambda^2 = -\frac{\mu_a <{\bf S_a^{(0)}}, {\bf S_a^{(0)}}> }{<{\bf S_a^{(0)}}, (\mathcal L_{+}^{-1})^{(0)} {\bf S_a^{(0)}}>}+O(\eps^2).
\eeq
The problem which remains is to find the leading order eigenvalues and eigenvectors of $\mathcal L_{-}$.
To compute 
%the leading order of eigenvalue of $\mathcal L_{-}$, 
this, we expand $\mathcal L_-$ as 
\bes
(\mathcal L_-)_{n,n} &=&  
\left( \begin{array}{cccc}
\gamma_a(a^2-a_n^2)  & -\eps C_1\\
-\eps C_1 & \gamma_b(b^2-b_n^2)
\end{array}\right)=(\mathcal L_-^{(0)})_{n,n}+\eps(\mathcal L_-^{(1)})_{n,n} + O(\eps^2) \\\nonumber
&=&\left( \begin{array}{cccc}
\gamma_a(a^2-(a_n^{(0)})^2)  & 0\\
0 & \gamma_b(b^2-(b_n^{(0)})^2)
\end{array}\right)+\eps
\left( \begin{array}{cccc}
-2\gamma_a a_n^{(0)}a_n^{(1)}  & - C_1\\
- C_1 & -2\gamma_b b_n^{(0)}b_n^{(1)}
\end{array}\right) +O(\eps^2)\\
%\ees
%\bes
%(\mathcal L_+)_{n,n} &=& 
%\left( \begin{array}{cccc}
%\gamma_a(a^2-3a_n^2)  & \eps C_  1\\
%\eps C_1 & \gamma_b(b^2-3b_n^2)
%\end{array}\right)=(\mathcal L_+^{(0)})_{n,n}+\eps(\mathcal L_+^{(1)})_{n,n} + O(\eps^2)  \\\nonumber
%&=&\left( \begin{array}{cccc}
%\gamma_a(a^2-3(a_n^{(0)})^2)  & 0\\
% 0 & \gamma_b(b^2-3(b_n^{(0)})^2)
%\end{array}\right)+\eps
%\left( \begin{array}{cccc}
%-6\gamma_a a_n^{(0)}a_n^{(1)}  &  C_1\\
% C_1 & -6\gamma_b b_n^{(0)}b_n^{(1)}
%\end{array}\right)+O(\eps^2)
%\ees
%Recall that 
%\bes
(\mathcal L_-)_{n,n+1} &=& {\bf 0}+ \eps
\left( \begin{array}{cccc}
 0 & -1\\
0 & 0
\end{array}\right) = (\mathcal L_-)_{n,n-1}^T.
\ees
and project $\mathcal L_{-} {\bf S_a} = \mu_a {\bf S_a}$ on to the kernel of $\mathcal L_{-}^{(0)}$,  we have
\bes
\mathcal L_{-}^{(0)} {\bf S_a^{(0)}} = \mu_a^{(0)} {\bf S_a^{(0)}} = 0,\\
\mathcal L_{-}^{(0)} {\bf S_a^{(1)}} + \mathcal L_{-}^{(1)} {\bf S_a^{(0)}} = \mu_a^{(1)} {\bf S_a^{(0)}}.
\ees
%\bes\label{mu1}
%\mathcal L_{-}^{(1)} {\bf S_a^{(0)}} = \mu_a^{(1)} {\bf S_a^{(0)}}.
%\ees
Here ${\bf S_a^{(0)}}$ can be written as combinations of basis
elements in the kernel of $\mathcal L_{-}^{(0)}$, i.e. ${\bf S_a^{(0)}} = \sum_{k=1}^{K}c_k { e_k}$, where the basis elements $ e_k$  can be chosen as  ${ e_k} = (\cdots ,0, \pm1, 0,\cdots)^T$, where  the ``$\pm 1$" is located at the $k$th excited site and the sign is chosen to be the same as the anti-continuum limit. 
Then, we obtain
\bes
{\bf M} c  = \mu^{(1)} c
\ees
where the $K\times K$ matrix $\bf M$  with elements
\bes\label{M}
{\bf M}_{m,n}  = <e_m, \mathcal L_{-}^{(1)} e_n>.
\ees
Once we have the eigenvalue and eigenvector of $\bf M$, then by Eqn. ($\ref{lambda}$) we get 
\beq\label{lambda2}
 \lambda^2 = -\frac{\mu_a^{(1)} <{\bf S_a^{(0)}}, {\bf S_a^{(0)}}> }{<{\bf S_a^{(0)}}, (\mathcal L_{+}^{-1})^{(0)} {\bf S_a^{(0)}}>}\eps+O(\eps^2).
\eeq

We will provide a series of concrete examples of this calculation 
for configurations with $2$-$4$ sites, in the section below titled Computations.

%\section{Continuation on $\eps$}

%
% and corresponds to the excited $A$ component, then 
%\bes
%\lim_{\eps\to 0} <{\bf S_a}, \mathcal L_+^{-1}{\bf S_a}> = <{\bf S_a}^{(0)}, (\mathcal L_+^{(0)})^{-1}{\bf S_a}^{(0)} > = \frac{1}{-2 \gamma_a a^2}
%\ees
%thus the leading order gives 
%\bes
%\lim_{\eps\to 0} \frac{\lambda_a^2}{\mu_a} = 2\gamma_a a^2. 
%\ees
%Similarly we have 
%\bes
%\lim_{\eps\to 0} \frac{\lambda_b^2}{\mu_b} = 2\gamma_b b^2. 
%\ees
%For $\eps\neq0$, due to gauge invariance there will still be a zero eigenvalue of $\mathcal L_{-} {\bf S} = {\bf 0}$, while other $K+M-1$ eigenvalues are non-zero.

%so $\mathcal L_{\pm}^{(0)}$, $\mathcal H^{(0)}$ are diagonal matrix, and
%\bes
%\mathcal H^{(1)}_{n,n+1} = \left( \begin{array}{cccc}
% 0  & 1 &0&0 \\
%  0&0&0&0\\
% 0 &0&0 & -1 \\
%  0&0&0 &0
%\end{array}\right), \ \ \ \ 
%\mathcal H^{(1)}_{n,n-1} = \left( \begin{array}{cccc}
% 0  & 0 &0&0 \\
%  1&0&0&0\\
% 0 &0&0 & 0 \\
%  0&0&-1 &0
%  \end{array}\right) 
%\ees

\section{Computations}
In the next three subsections we will discuss setups with two, three and 
four excited sites, respectively. We denote excited nodes of  
$a_n^{(0)}$ and $b_n^{(0)}$ as ``$+$" if they are positive, ``$-$" if they are negative, and  use parameters $C_1 = 2,\ a=1,\ b =1, \Delta \beta =1,\ \gamma_a =1,\ \gamma_b = 2$ unless  stated otherwise.  We will first identify  the stationary solutions using Eqn. (\ref{sab}) by means of fixed-point
Newton iterations and parameter continuation on $\eps$ starting from  $\eps = 0$. Then, we will identify the spectrum of the linear stability 
analysis for these stationary solutions numerically, by solving
the full matrix eigenvalue problem of Eqn. (\ref{lambdaJH}). Lastly, we will compare the eigenvalues that bifurcate from zero
--these are the ones that are potentially responsible for emerging
instabilities-- with their leading order theoretical
approximations provided by Eqn. ($\ref{lambda2}$). 
Finally, when the solutions are identified to be spectrally unstable,
we will also
 study their evolutionary dynamics, to explore the outcome of this
instability upon propagation in $z$.  

\subsection{Two excited sites}
Let us first consider the case where
the only nonzero entries of  $(a_n^{(0)}, b_n^{(0)})$ are 
\beq
(a_0^{(0)}, b_0^{(0)}) = %\left\{
%\begin{array}{ll}
(ae^{ic_0}, be^{id_0}),   %\\
%(0,0),& n\neq0
%\end{array}
%\right.
\eeq
with $c_0, d_0 \in \{0, \pi\}$. From Eqn. (\ref{ab1}), we get the nonzero entries for  $(a_n^{(1)}, b_n^{(1)})$ are
\beq
(a_n^{(1)}, b_n^{(1)}) = \left\{
\begin{array}{lr}
( \frac{be^{id_0}}{\gamma_a a^2} ,0 ),& n=-1\\
(\frac{ C_1 be^{id_0} }{-2\gamma_a a^2} , \frac{ C_1 ae^{ic_0}}{-2\gamma_b b^2} ),& n=0\\
(0, \frac{ae^{ic_0}}{\gamma_b b^2} ),& n=1.%\\
%(0,0),& \text{otherwise}
\end{array}
\right.
\eeq
%from Eqn. (\ref{ab1}), one can get the nonzero entries $a_0^{(1)} =\frac{ C_1 b}{-2\gamma_a a^2} $, $ b_0^{(1)} =\frac{ C_1 a}{-2\gamma_b b^2} $, $a_{-1}^{(1)} = \frac{b}{\gamma_a a^2}$, $b_{1}^{(1)} = \frac{a}{\gamma_b b^2}$.
From Eqn. (\ref{M}), we obtain
\bes
{\bf M} =C_1e^{i(d_0-c_0)} \begin{pmatrix}
\frac{b}{a}& -1\\
-1 & \frac{a}{b}
\end{pmatrix}.
\ees
$\bf M$ has zero eigenvalue with eigenvector $(a, b)^T$,
which will map to a double zero of $\lambda$. This corresponds 
to the phase (or gauge) invariance of the model, associated with the
U$(1)$ symmetry. The nonzero eigenvalue $C_1\cos(d_0-c_0)(\frac a b + \frac b a)$ has eigenvector $(b, -a)^T$, so the leading order of $\lambda^2$ is 
\beq
\lambda^2 \approx
%\frac{2C_1a_0^{(0)}b_0^{(0)}(a^2+b^2)^2}{\frac{b^4}{\gamma_a}+\frac{a^4}{\gamma_b}}\varepsilon=
 \frac{C_1(\frac b a +\frac a b)(a^2+b^2)}{\frac{b^2}{2\gamma_aa^2}+\frac{a^2}{2\gamma_b b^2}} \cos(c_0-d_0)\eps%= \frac{2C_1\gamma_a\gamma_bab(a^2+b^2)^2}{\gamma_bb^4+\gamma_aa^4}\cos(c_0-d_0)\eps %= \sqrt{\frac{32\eps}{3}}
\eeq
That means if $\gamma_a = \gamma_b=\gamma$, $\lambda^2 \approx \eps \cos(c_0-d_0) \gamma C_1 c$, where $c = \frac{2(b/a+a/b)(a^2+b^2)}{b^2/a^2+a^2/b^2}$ is a positive number. So if $\cos(c_0-d_0) \gamma C_1>0$ we will have an
unstable stationary solution.  

\begin{figure}[!htbp]
 \centering
 \includegraphics[width = 0.35\textwidth]{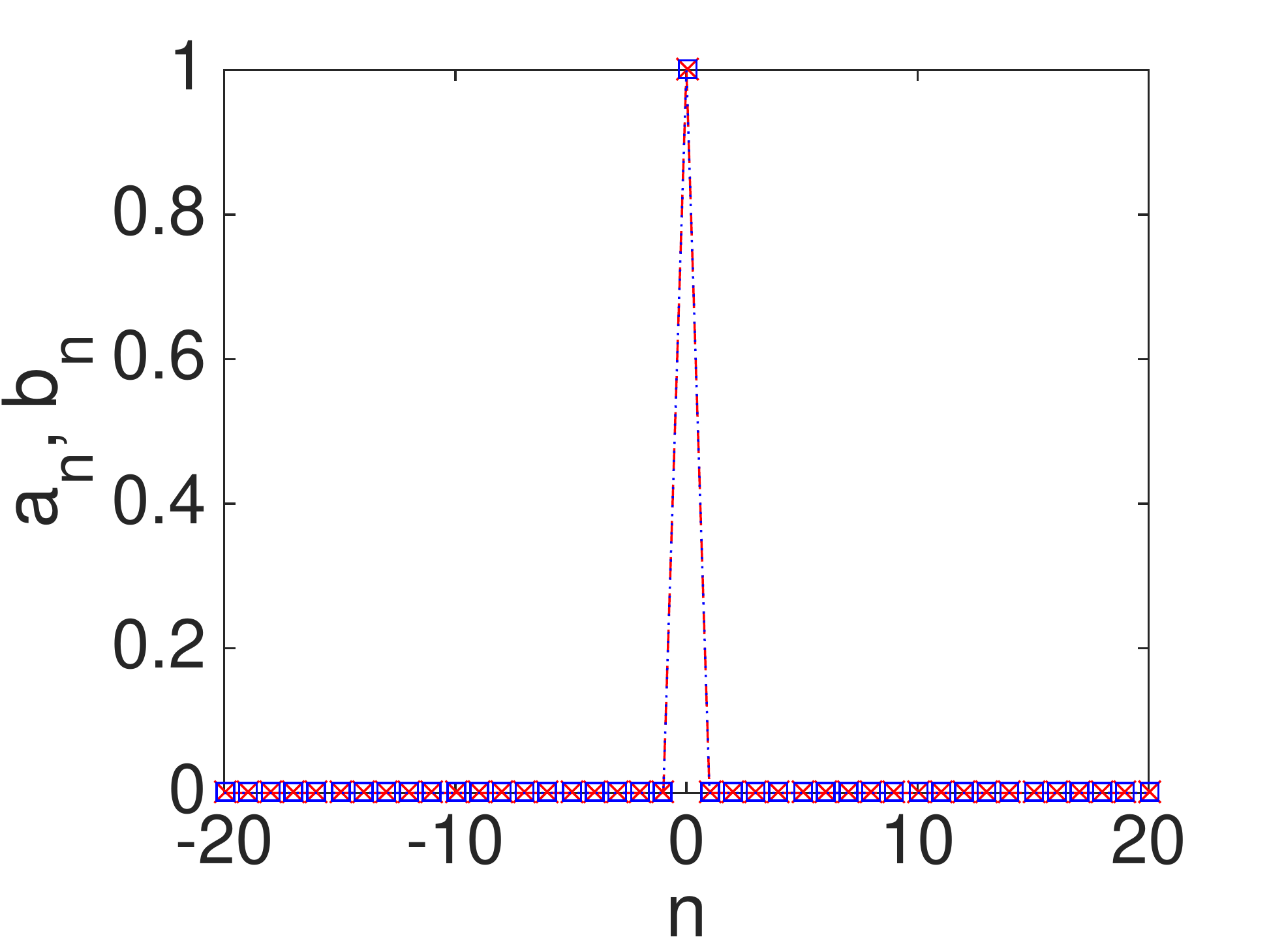}
 \includegraphics[width = 0.35\textwidth]{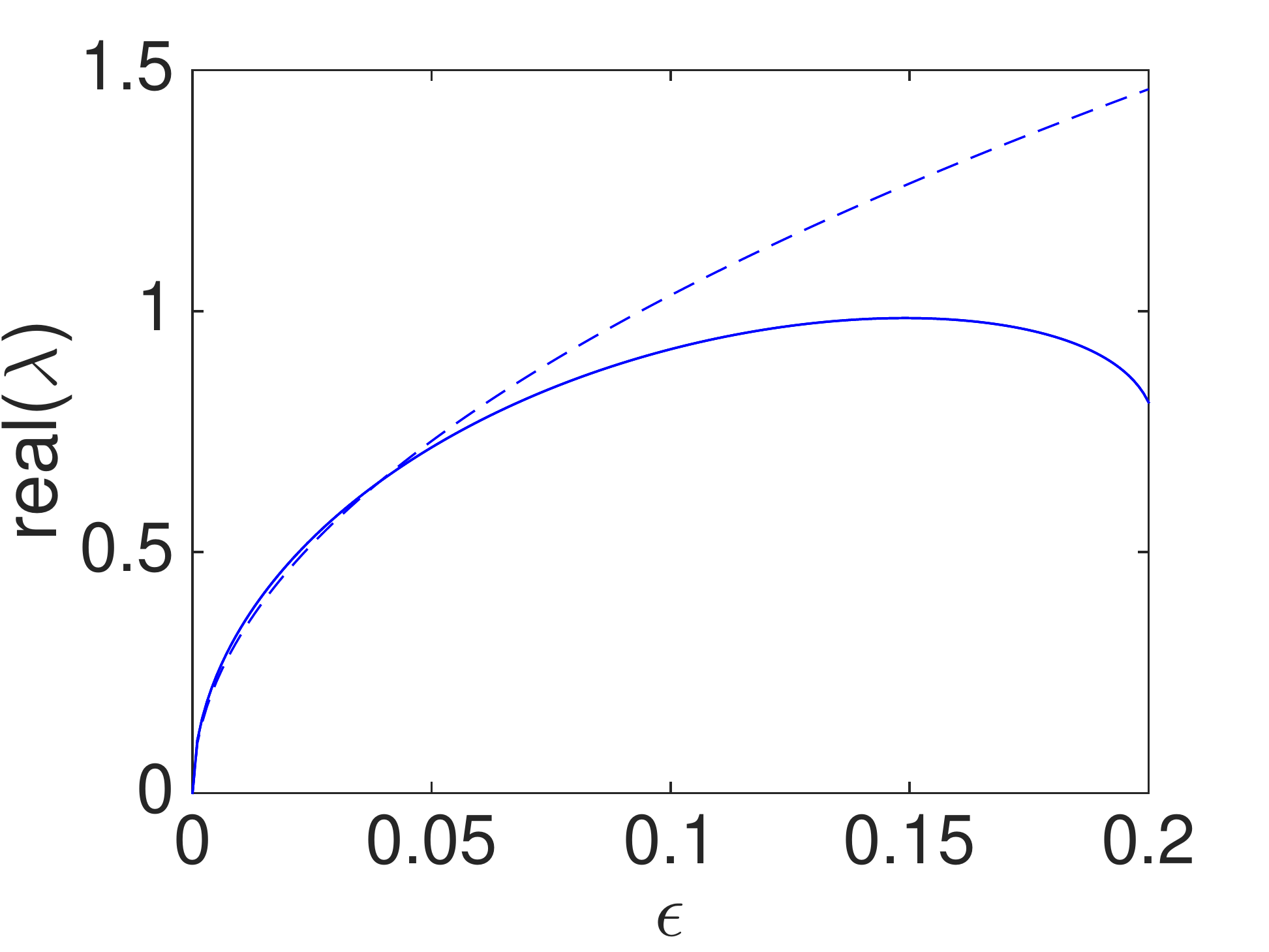}
 \includegraphics[width = 0.3\textwidth]{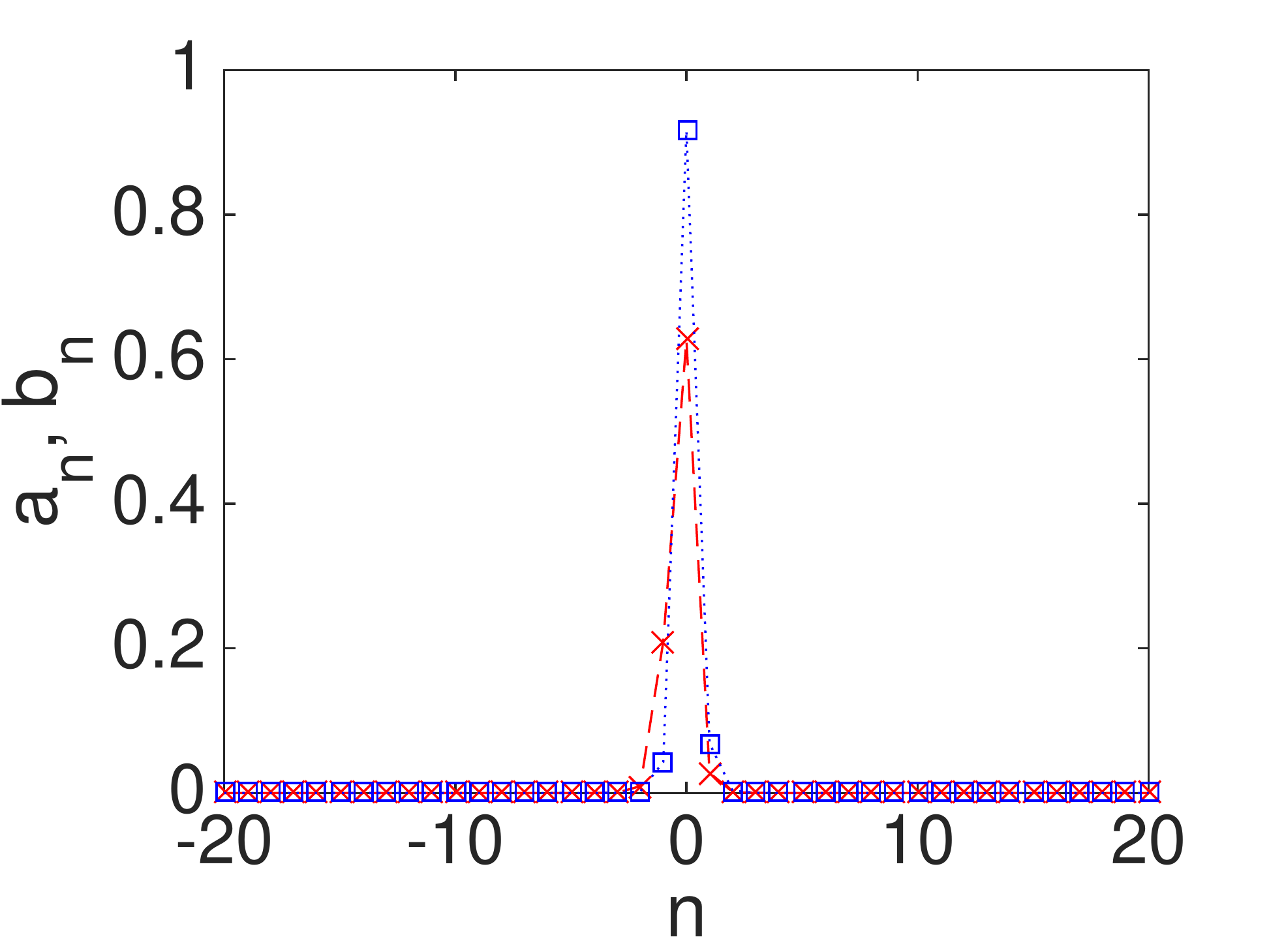}
 \includegraphics[width = 0.3\textwidth]{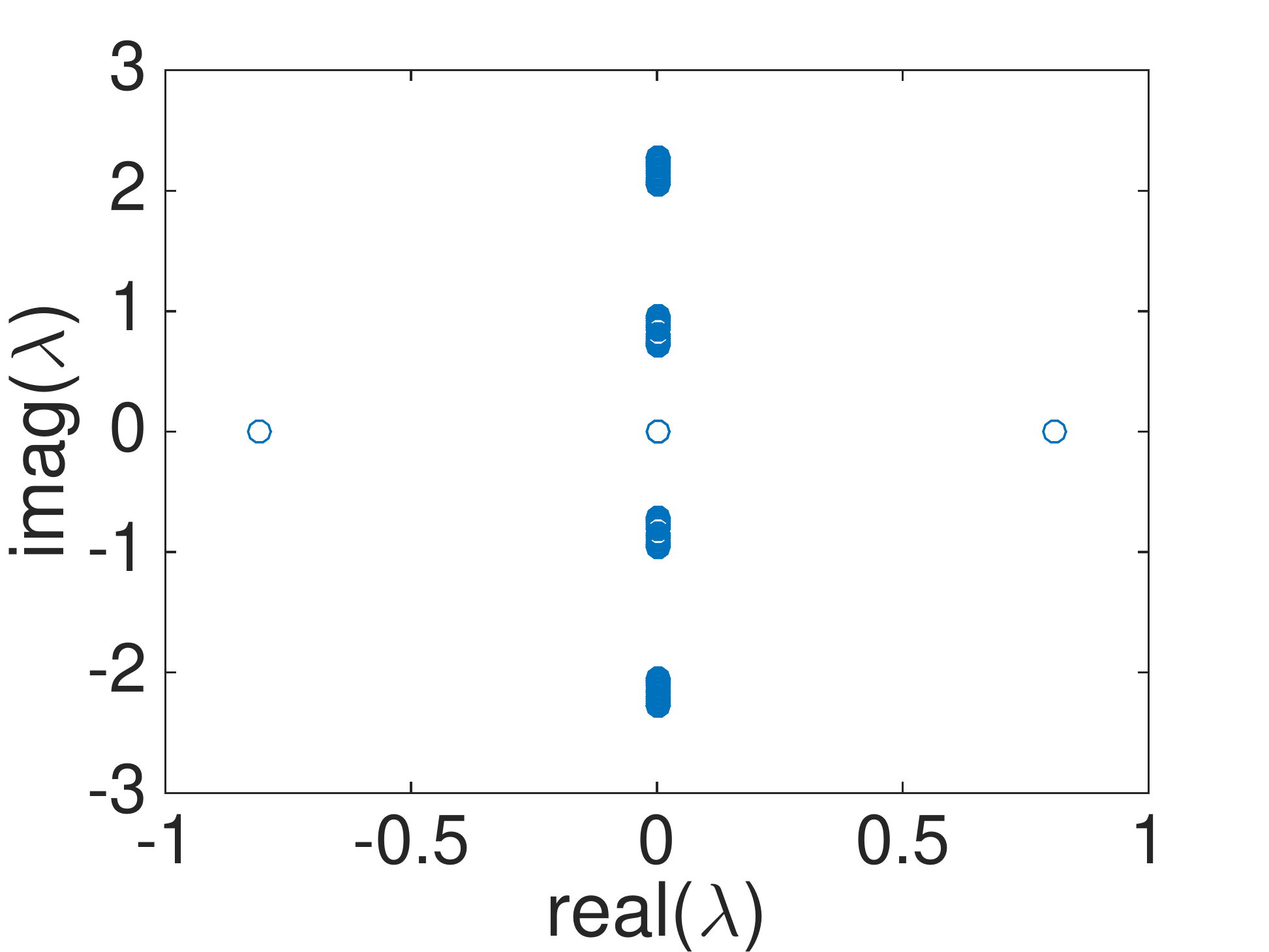}
 \includegraphics[width = 0.3\textwidth]{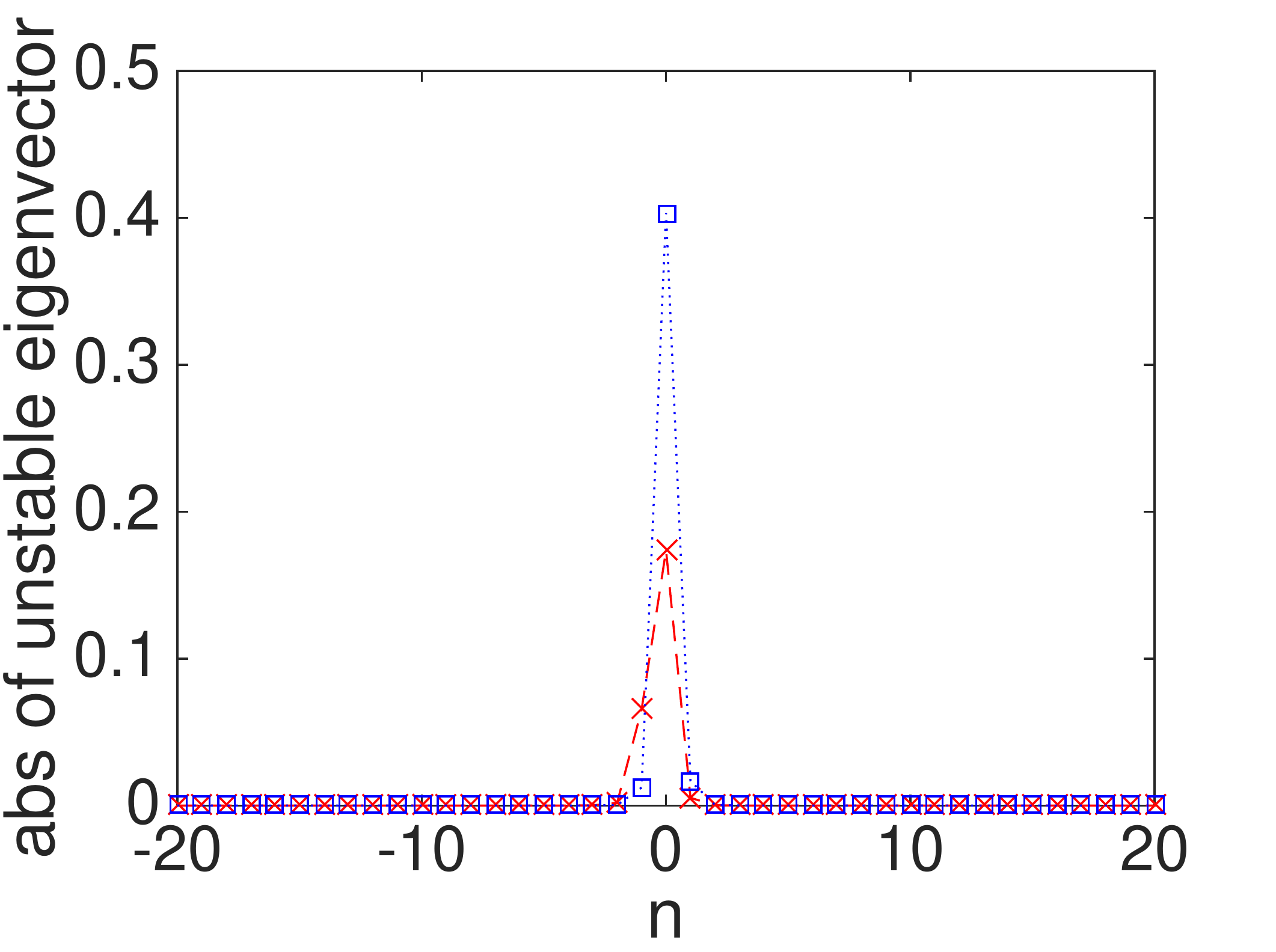}
 \includegraphics[width = 0.3\textwidth]{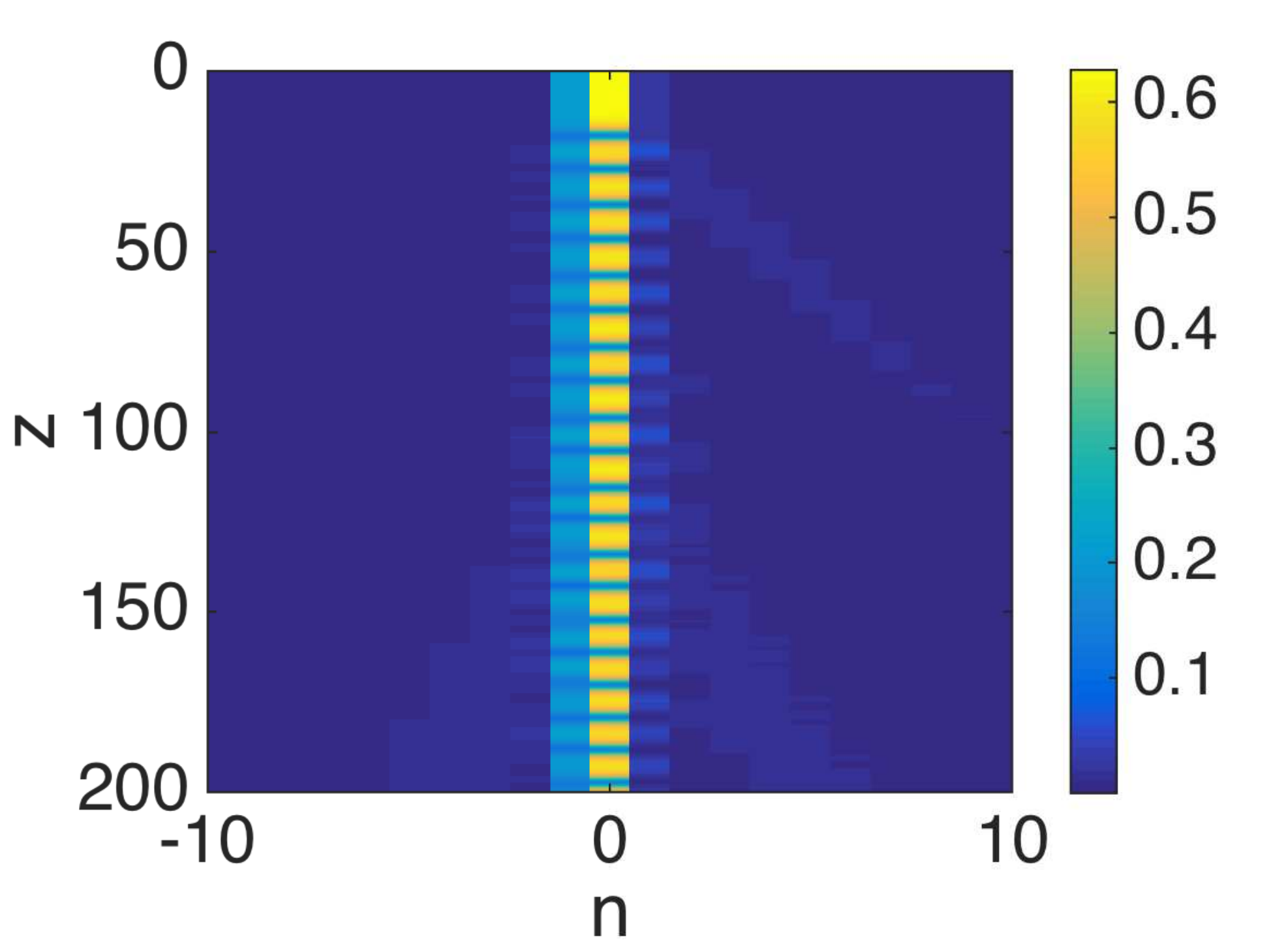}
 \includegraphics[width = 0.3\textwidth]{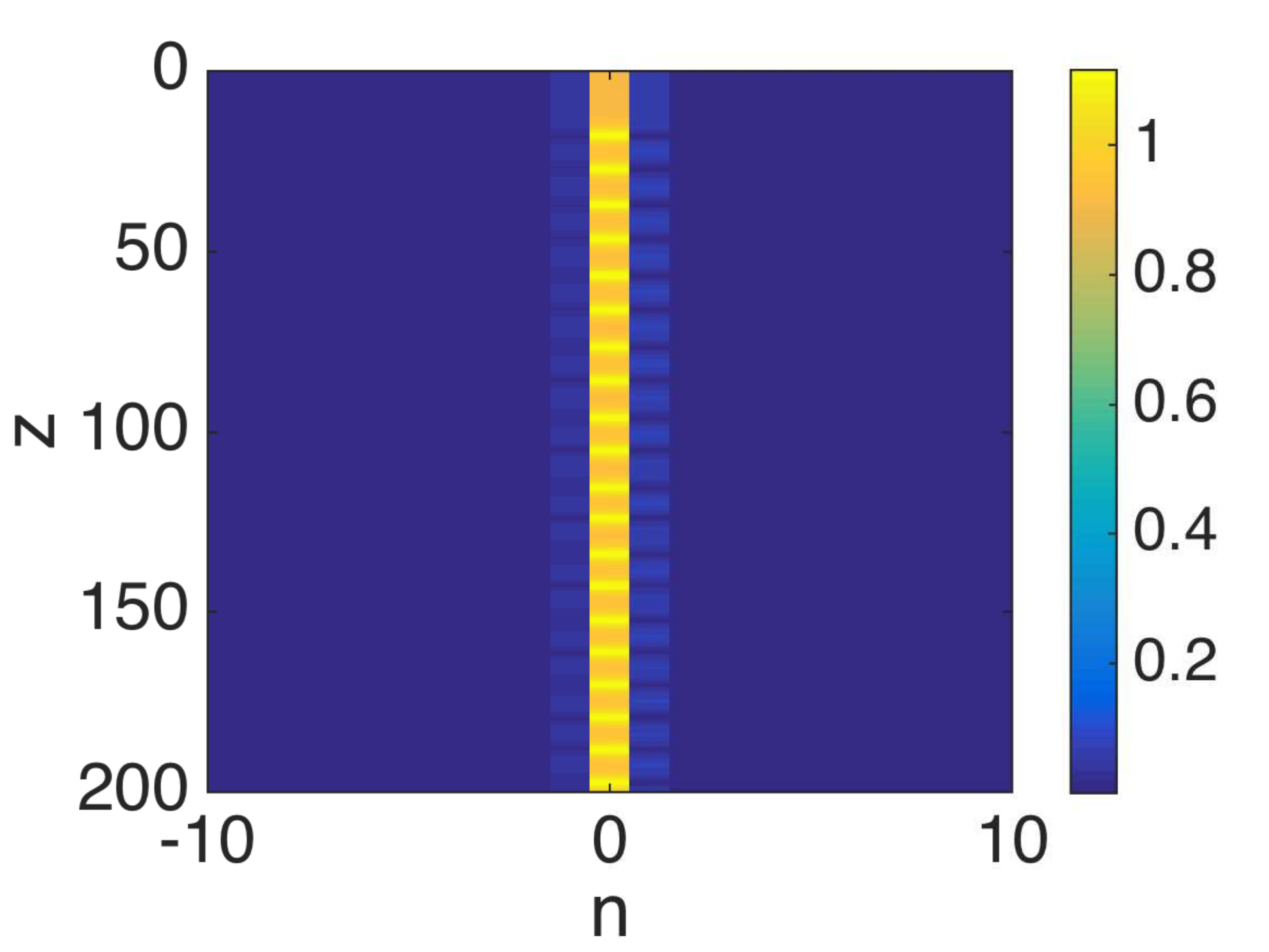}
 \includegraphics[width = 0.3\textwidth]{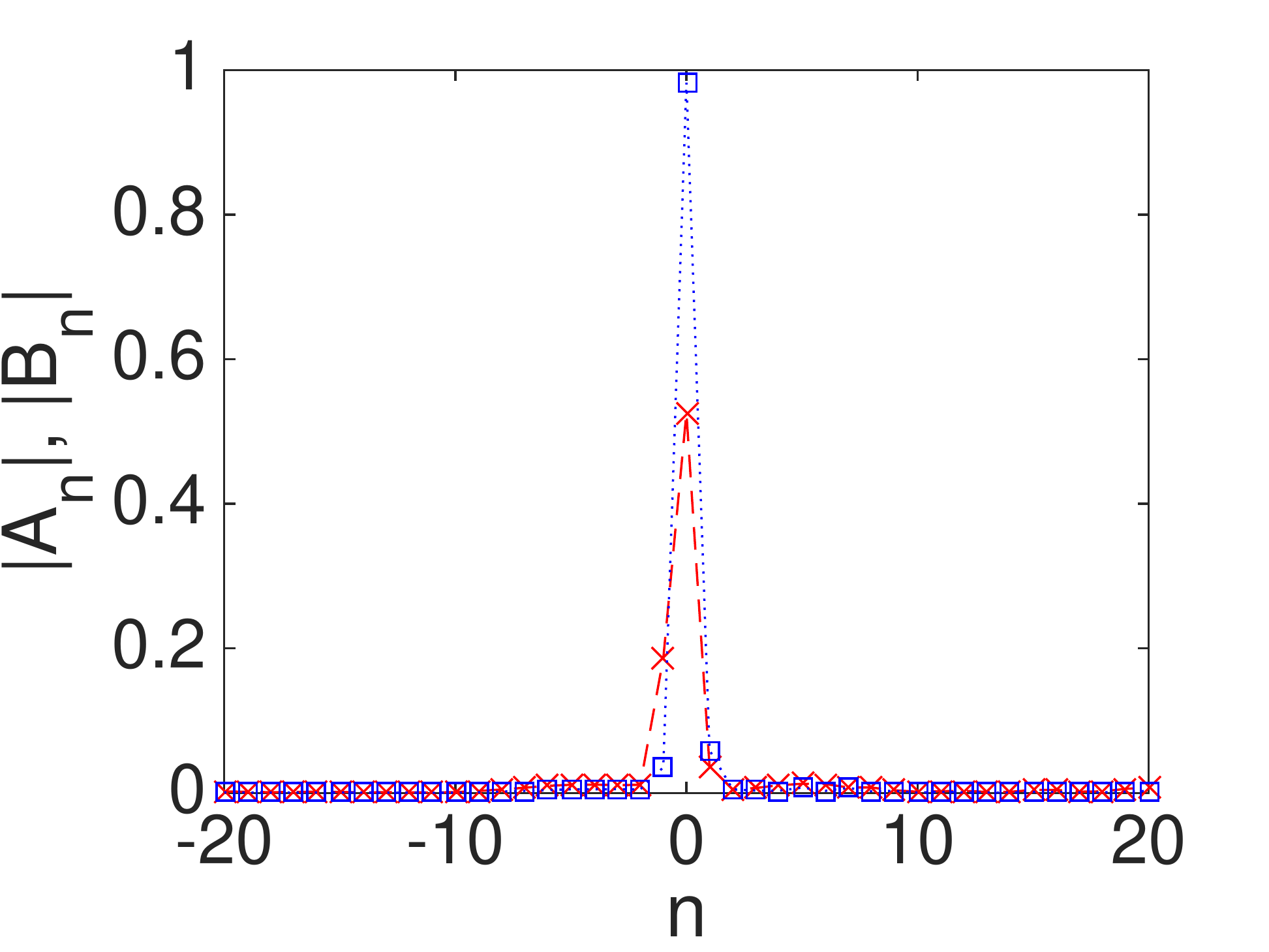}
 \caption{The stationary solution profiles of $a_n$ in red crosses and $b_n$ in blue squares is shown.  Starting from $\eps = 0$ on the top left panel, %with phase $(\cdots, c_{-1}, d_0, c_0, d_{+1},\cdots)= ( \cdots, 0, \pi, \pi,0,\cdots )$, 
 continuously increasing  up to $\eps=0.2$, we find the stationary solution in the middle left panel and its linear stability on the central middle 
row panel.  In top right panel, the real part of the relevant 
unstable eigenvalue is shown as a function of $\eps$ by the solid line, and the dashed line indicates its leading order theoretical approximation from
the analysis of section IV. At $\eps = 0.2$, if we perturb the stationary solution with the unstable eigenvector (in middle right panel) with an amplitude of $0.1\%$ of the stationary solution, we find the evolution of $|A_n|$ and $|B_n|$ 
%breath into a relative stable periodic pulse as show in the bottom left and middle panel respectively, and 
shown as a contour plot in the spatial ($n$) and evolution ($z$) variables 
in the bottom left and bottom middle panels.
The profiles at $z=200$ are shown in the bottom right panel.}
 \label{1in}
 \end{figure}

\begin{figure}[!htbp]
 \centering
 \includegraphics[width = 0.3\textwidth]{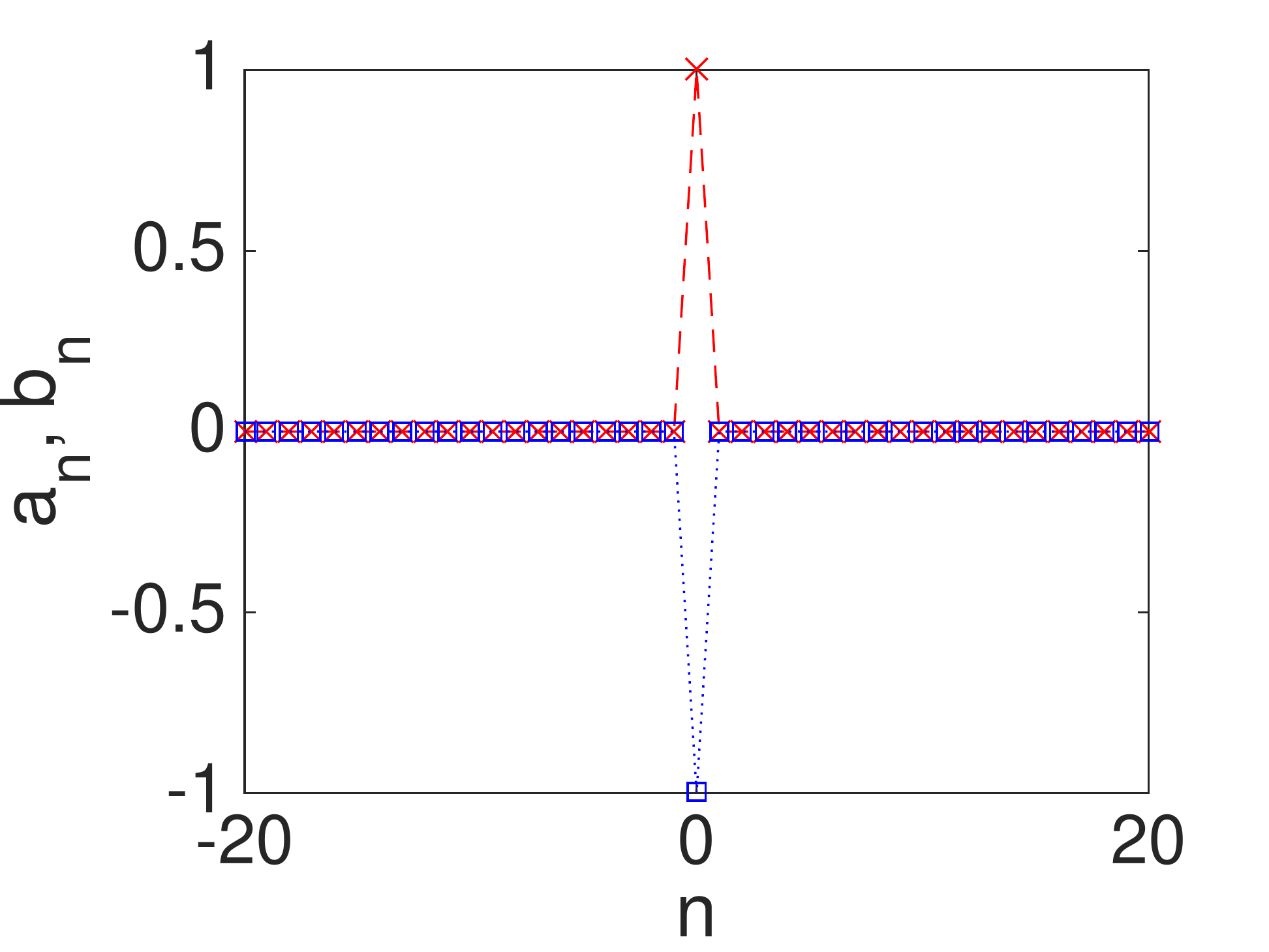}
 \includegraphics[width = 0.3\textwidth]{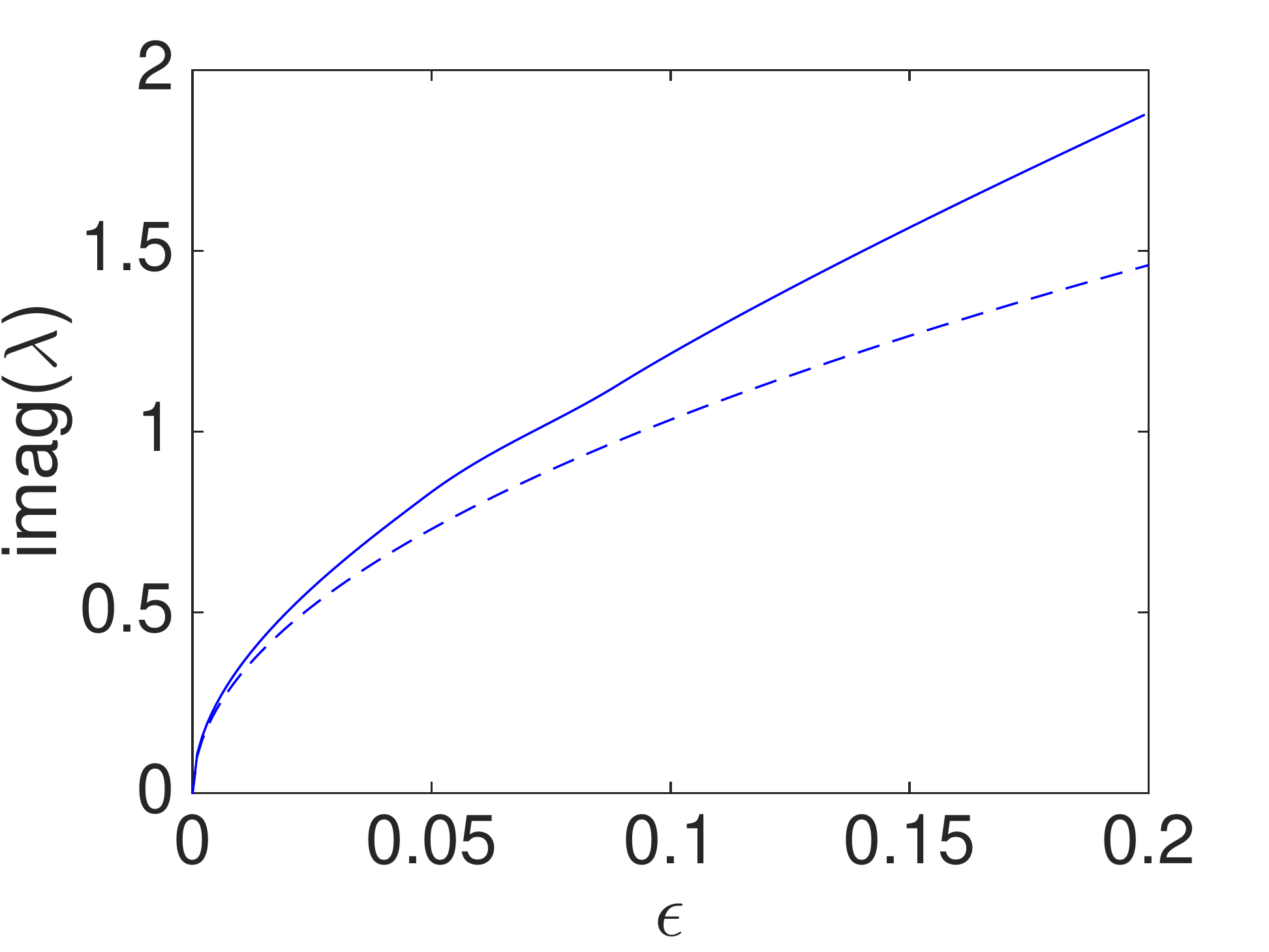}
 \includegraphics[width = 0.3\textwidth]{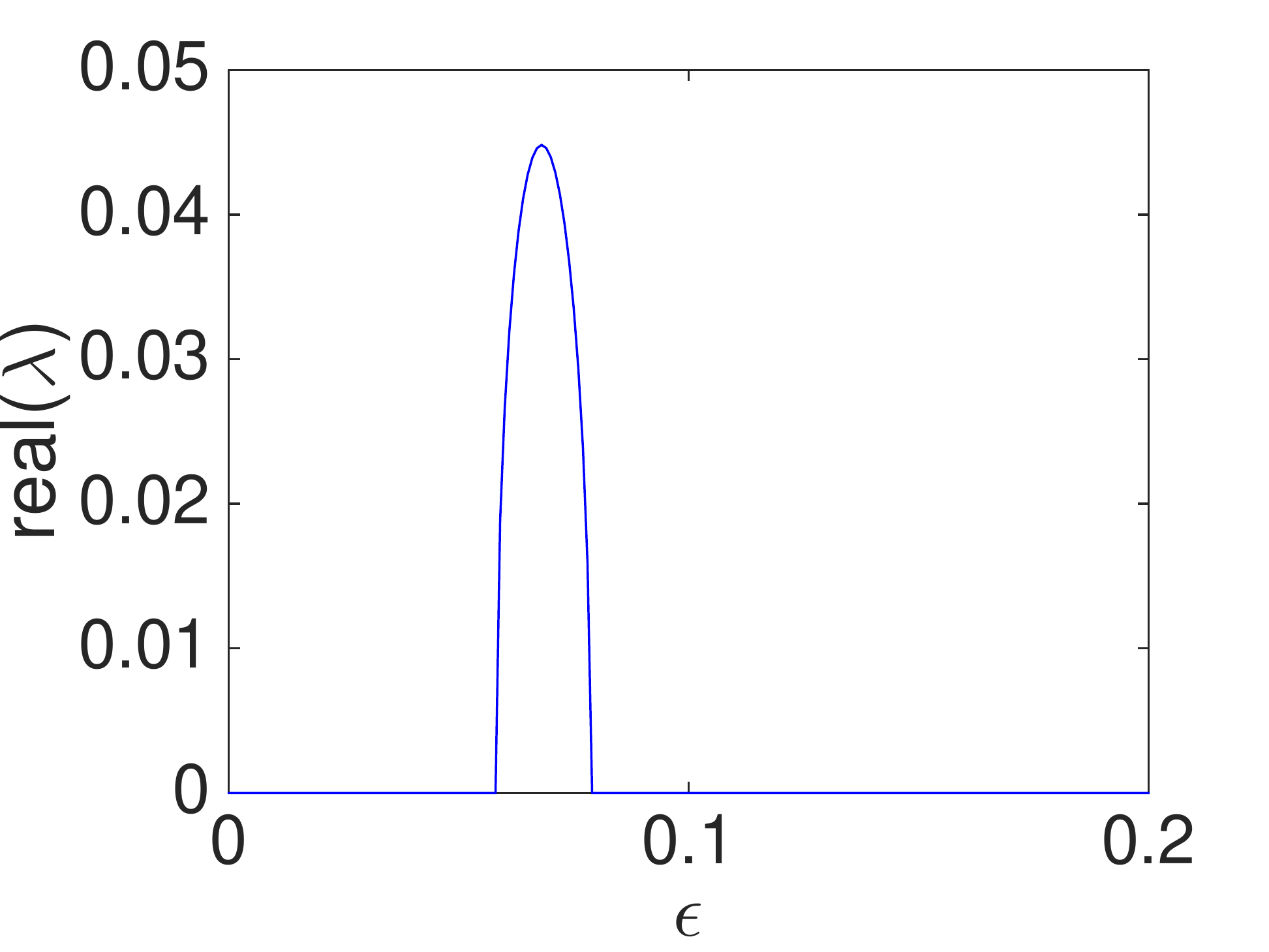}
 \includegraphics[width = 0.3\textwidth]{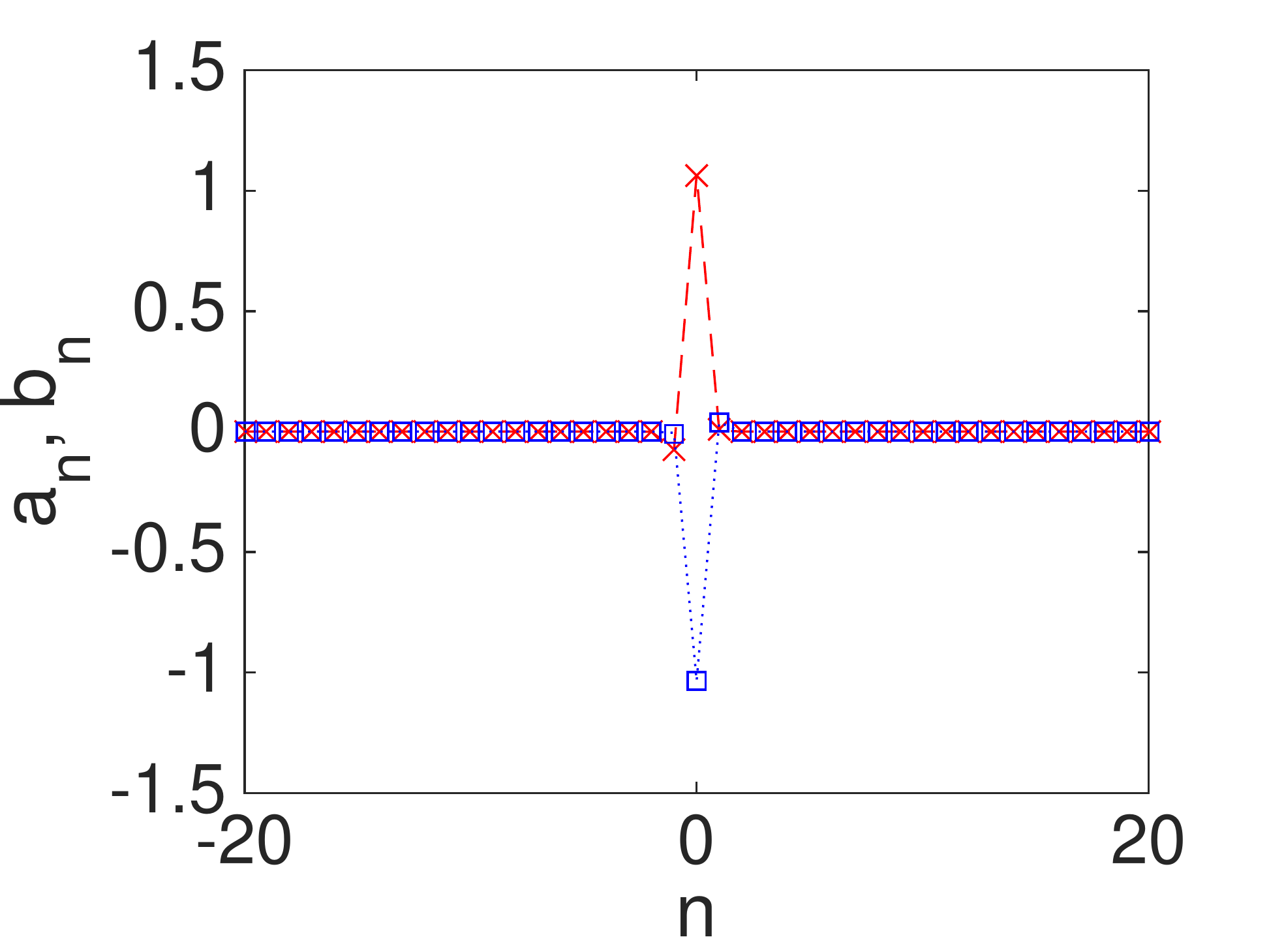}
 \includegraphics[width = 0.3\textwidth]{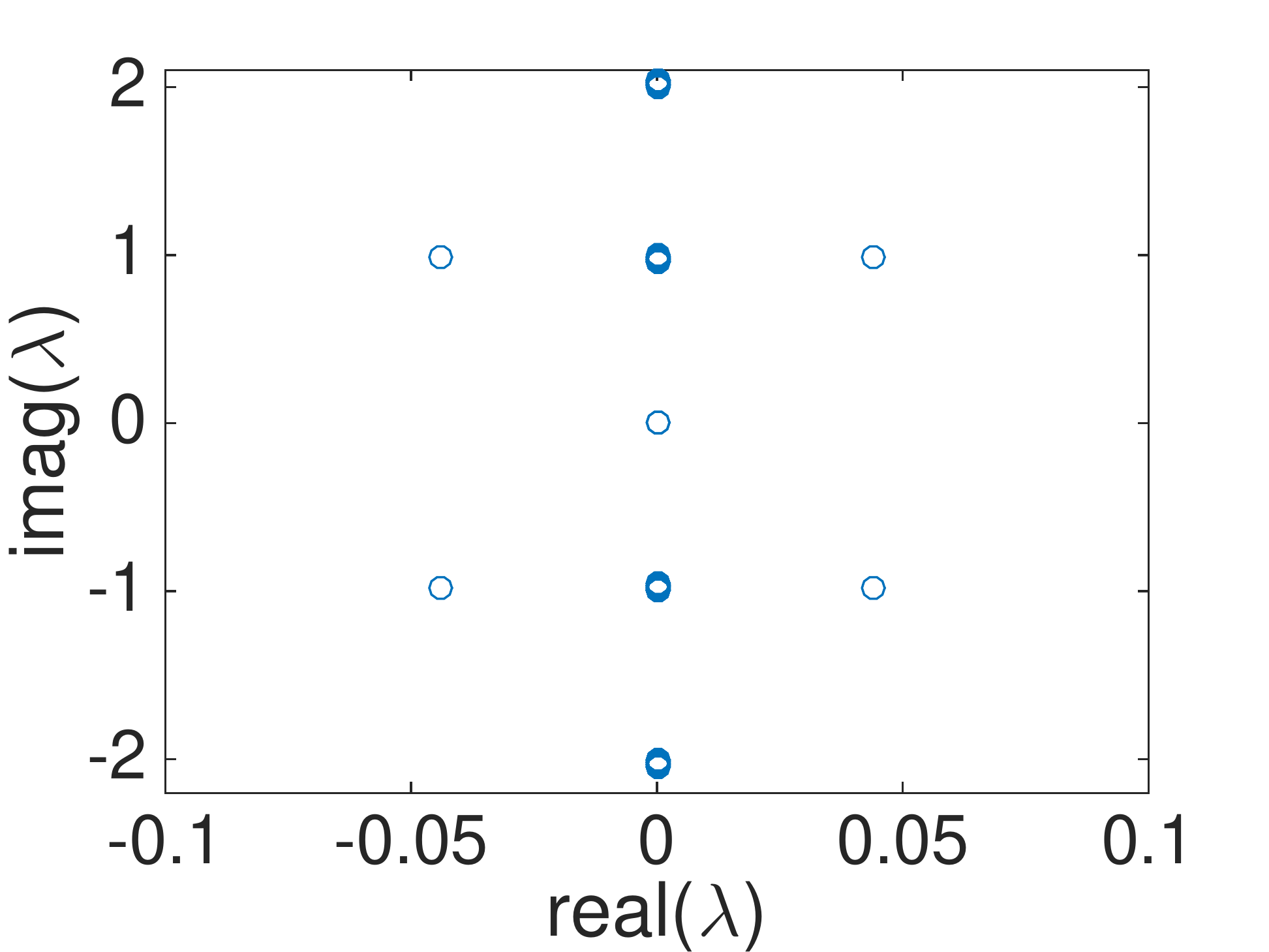}
\includegraphics[width = 0.3\textwidth]{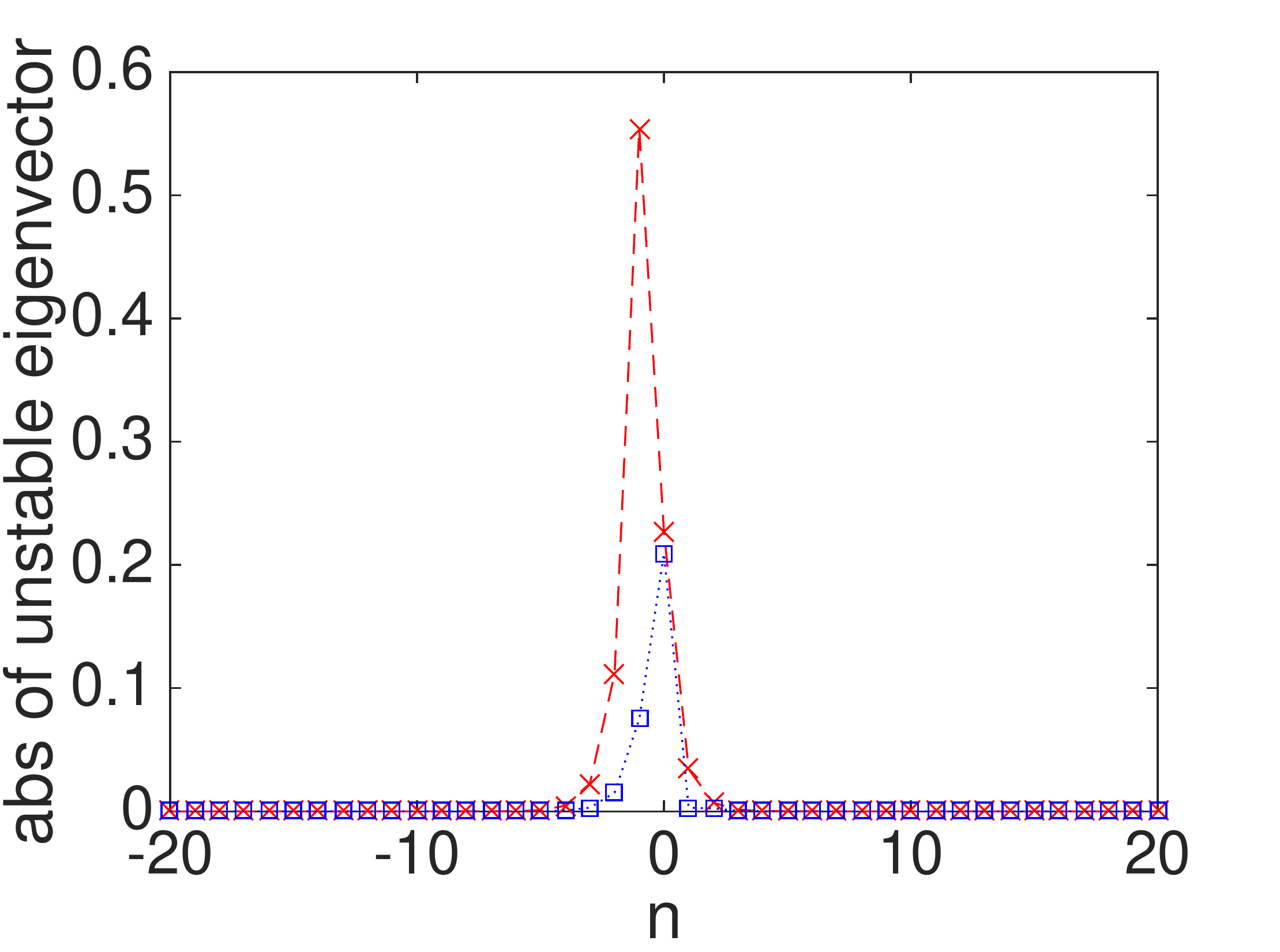}
 \includegraphics[width = 0.3\textwidth]{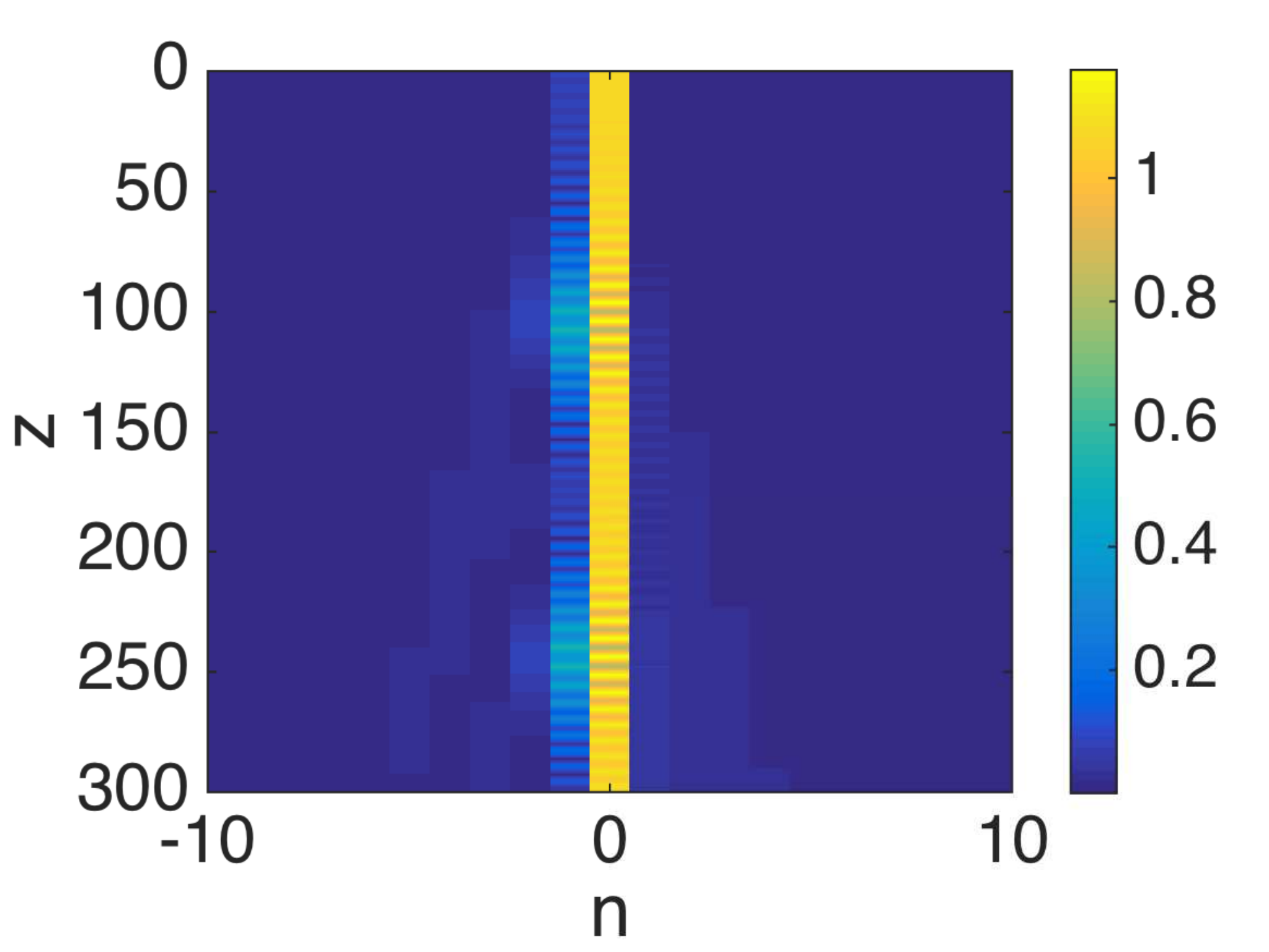}
\includegraphics[width = 0.3\textwidth]{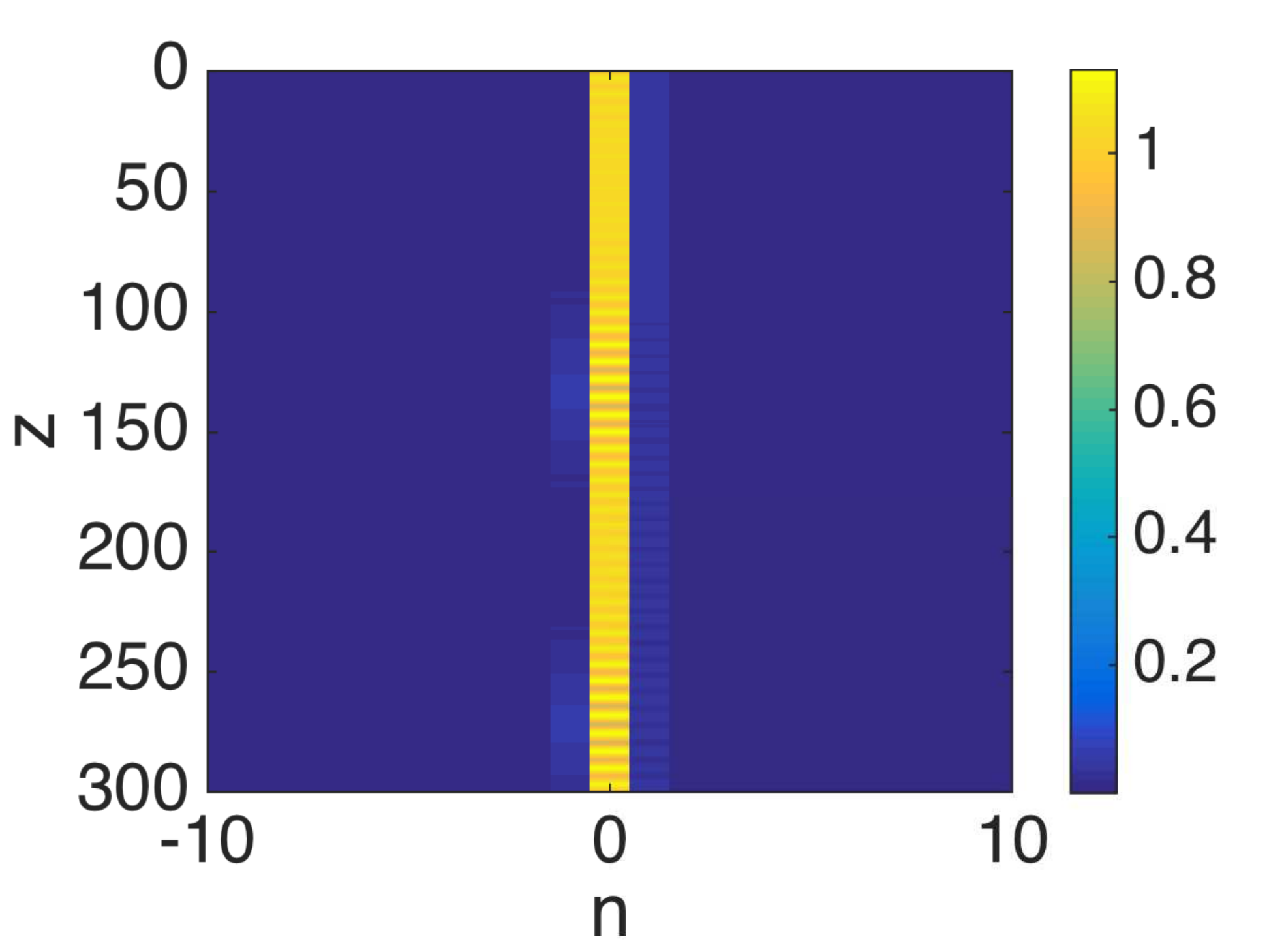}
\includegraphics[width = 0.3\textwidth]{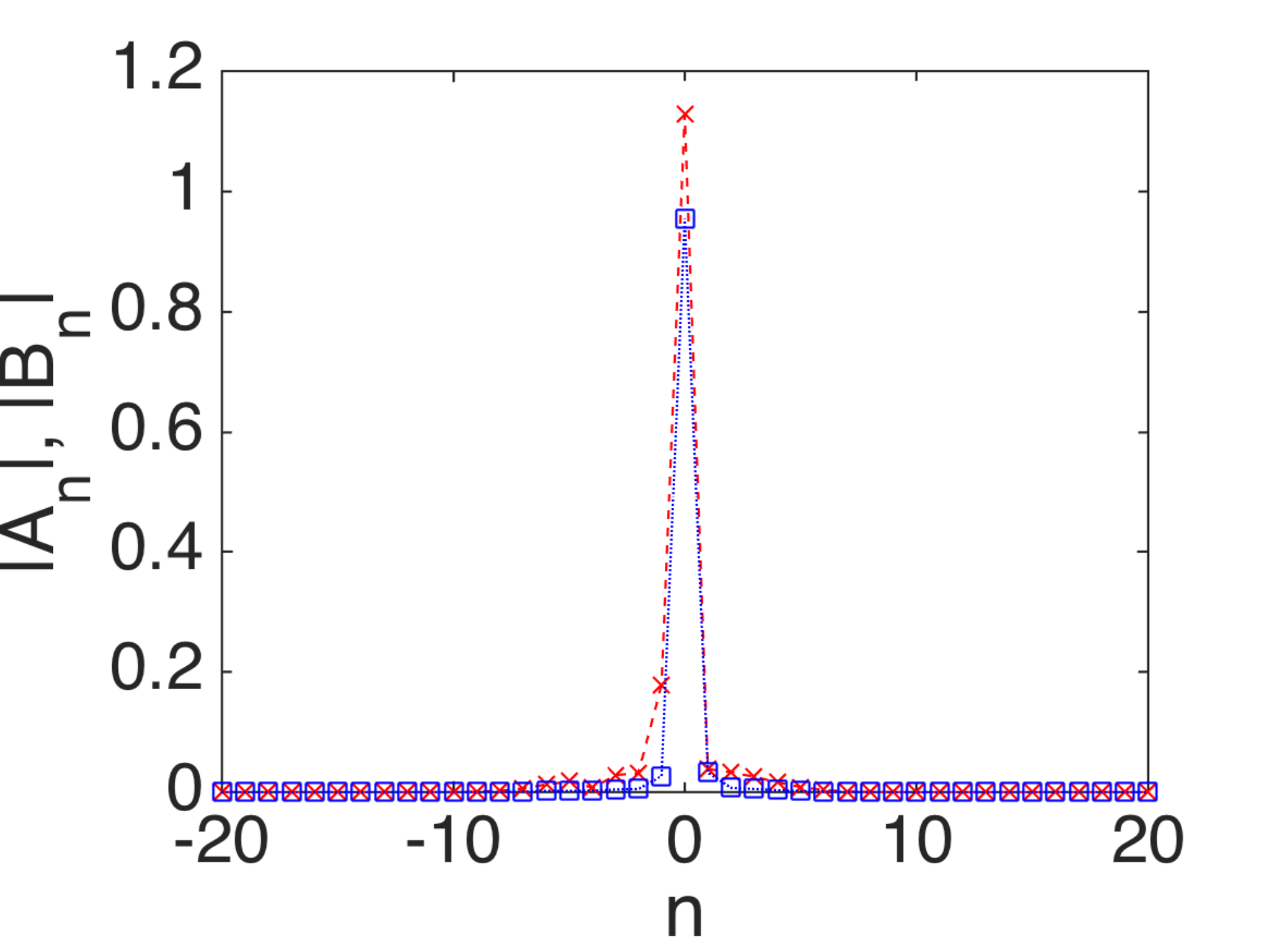}
 \caption{Similar to Fig.~\ref{1in}, but now for the case
where the two excited sites for $a_n$ and $b_n$ are out of
phase.  Starting from $\eps = 0$ on the top left panel, we
continuously increase $\eps$ up to $\eps=0.2$. The real and imaginary part 
of the relevant eigenvalue that bifurcates from zero are shown in top middle and top right panels respectively, where the dash line is the leading order theoretical approximation of Section IV. As $\eps$ increases, 
a pair of eigenvalues bifurcates from zero and moves along the imaginary axis.  For $0.058<\eps<0.079$, the imaginary eigenvalues collide with the continuous spectrum and yield a complex quartet (associated with an oscillatory
instability --hence the existence of a nonzero real part in the top right panel
within this interval--). The stationary solution at $\eps=0.07$  is shown in middle left panel. The spectrum of  linear stability and the most unstable
eigenmode are observed in the central and the right panel of the middle
row. Finally, with  perturbation amplitude  $10\%$ of the stationary state, the unstable dynamics of the state is shown in the bottom panels via contour plots (as in Fig.~\ref{1in}), and the  
modulus profile at the final propagation distance of $z=300$.}
 \label{1out}
 \end{figure}

As shown in both Fig. \ref{1in} and Fig.~\ref{1out}, 
a pair of eigenvalues will bifurcate from zero as we turn on the 
coupling $\eps$. In the case in which the two excited sites 
have the same sign 
when $\eps=0$ in Fig. \ref{1in},  as we increase $\eps$, the pair of eigenvalues move along the real axis as predicted by the leading order and 
the stationary solution becomes unstable. This is reminiscent of the 
standard DNLS
case of $C_1=1$~\cite{Kevrekidis}, 
although we should point out based on the above results that this
would change via a change of the sign of $C_1$. The middle 
panels of Fig.~\ref{1in} from left to right show, for $\eps = 0.2$, 
the stationary solution, its corresponding linearization spectrum and the 
unstable eigenvector respectively. 
Here, it is evident that in addition to the zero eigenvalue-pair 
(due to the phase invariance) and the real pair 
(the instability discussed above), two bands of continuous spectrum are 
forming, per the dispersion relation discussed in Eqn. ({\ref{cs}}).

If we now perturb the stationary solution along the unstable eigendirection 
by initially adding to it a small (amplitude of 0.1\%) contribution of 
the unstable eigenvector, we obtain the dynamical evolution shown 
in the bottom panels
of Fig.~\ref{1in}. The contour plots in the spatial ($n$) and
the propagation variable ($z$), clearly shows that the instability evolves
into a robust periodic orbit. Since we are only
showing the modulus of the waveform here (absorbing an additional
$e^{-i \beta z}$ factor), this suggests that the nature of the solution
is quasi-periodic. %This is a generic finding in our results in what follows.
%get $A_n$, $B_n$ actually render itself to a periodic solution \textbf{(I have tried smaller perturbations all the way to the order of $10^{-4}$, we get similar results)}.  

On the other hand, in Fig.~\ref{1out} the two components are out of phase when uncoupled. As we increase $\eps$, a pair of eigenvalues bifurcates 
from zero and moves along the imaginary axis,  so the stationary solution is stable; again this is in line with the DNLS result of~\cite{Kevrekidis}
which is hereby generalized for $C_1 \neq 1$. 
At $\eps\approx 0.058$ the purely imaginary discrete pair of
eigenvalues collides with the first band edge of the continuous 
spectrum and becomes a complex quartet, while the stationary solution becomes 
unstable, via the resulting oscillatory instability.  As the eigenvalue 
pairs return to the imaginary axis for $\eps \approx 0.079$, 
the solution becomes stable again until it reaches the second band edge of the continuum spectrum. So instead of being purely imaginary for
all $\eps$, here we see that 
the discrete eigenvalues become resonant with the continuous spectrum and
yield intervals of oscillatory instability, associated with complex
eigenvalue quartets, such as the one for $0.058<\eps<0.079$.  
Perturbing the stationary solution with $10\%$ of its amplitude in the 
unstable direction results in the dynamics shown in the bottom panels of 
Fig. \ref{1out}. Importantly,
the instability dynamics still results in a 
robust quasi-periodic breather type of waveform eventhough the behavior 
of the eigenvalues manifests. in a different manner.

\subsection{Three excited sites}
\begin{figure}[!htbp]
 \centering
 \includegraphics[width = 0.35\textwidth]{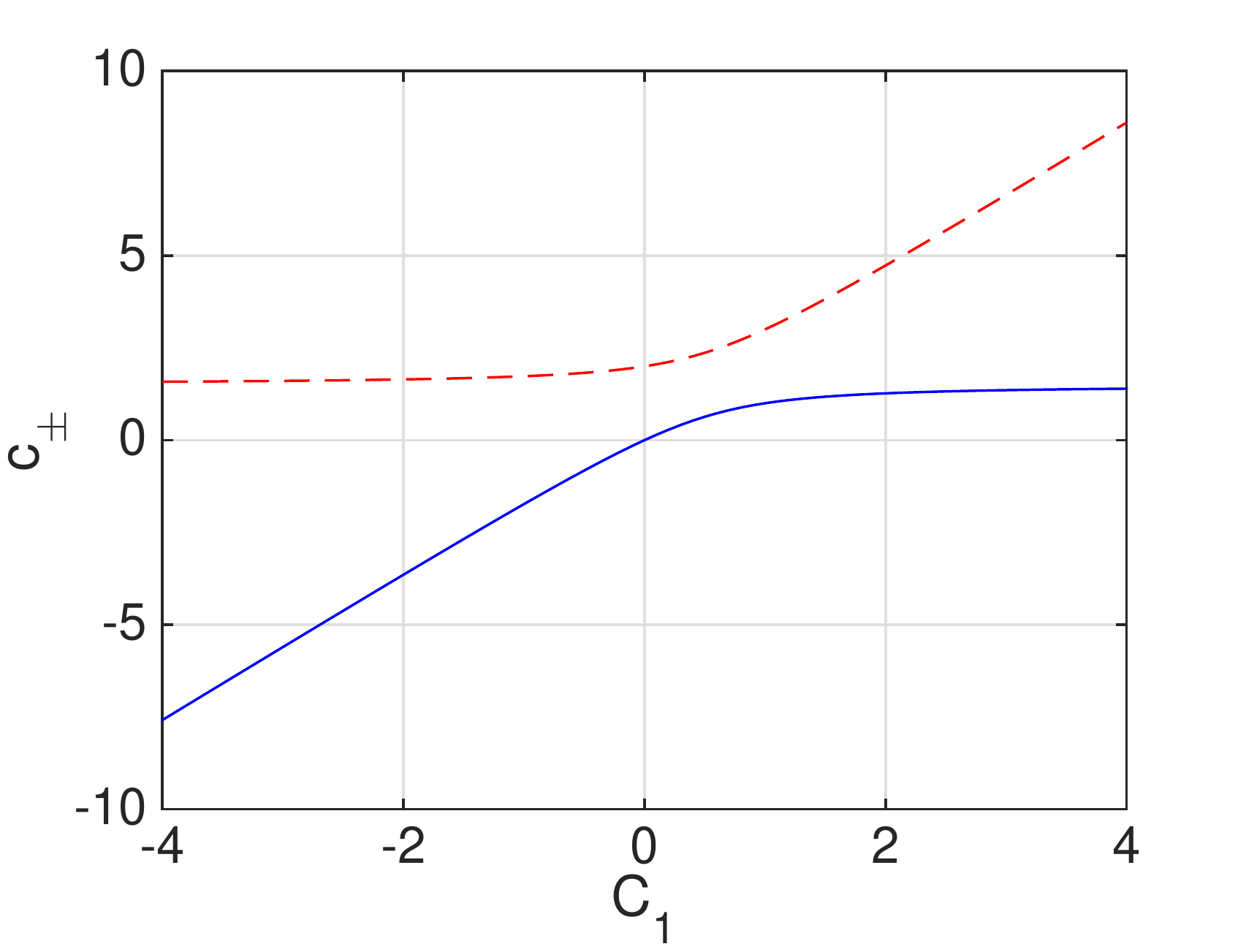}
 \includegraphics[width = 0.35\textwidth]{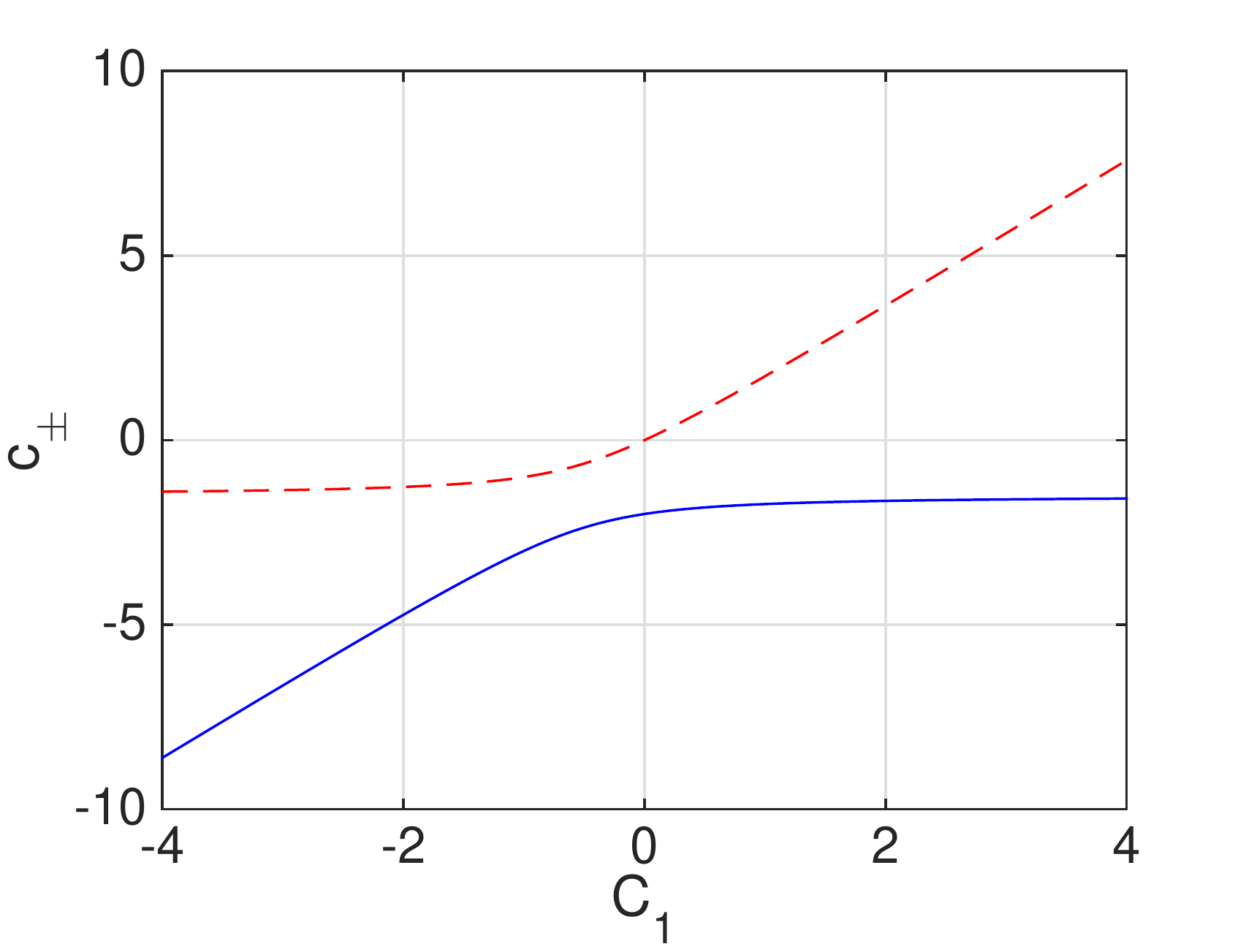}
 \caption{$c_{\pm}$ entering the expression of Eq.~(\ref{homoe}), is
given as a function of $C_1$. The left panel is for $s_2= 1$ and the right panel is for $s_2 = -1$. The red dashed line indicates $c_+$, while
the blue solid indicates $c_-$.}
 \label{cpm}
 \end{figure}

We now explore the case with three excited sites,
i.e., at the AC limit $(a_n, b_n)$ of the form:  
\beq
(a_n^{(0)}, b_n^{(0)}) = \left\{
\begin{array}{lll}
(ae^{ic_{-1}}, &0),& n=-1, \\ 
(ae^{ic_0},& be^{id_0}),& n=0, 
\end{array}
\right.
\eeq
From Eqn. (\ref{ab1}), the nonzero entries for the first order 
expansion of $(a_n, b_n)$ will be 
\beq
(a_n^{(1)}, b_n^{(1)}) = \left\{
\begin{array}{lll}
(\frac{be^{id_0}}{-2\gamma_a a^2},&\frac{C_1ae^{ic_{-1}}}{\gamma_b b^2}),& n=-1, \\
(\frac{C_1be^{id_0}}{-2\gamma_a a^2}, &\frac{ae^{ic_{-1}}+C_1a e^{ic_0}}{-2\gamma_b b^2}),& n=0, \\
(0,&\frac{ae^{ic_0}}{\gamma_bb^2}),& n=1.
\end{array}
\right.
\eeq
Using Eqn. (\ref{M}), denoting $s_1 = e^{i(d_0-c_0)} $, $s_2 = e^{i(c_0-c_{-1})}$, we find
\bes
{\bf M} = s_1
\begin{pmatrix}
\frac{b}{a}s_2& 0&-s_2\\
0 & \frac{C_1b}{a}&-C_1\\
-s_2&-C_1&\frac{a}{b}(C_1+s_2)
\end{pmatrix}.
\ees
 Again, as in the case of two excited cases,  $\bf M$ has a $0$ eigenvalue with eigenvector $(a,a, b)^T$ that corresponds to the phase invariance
of the underlying binary waveguide model. The other two eigenvalues are 
 \beq \mu_\pm =s_1 \frac1  2 \Big[(a/b + b/a)(C_1+s_2)\pm \sqrt{(a/b + b/a)^2(C_1+s_2)^2- 4s_2C_1( b^2/a^2+2) }\Big]\eeq
  with corresponding eigenvectors
 $(\frac{1}{b/a-s_2 \mu_\pm}, \frac{C_1}{C_1 b/a -\mu_\pm},1) $.
So according to Eqn. (\ref{lambda2}), the leading order of $\lambda^2$ will be
\beq\label{3ll}
\lambda^2 \approx \eps \frac{\mu_\pm (\frac{1}{(b/a-s_2\mu_\pm)^2}+\frac{C_1^2}{(C_1b/a-\mu_\pm)^2}+1)}{\frac{1}{(b/a-s_2\mu_\pm)^2(2\gamma_a a^2)}+\frac{C_1^2}{(C_1b/a-\mu_\pm)^2(2\gamma_a a^2)}+\frac{1}{2\gamma_b b^2}}.
\eeq
In the simpler case of $\gamma_a = \gamma_b = \gamma$, and $a = b$, 
\beq\label{homoe}
\lambda^2\approx\eps2a^2 \gamma  s_1[(C_1+s_2)\pm\sqrt{(C_1+s_2)^2-3s_2C_1}] =:\eps2a^2 \gamma  s_1 c_{\pm},
\eeq
 where  $\lambda^2$ depends on $c_{\pm}$, in addition to
the product $\eps2a^2\gamma s_1$. In order to gauge the role of
$C_1$ in modifying the relevant eigenvalues from the DNLS limit
of $C_1=1$, 
Fig. \ref{cpm} shows $c_{\pm}$ as increasing functions of $C_1$. The graphs have an asymptotic behavior described by the straight 
lines $y = s_2$ and $y = 2x+s_2$. It can again be seen that $C_1$ plays
a critical role in the stability properties, with its sign variation inducing
a change from real to imaginary of one of the relevant eigenvalue
pairs. It is interesting to point out that generally, increasing the  magnitude of
the parameter $C_1$ leads to an increase of $c_{\pm}$ rendering
the configuration more prone to potential instabilities.

%which us the idea of how the stability changes w.r.t $C_1$. 

We now discuss the four different potential combinations of
signs of $s_1$, $s_2$.  We will denote  
$(a_{-1}^{(0)}a_0^{(0)},b_{-1}^{(0)}b_0^{(0)})$ only by the sign of each 
of the elements.
\begin{itemize}
\item $(++,0+)$: 

The stationary solution for $\eps=0$ is shown in top left panel of 
Fig.~\ref{31}. 
In this case $s_1=s_2=+1$.
This in phase configuration results into two pairs
of real eigenvalues.   
%As predicted by Eqn. (\ref{3ll}), there will be two unstable eigenvalues  bifurcate from zero. 
This can be confirmed by the top right panel of Fig. (\ref{31}), where 
we get two unstable eigenvalue pairs as functions of $\eps$ that are well 
predicted by their leading order approximations in dashed line, at least
for small values of $\eps$. 
When $\eps$ gets to be large, then the eigenvalues show a decreasing tendency
in their variation over $\eps$ suggesting that higher order terms become
significant.
At $\eps=0.2$, we find the stationary solution and its linear stability spectrum in the middle left and center panels respectively. If we perturb this stationary solution with the most unstable eigenvector in middle right panel, of amplitude $1\%$ of the stationary state, we get the evolution of $|A_n|$, $|B_n|$ shown in the bottom left and middle panels respectively. The profile of $|A_n|$, $|B_n|$ at $z = 200$ is 
shown in bottom right panel. Once again, we can observe the formation
of a robust periodic breathing state, in the (modulus) evolution dynamics.
%In fact, given the similarity of the rest of the evolution examples that we have considered to the ones shown above, we do not show any more surface plots.
%\bes
%M =  
%\begin{pmatrix}
%\frac{b}{a}& 0&-1\\
%0 & \frac{C_1b}{a}&-C_1\\
%-1&-C_1&\frac{a}{b}(C_1+1) 
%\end{pmatrix}
%\ees
%with eigenvalues $0$ and $\mu_\pm = \frac1  2 \Big[(a/b + b/a)(C_1+1)\pm \sqrt{(a/b + b/a)^2(C_1+1)^2- 4C_1( b^2/a^2+2) }\Big]$, the corresponding eigenvectors will be
%$(a,a, b)^T$, $(\frac{a}{b-a\mu_\pm}, \frac{C_1a}{C_1 b -a\mu_\pm},1) $.
%So the leading order of $\lambda^2$ will be
%\beq
%\lambda^2 \approx \eps \frac{\mu_\pm (\frac{a^2}{(b-a\mu_\pm)^2}+\frac{C_1^2a^2}{(C_1b-\mu_\pm)^2}+1)}{\frac{a^2}{(b-a\mu_\pm)^2(2\gamma_b b^2)}+\frac{C_1^2a^2}{(C_1b-\mu_\pm)^2(2\gamma_a a^2)}+\frac{1}{2\gamma_b b^2}}
%\eeq
%

\begin{figure}[!htbp]
 \centering
 \includegraphics[width = 0.35\textwidth]{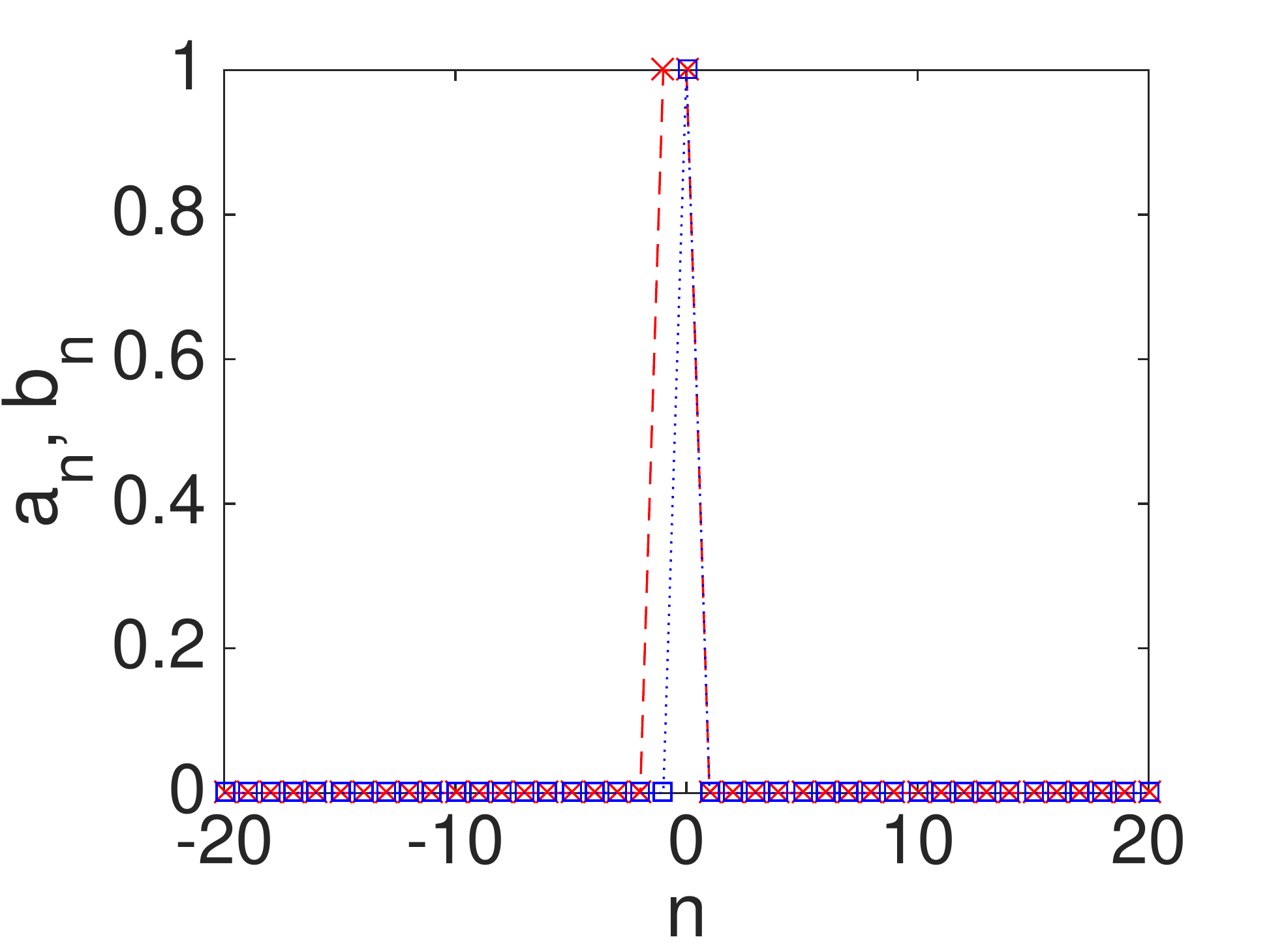}
 \includegraphics[width = 0.35\textwidth]{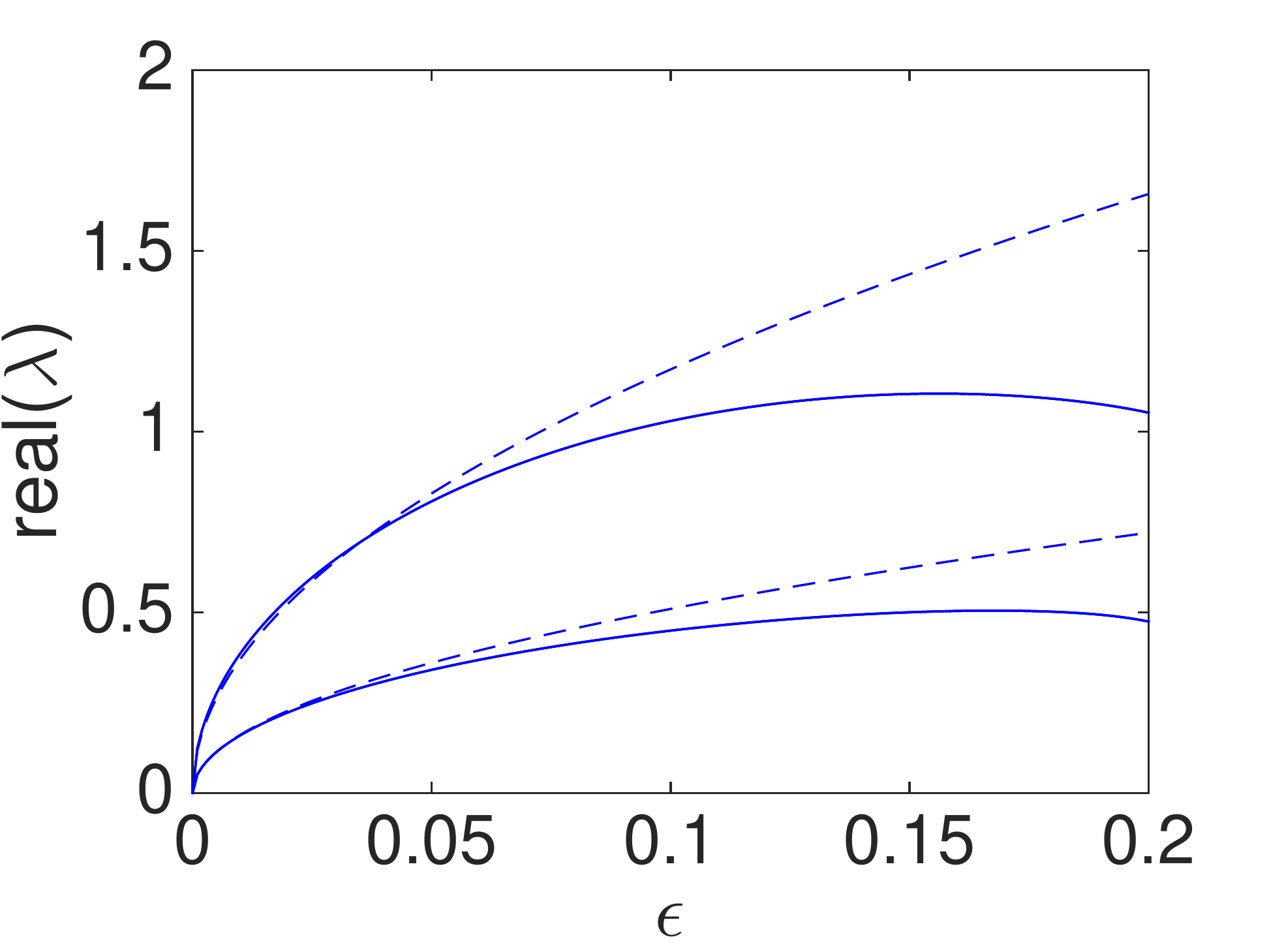}
 \includegraphics[width = 0.3\textwidth]{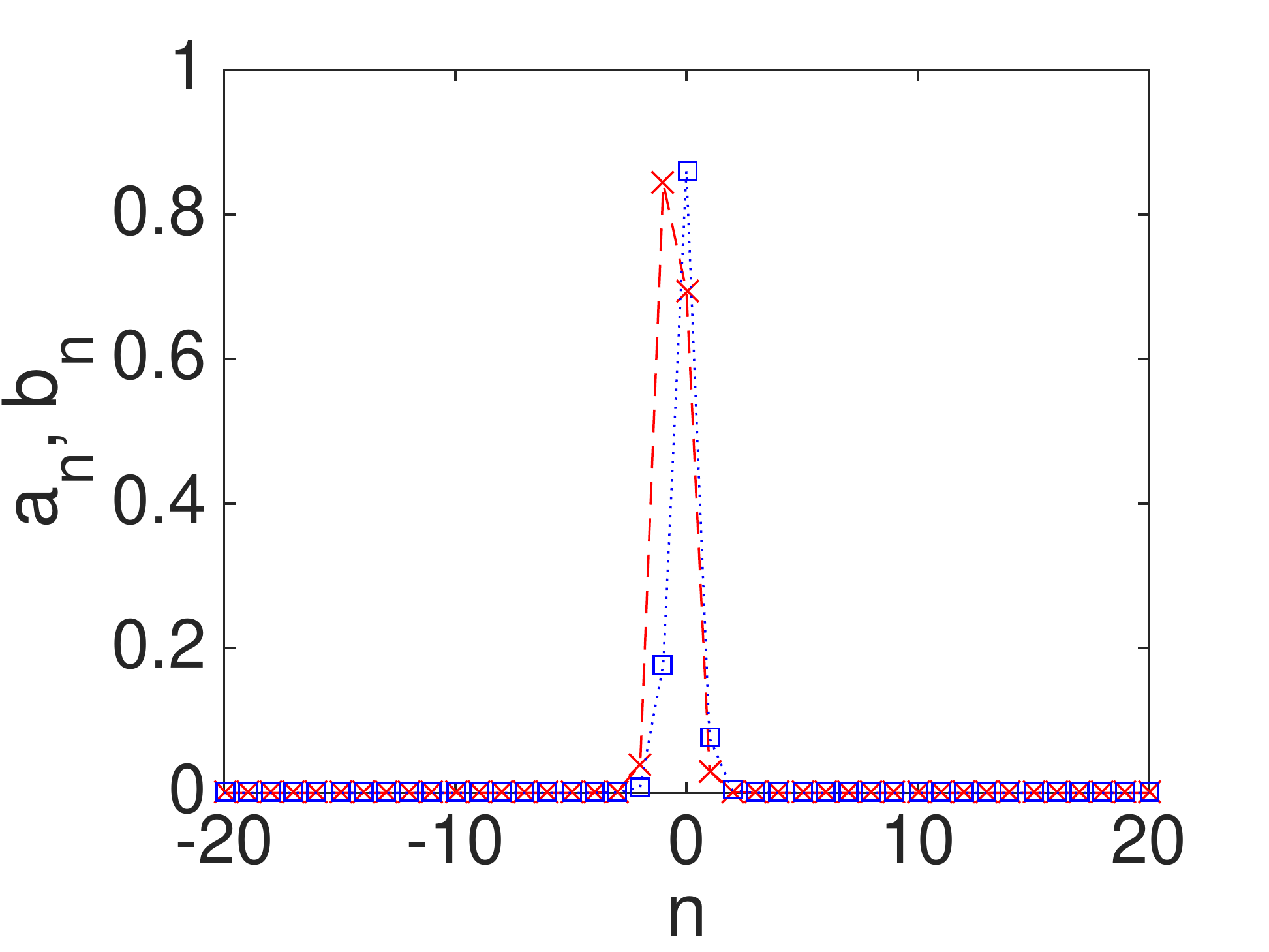}
 \includegraphics[width = 0.3\textwidth]{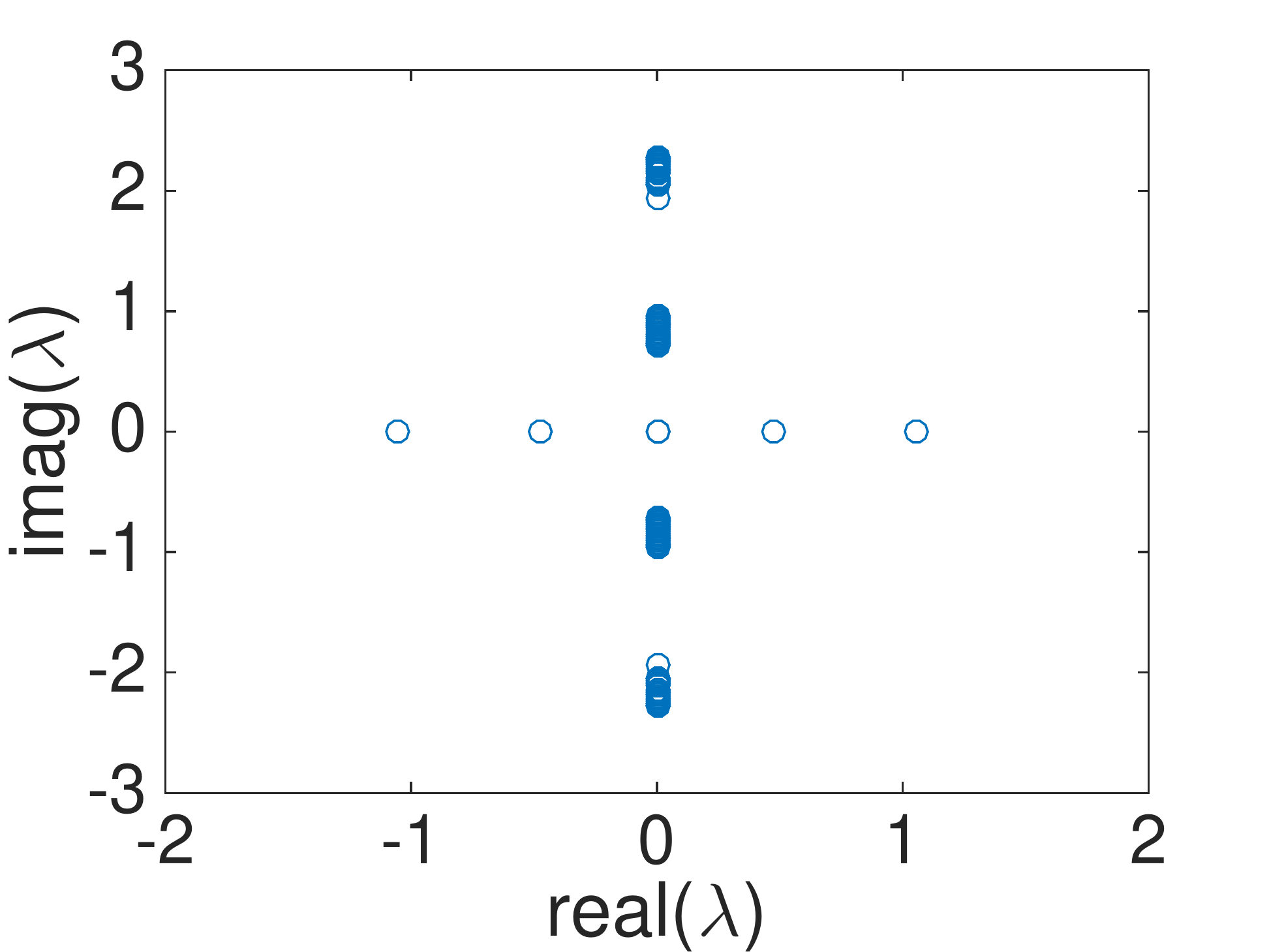}
 \includegraphics[width = 0.3\textwidth]{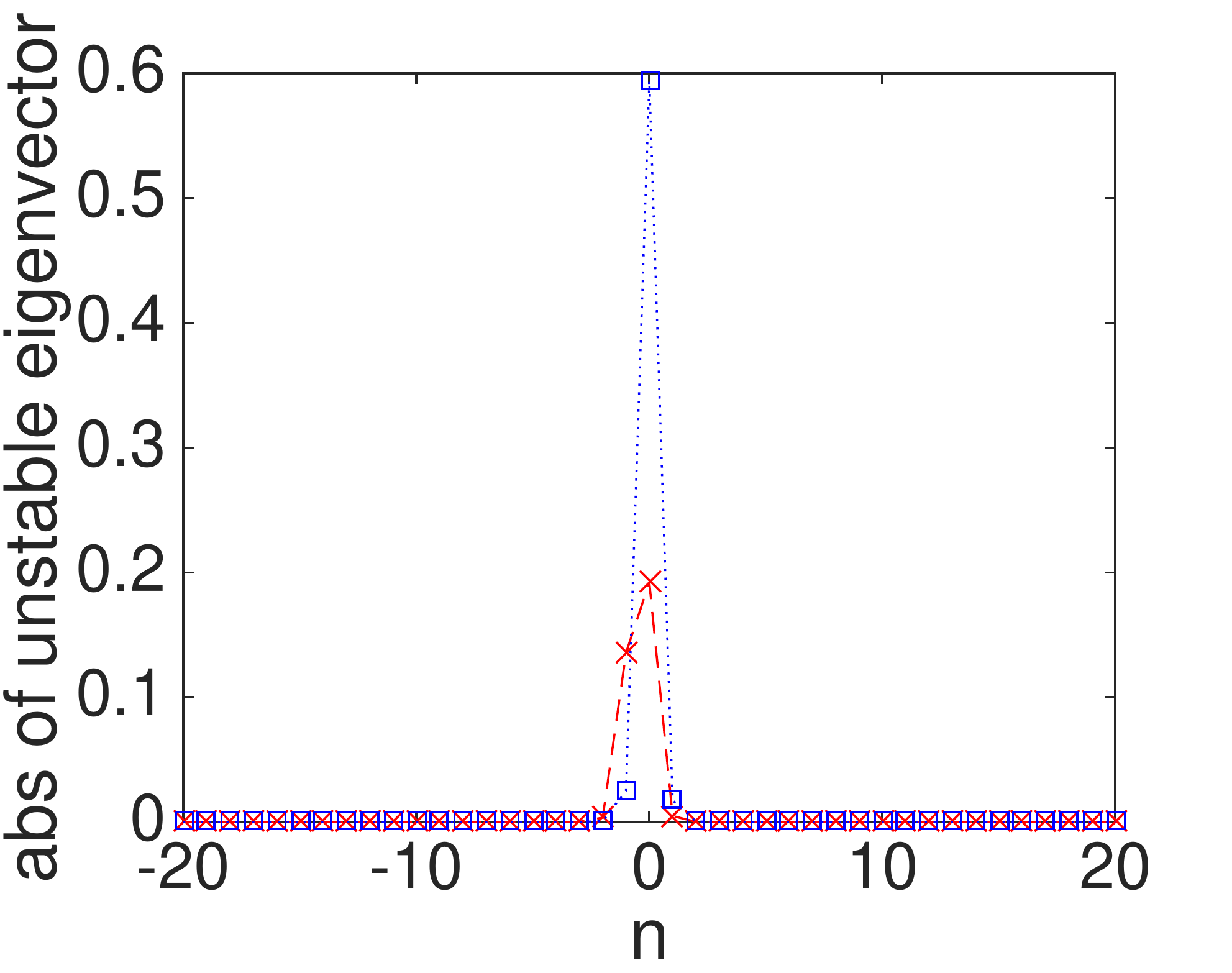} \\
 \includegraphics[width = 0.3\textwidth]{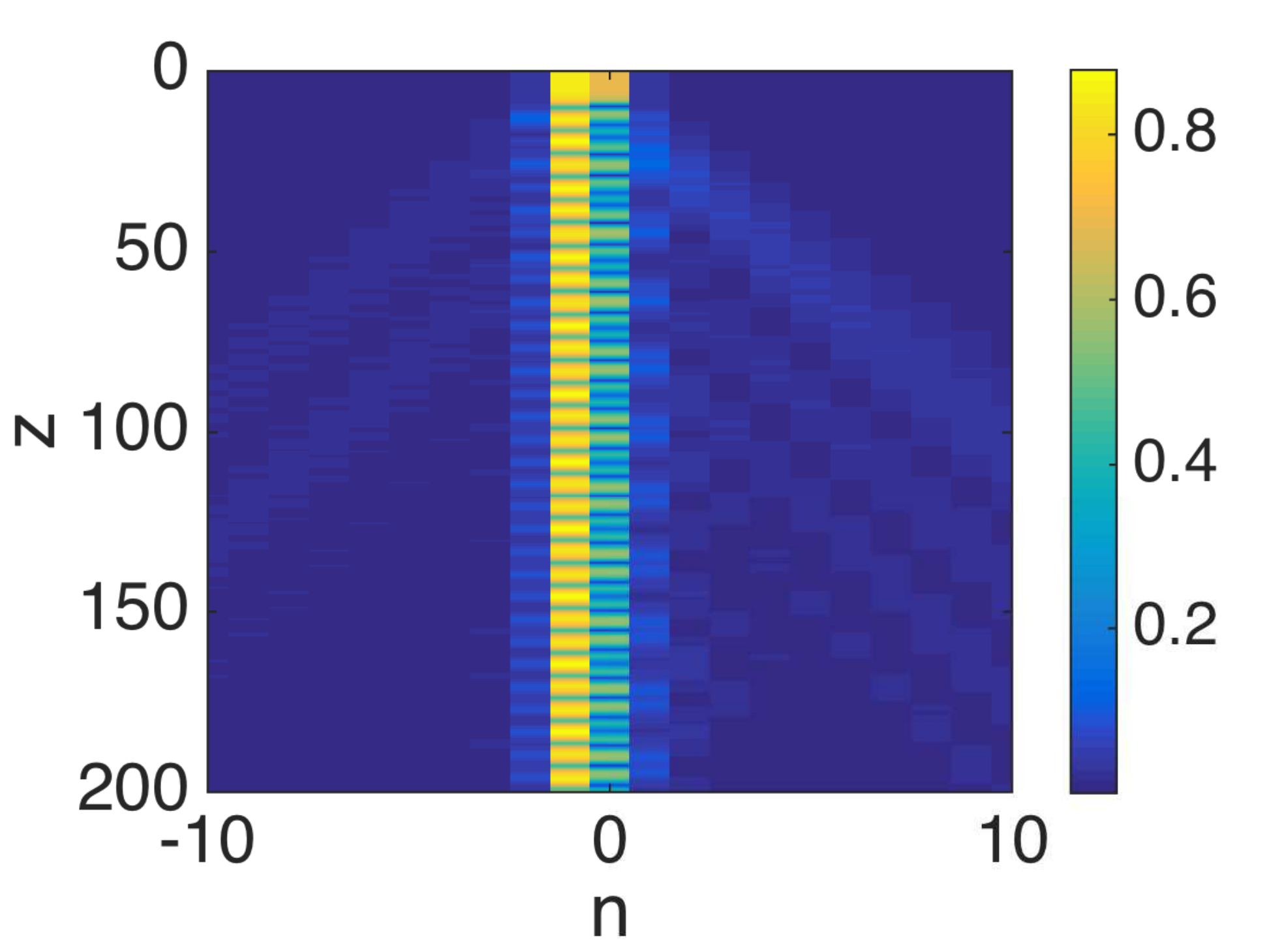}
 \includegraphics[width = 0.3\textwidth]{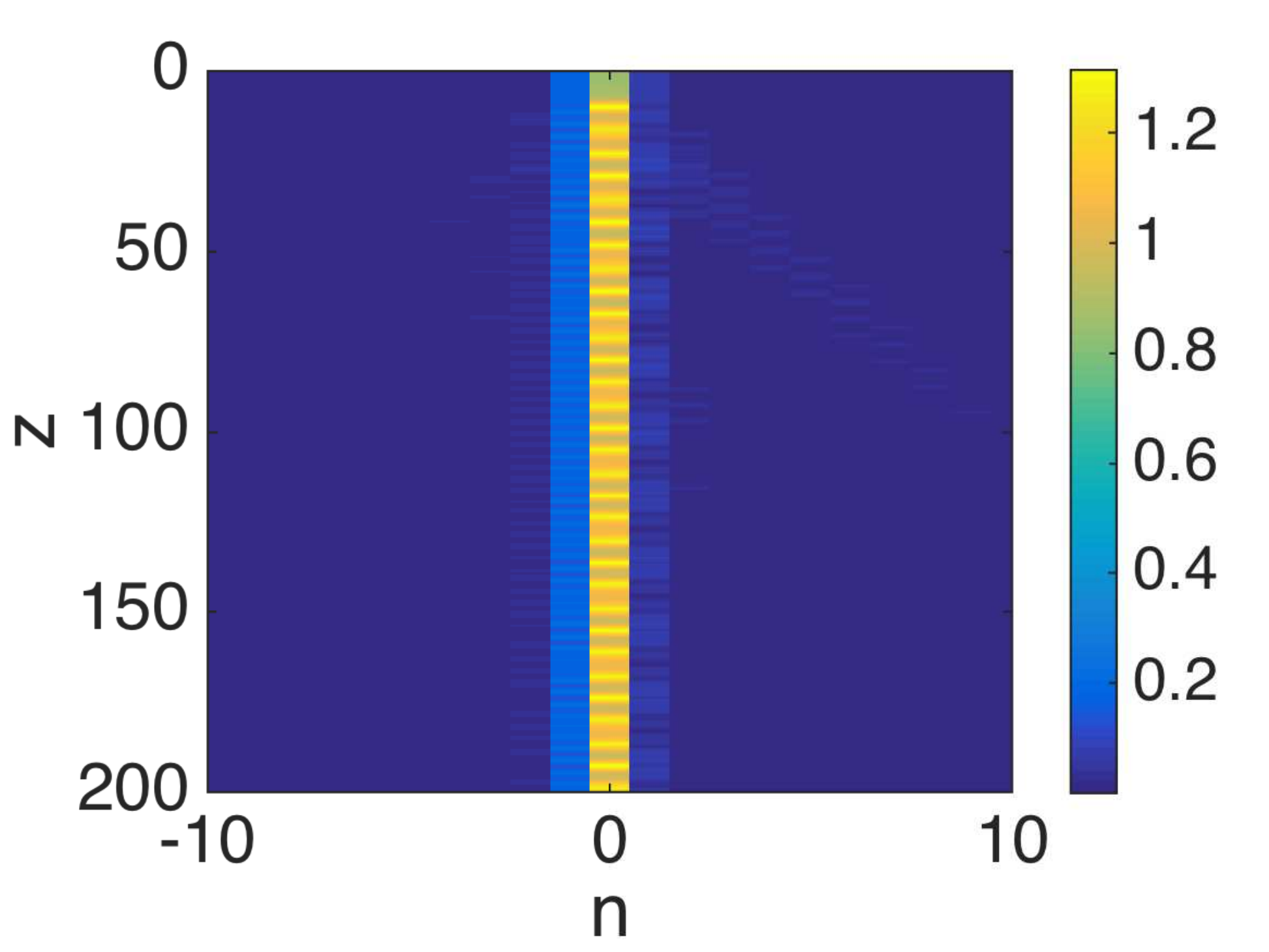}
 \includegraphics[width = 0.3\textwidth]{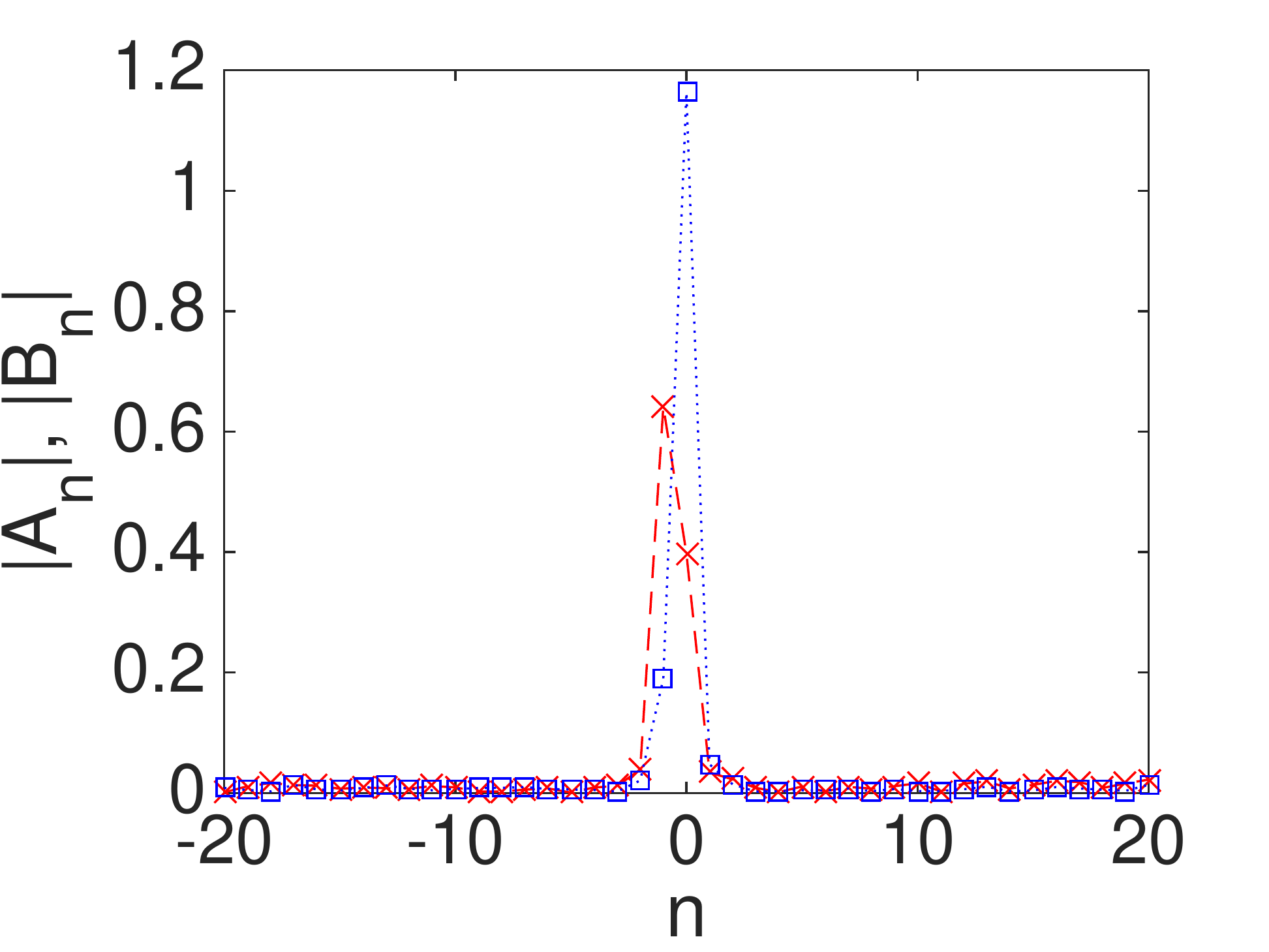}
 \caption{ 
%Absolute value of stationary solutions of $a_n$ in red crosses and $b_n$ in blue squares.  Starting from $\eps = 0$ on the top left panel,  continuously increasing  to $\eps=0.2$. We get at $\eps=0.2$, the stationary solution  (middle left panel), its linear stability (middle middle panle) and most unstable eigenvector (middle right) panel. If we use perturbation of $1\%$ of amplitude of the stationary solution in the most unstable direction, we get the evolution of $|A_n|$, $|B_n|$  in bottom left and middle panels respectively and their profile $t=100$ are shown in bottom right panel.  
Similar to Fig.~\ref{1in} but for the case of the 3-excited sites
in the form $(++,0+)$. Here the main difference is that there are 
two real eigenvalue pairs (as opposed to in Fig.~\ref{1in}) responsible
for the instability.}
 \label{31}
 \end{figure} 
 
%If we set $C_1=1$, with other parameter unchanged, as shown in Fig. (\ref{homo}) we find that, the unstable eigenvalues are smaller than when $C_1 =2$, for small $\eps$. 

% \begin{figure}[!htbp]
% \centering
% \includegraphics[width = 0.35\textwidth]{homoComp}
% \caption{Same as in top right panel of Fig. (\ref{31}), but with $C_1 = 1$. Fo%r comparison, we indicate the unstable eigenvalues when $C_1=2$ with black dot %lines.}
% \label{homo}
% \end{figure} 

\item{$(-+,0+)$}: 

In this case $s_1= 1$, $s_2=-1$, the two pairs of eigenvalues that bifurcate from zero will consist of one pair moving along the real axis and the other pair moving along the imaginary axis. As shown in Fig.~\ref{32}, both the one moving along real axis in the top middle panel and the one moving along imaginary axis in top right panel are well predicted by their leading order approximations in dashed lines
for small values of $\eps$. The stationary solution at $\eps =0.1$, its spectrum and the unstable eigenvector are shown in the middle panels. For larger values of $\eps$, the pair of eigenvalues that moves along the imaginary axis will collide with the continuous spectrum, yielding additional potential oscillatory instabilities (not considered
herein, given the instability of this state immediately off of the AC limit). { The dynamics with $1\%$ perturbation in the unstable direction are shown 
in the bottom panels, where  the amplitude of the localized excited cites oscillate but this time in a less regular fashion.  }
 
\begin{figure}[!htbp]
 \centering
 \includegraphics[width = 0.3\textwidth]{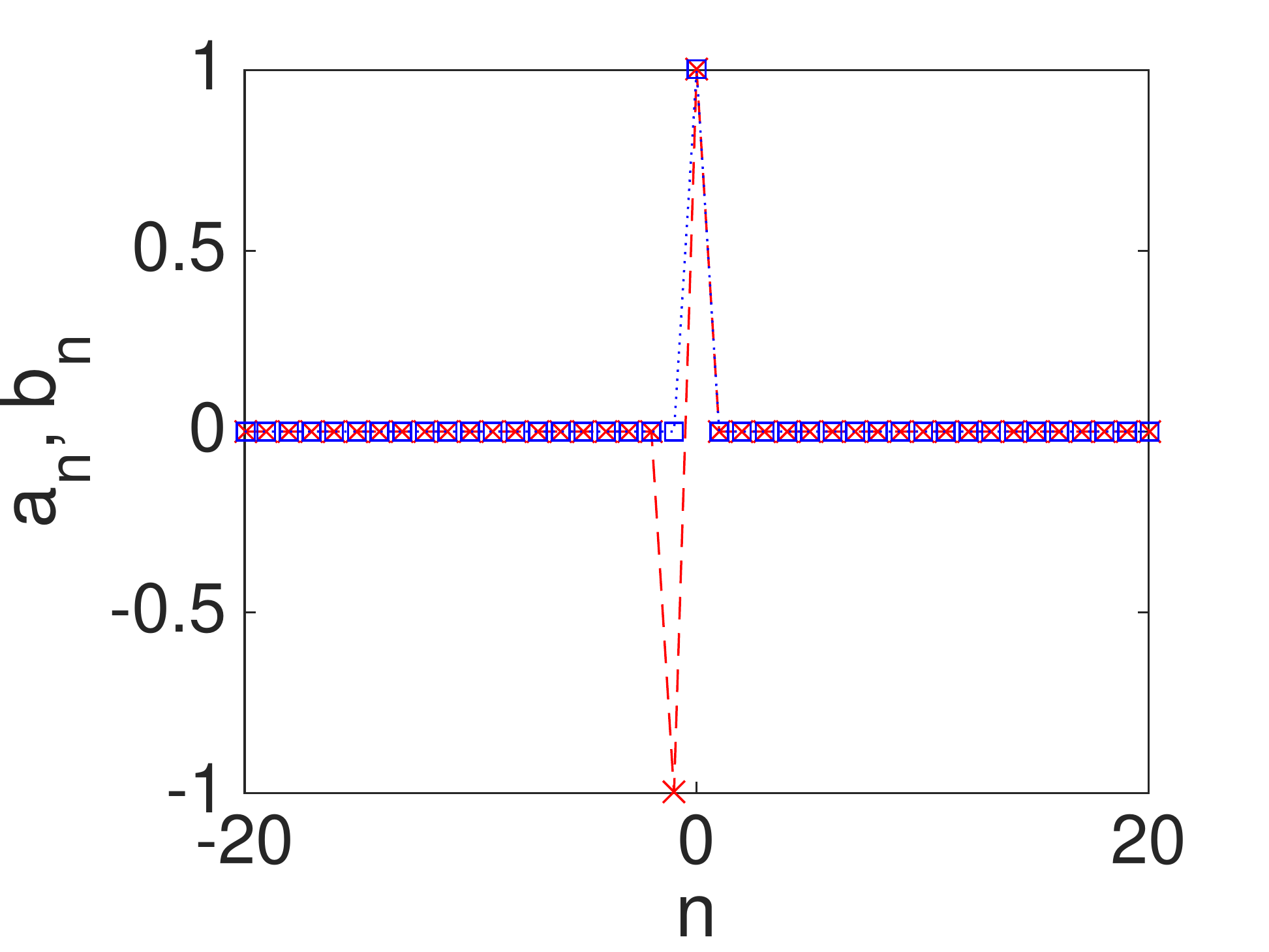}
 \includegraphics[width = 0.3\textwidth]{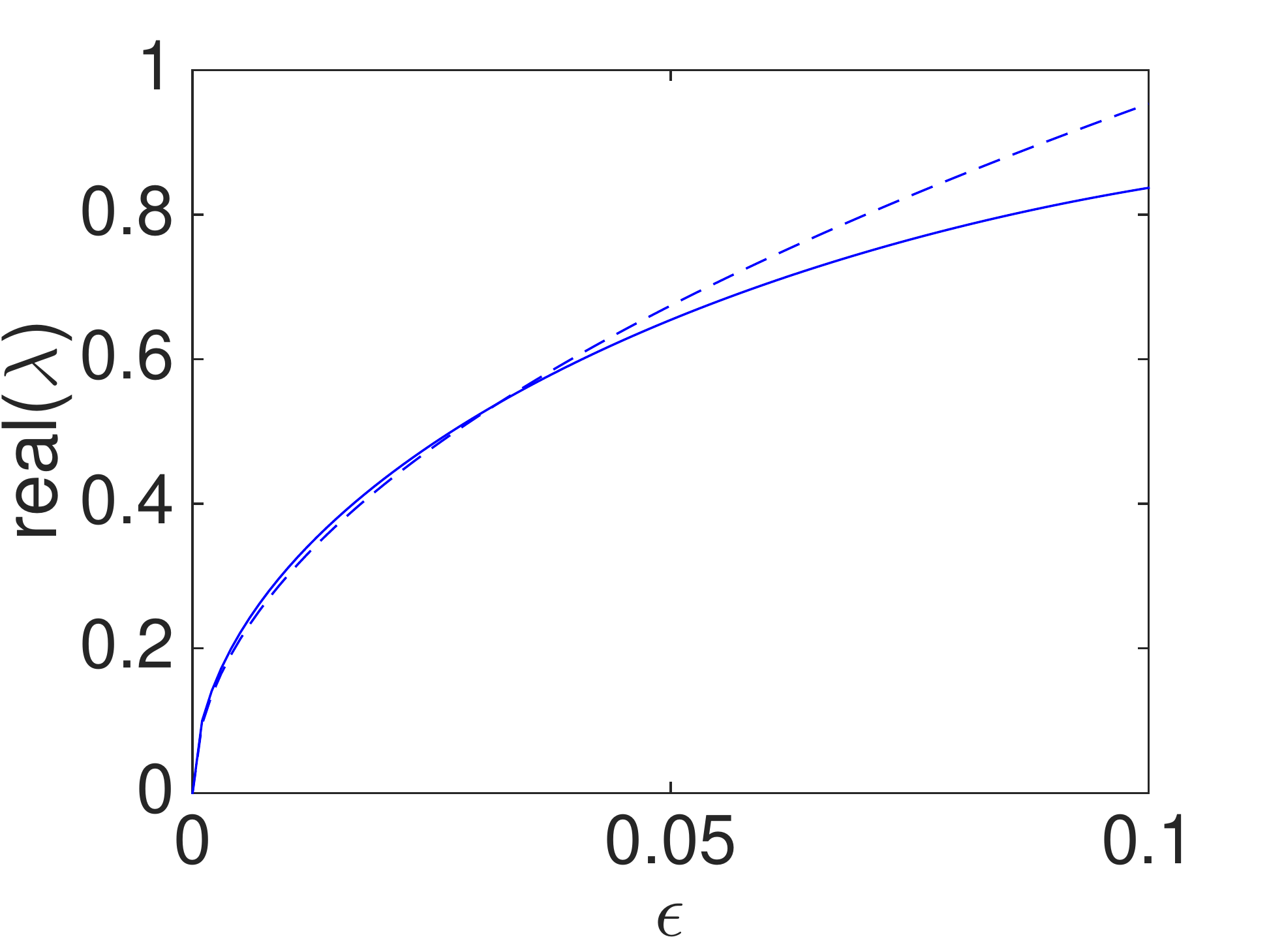}
 \includegraphics[width = 0.3\textwidth]{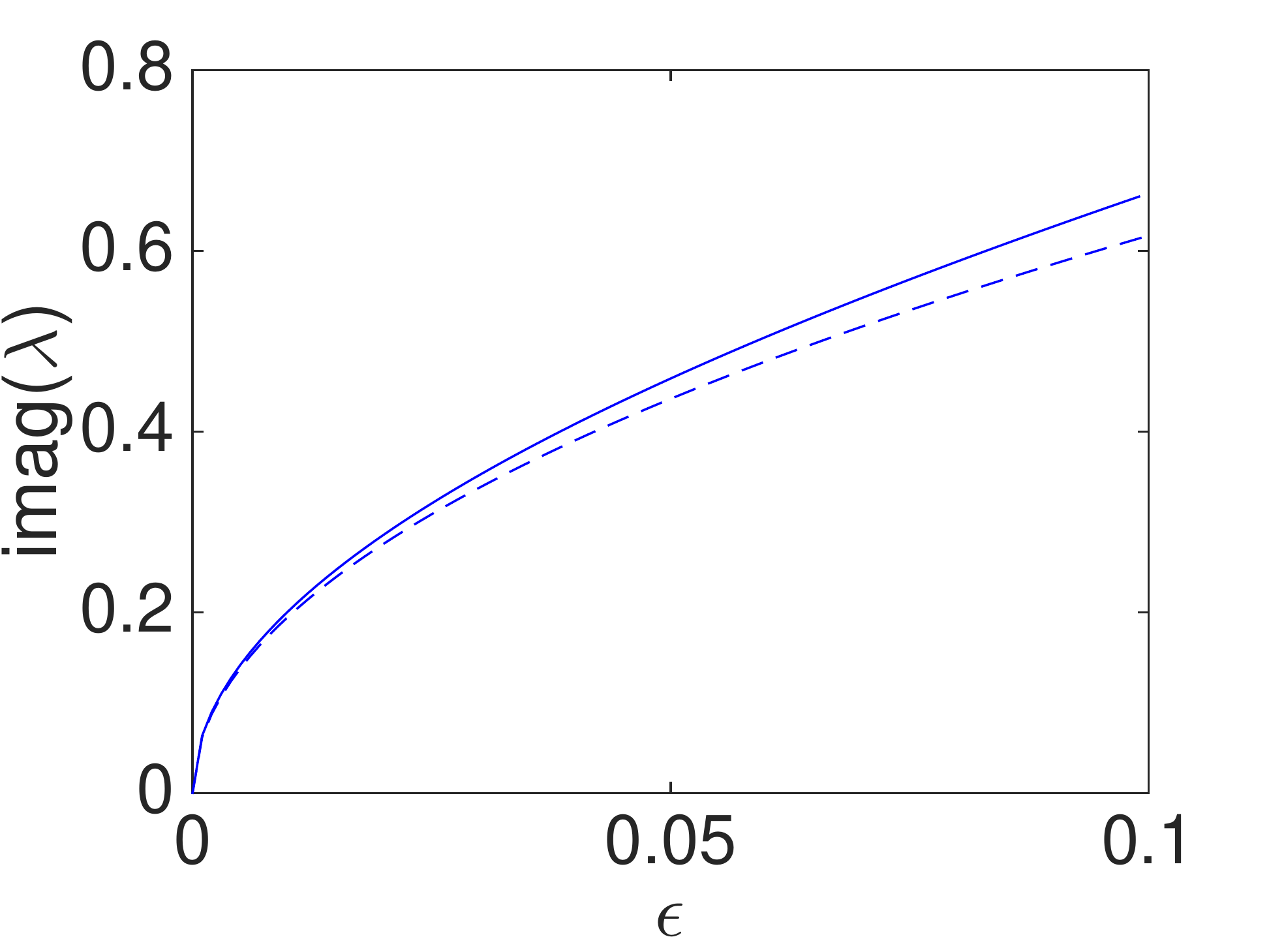}
 \includegraphics[width = 0.3\textwidth]{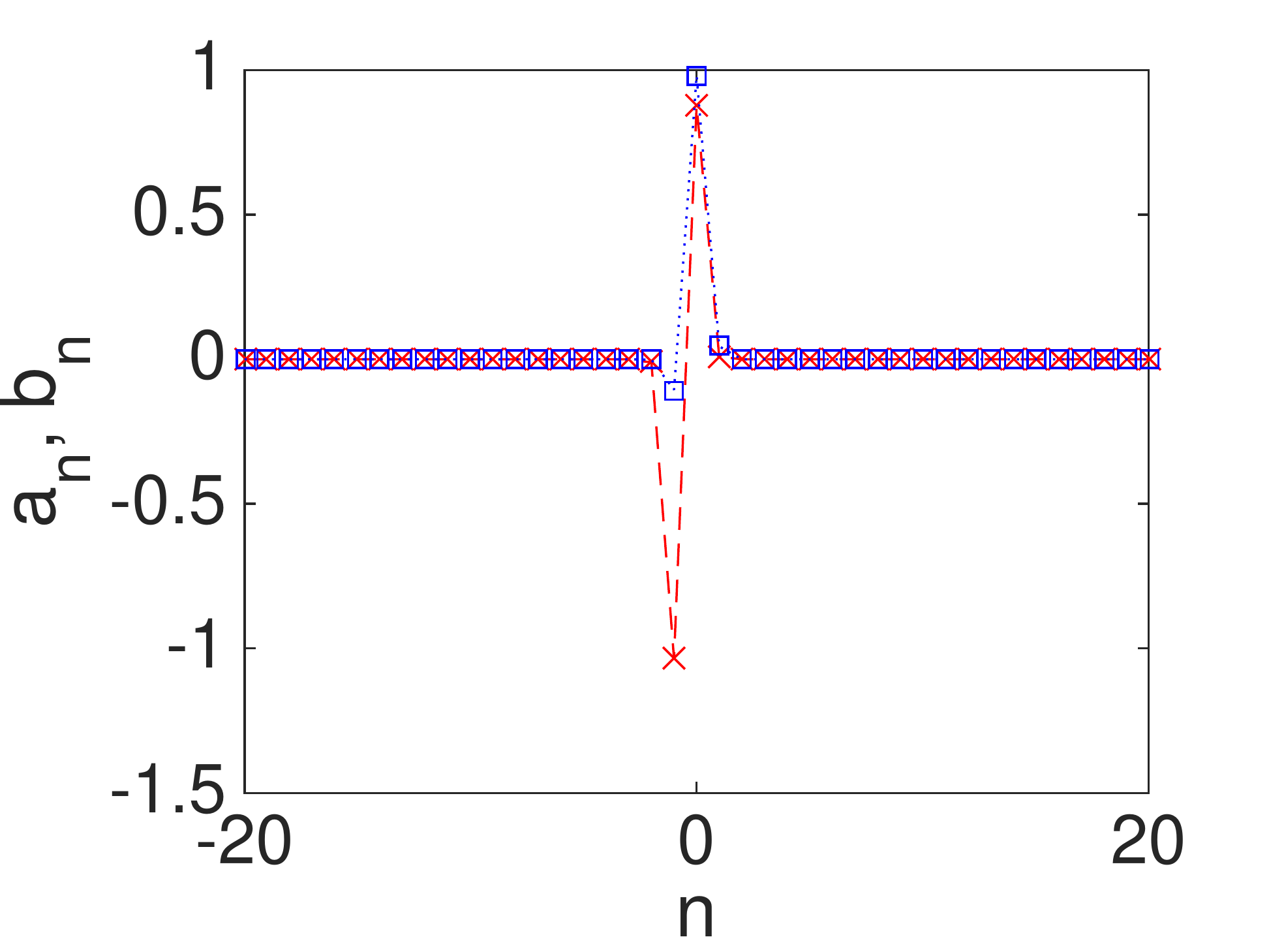}
 \includegraphics[width = 0.3\textwidth]{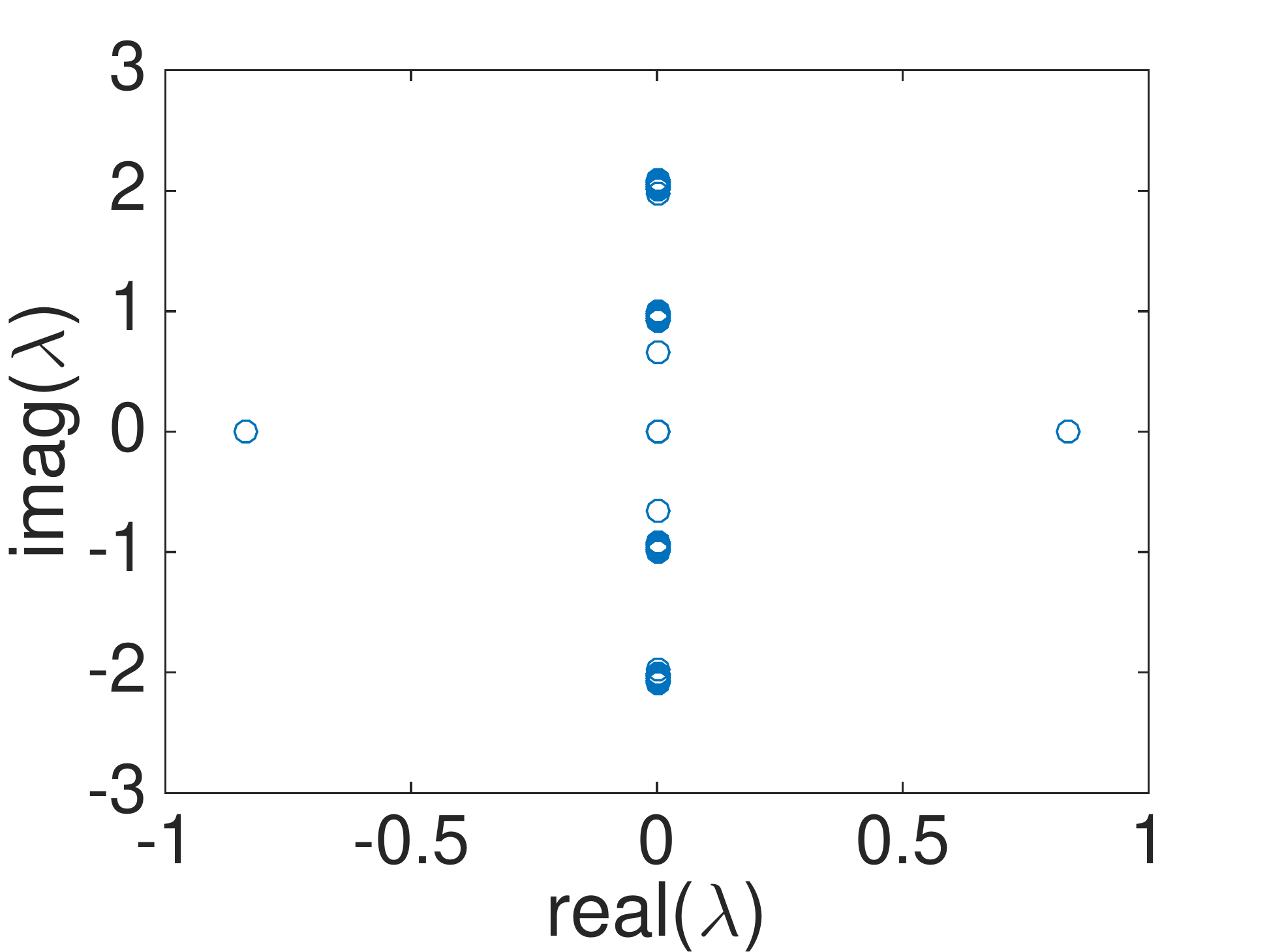}
 \includegraphics[width = 0.3\textwidth]{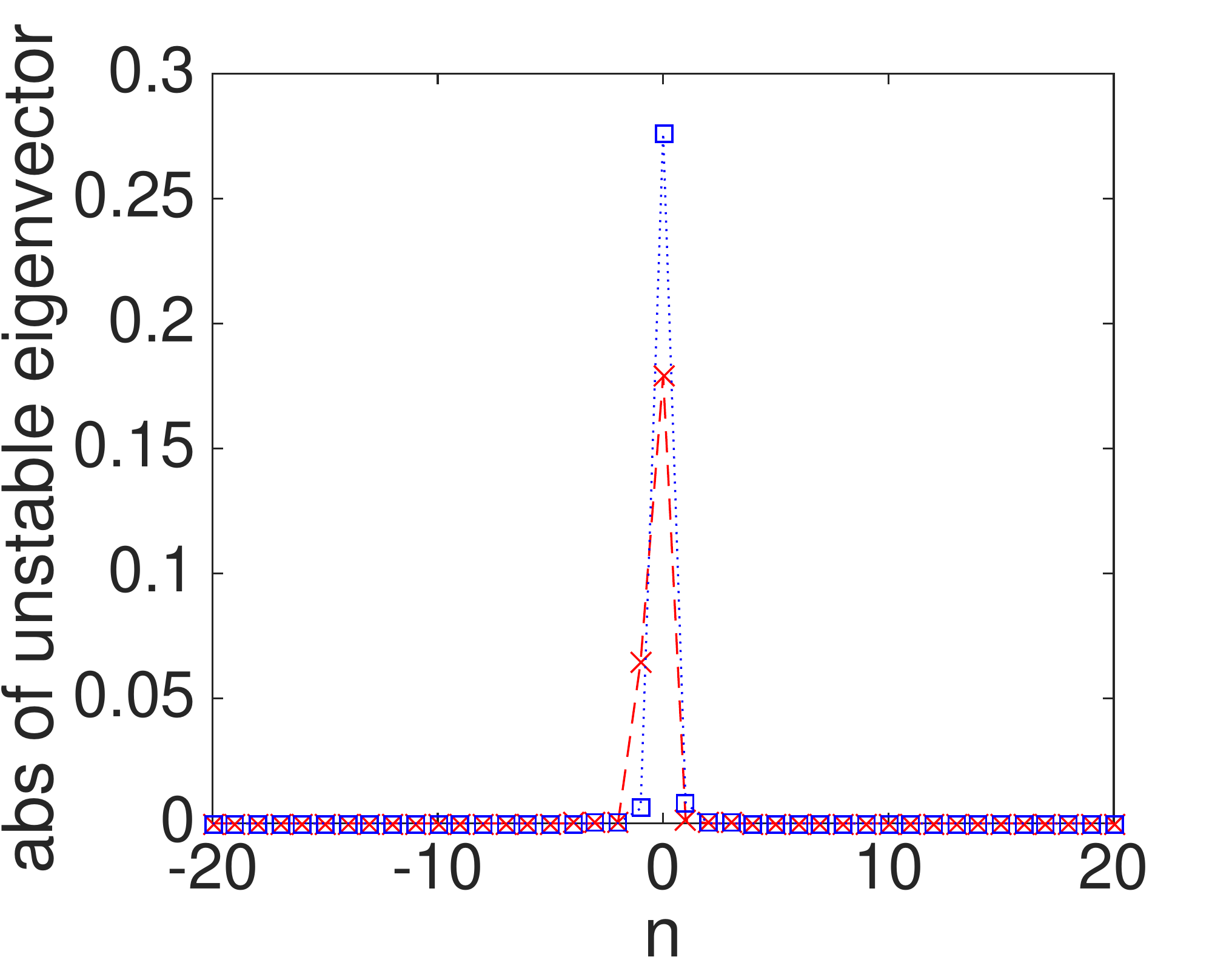}
 \includegraphics[width = 0.3\textwidth]{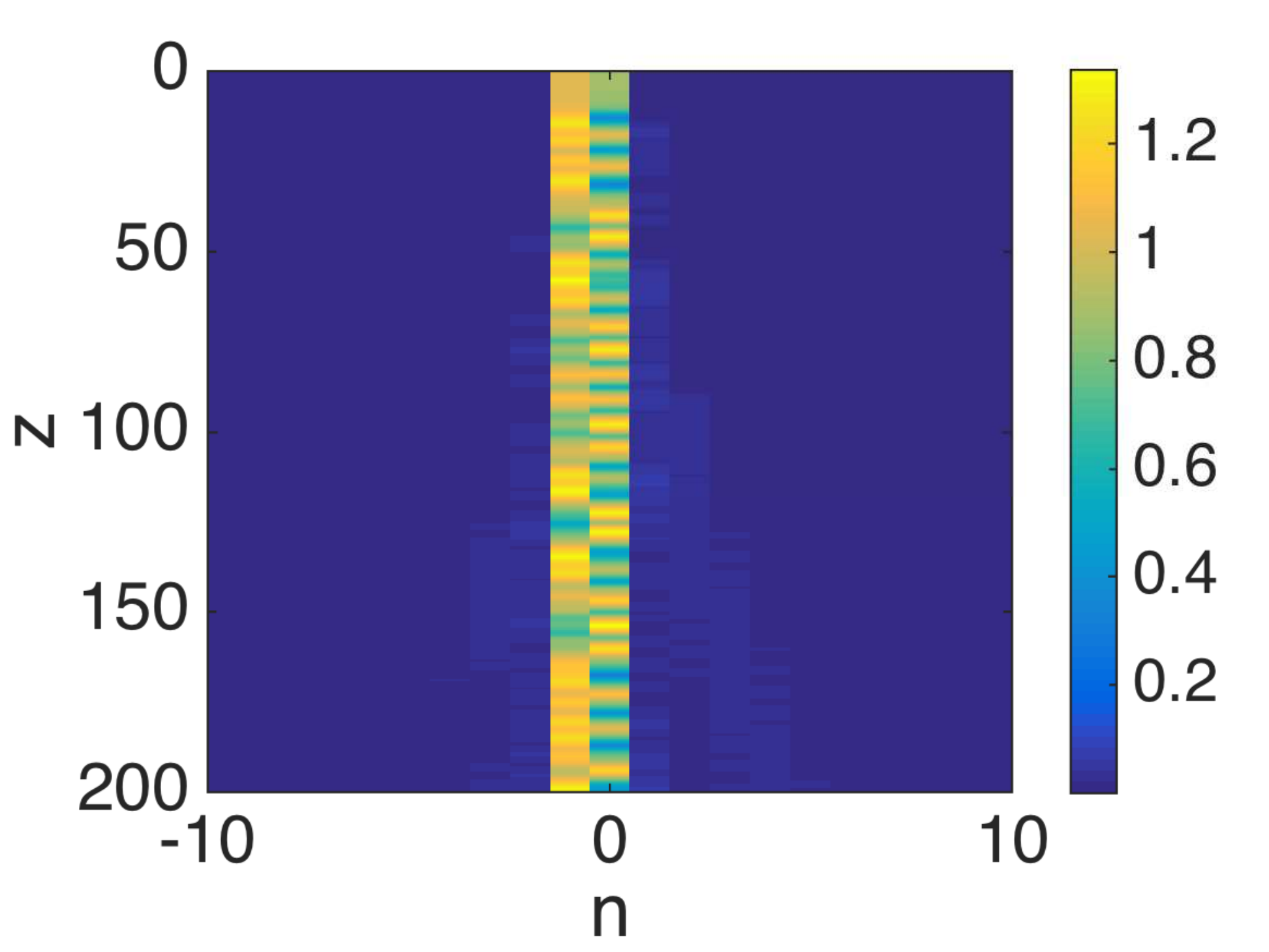}
 \includegraphics[width = 0.3\textwidth]{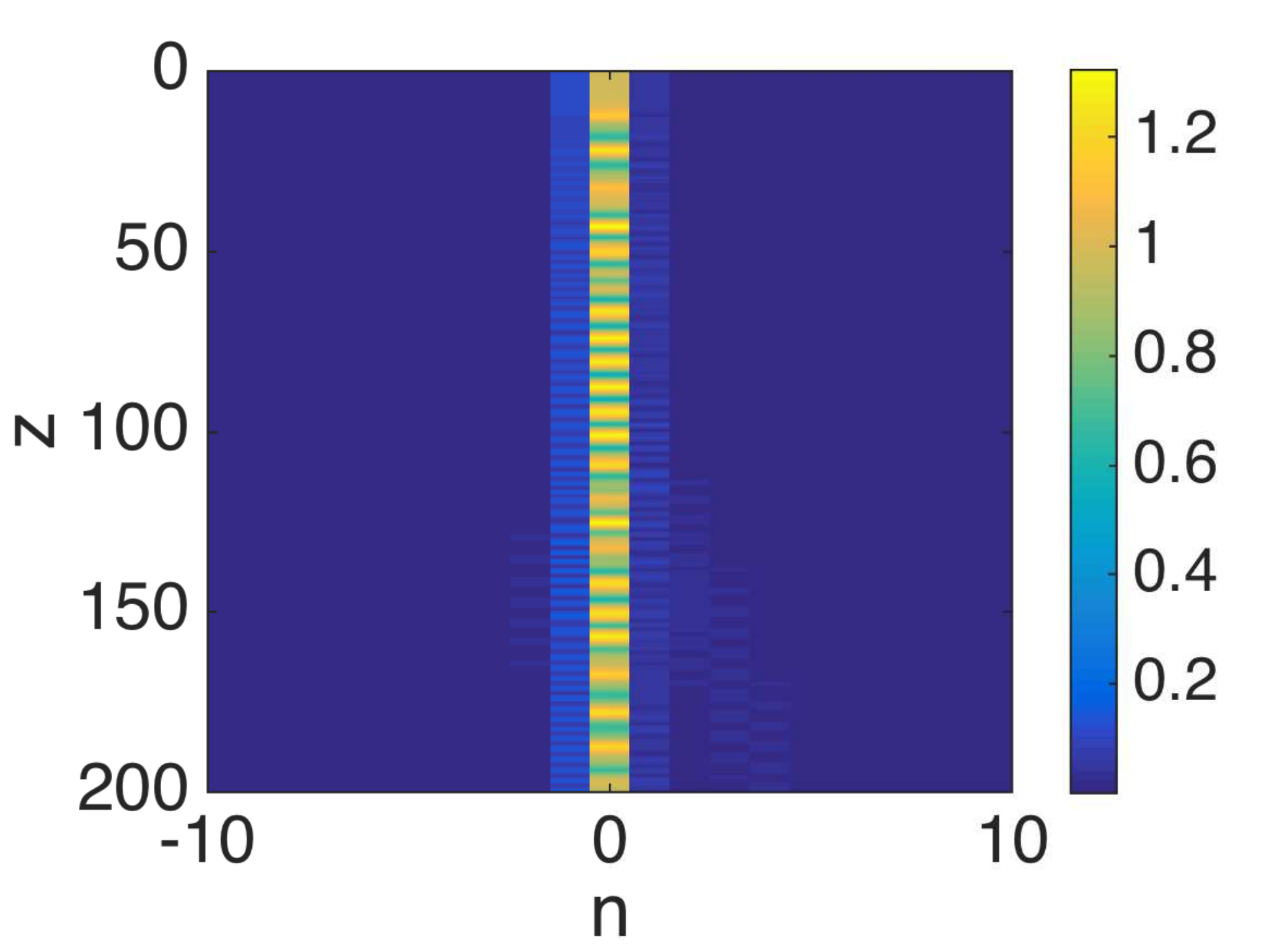}
 \includegraphics[width = 0.3\textwidth]{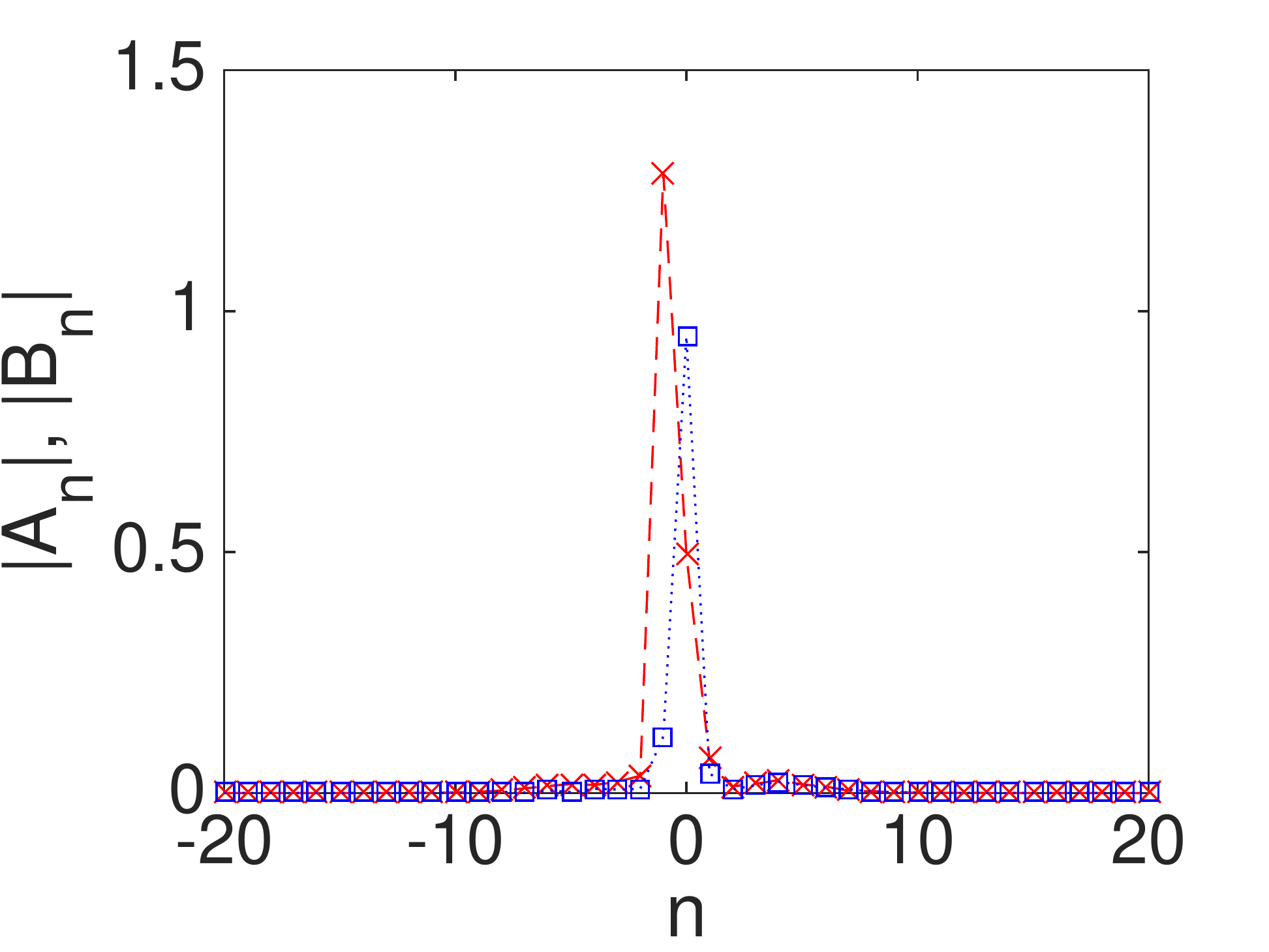}
 \caption{The $(-+,0+)$ solution profile
is shown at the AC limit (top left panel) and
for $\eps=0.1$ (middle left panel). The dependence of the one real
and one imaginary pair emerging in this case as a function of 
$\eps$ is shown in the top middle and top right panels while the 
full spectral plane for $\eps=0.1$ is shown in the middle 
panel of the second row. The bottom
panels show the unstable dynamics and the final propagation distance of $z=200$. }
 \label{32}
 \end{figure} 

\item{$(+-,0+)$}: 

In this case $s_1 = -1$, $s_2=-1$. This is a scenario quite
similar to the previous one, leading to a real and an imaginary
eigenvalue pair and a configuration immediately unstable
off of the AC limit. For this reason, we do not focus on it
further here.

%Similar as previous case, see Fig. (\ref{33}). We get one pair of eigenvalues moving along real axis (top middle panel ) and the other pair moving along imaginary axis (top right panel), both are well predicted by their leading order approximations in dash lines. At $\eps = 0.2$, we have the stationary solution in bottom left panel and its spectrum in bottom right panel. 
%\begin{figure}[!htbp]
% \centering
% \includegraphics[width = 0.3\textwidth]{C2_eps0_a1_a1_b1_beta1_3}
% \includegraphics[width = 0.3\textwidth]{C2_a1_a1_b1_beta1_3_R}
% \includegraphics[width = 0.3\textwidth]{C2_a1_a1_b1_beta1_3_I}
% \includegraphics[width = 0.35\textwidth]{C2_eps02_a1_a1_b1_beta1_3}
% \includegraphics[width = 0.35\textwidth]{C2_eps02_a1_a1_b1_beta1_stability_3}
% \caption{  Absolute value of statoinary solutions of $a_n$ in red crosses and $b_n$ in blue squares.  Starting from $\eps = 0$ on the top left panel,  continuously increasing  to $\eps=0.2$. We get the stationary solution for $\eps =0.2$ in bottom left panel and its spectrum in bottom right panel. The eigenvalue bifurcate from zero are shown as a function of $\eps$ in the top middle and left panels.}
% \label{33}
% \end{figure} 
 
\item{$(++,0-)$}: 

In this case $s_1 = -1$, $s_2=1$. The two pairs of eigenvalues are both moving along the imaginary axis, as shown in Fig.~\ref{34}. Again this is well captured by the leading order approximations, for small values
of $\eps$. This effective ``out of phase'' configuration (i.e.,
with adjacent waveguides bearing alternating $0$ and $\pi$ phases)
is spectrally stable for small values of $\eps$, similarly to its
corresponding DNLS cousin~\cite{Kevrekidis}. Yet, it is subject
to up to 2 quartets of oscillatory instabilities, as $\eps$ is
increased due to collisions of the relevant pairs of imaginary
eigenvalues growing from $0$ with the continuous spetrum.

It is relevant to note that we have confirmed the results of 
Fig.~\ref{cpm} for the full system. For instance, in the present
setting, we also considered the case of
 $C_1 = -1,\ a=1,\ b =1, \Delta \beta =0,\ \gamma_a =1,\ \gamma_b = 1$,
observing that in line with the results of the latter figure,
instead of 2 imaginary eigenvalue pairs, in that case a real
and an imaginary pair arise. The predictions of the theory
for small values of $\eps$ were again found to be in
good agreement with the full numerical results.

\begin{figure}[!htbp]
 \centering
 \includegraphics[width = 0.35\textwidth]{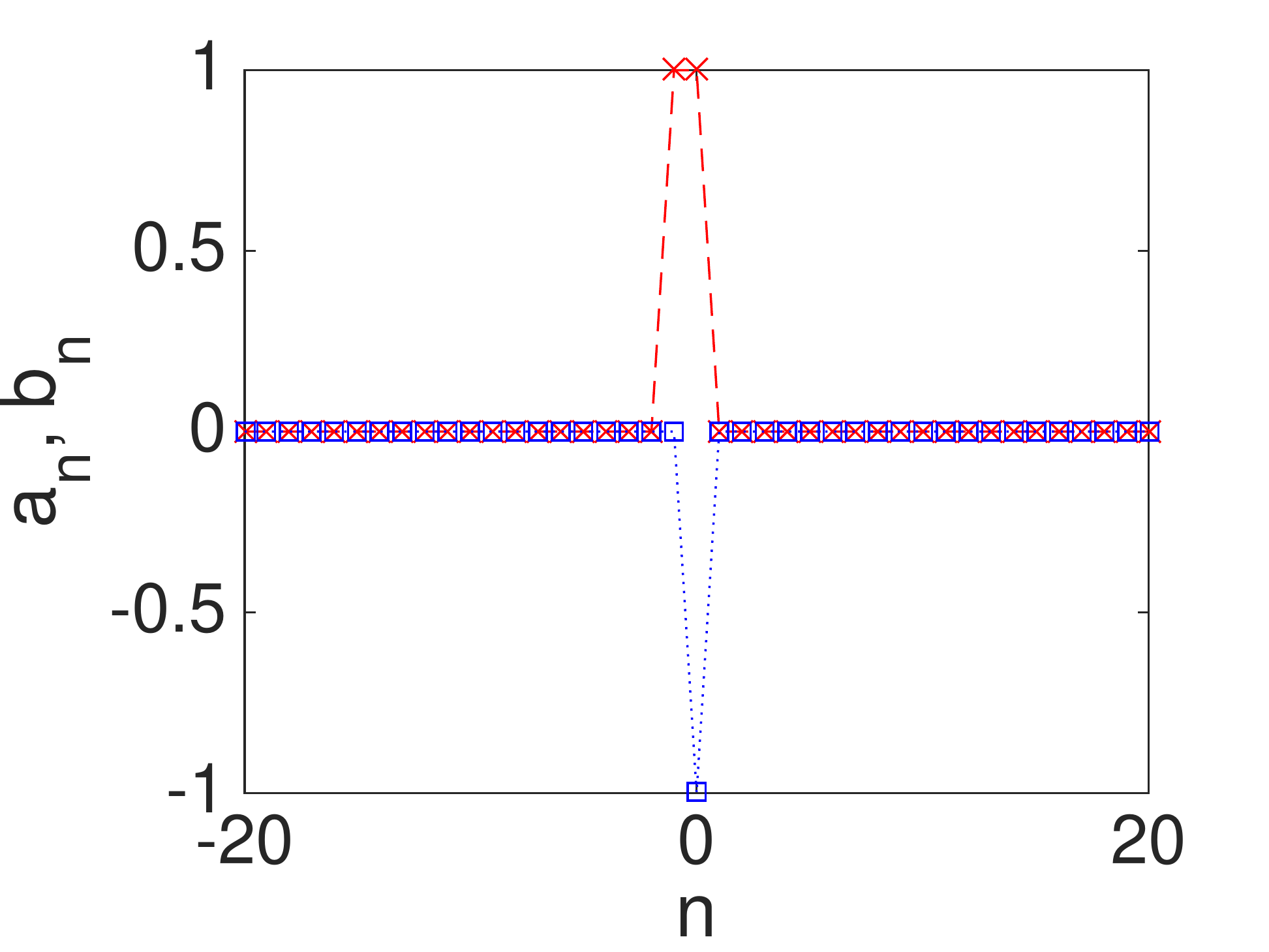}
 \includegraphics[width = 0.35\textwidth]{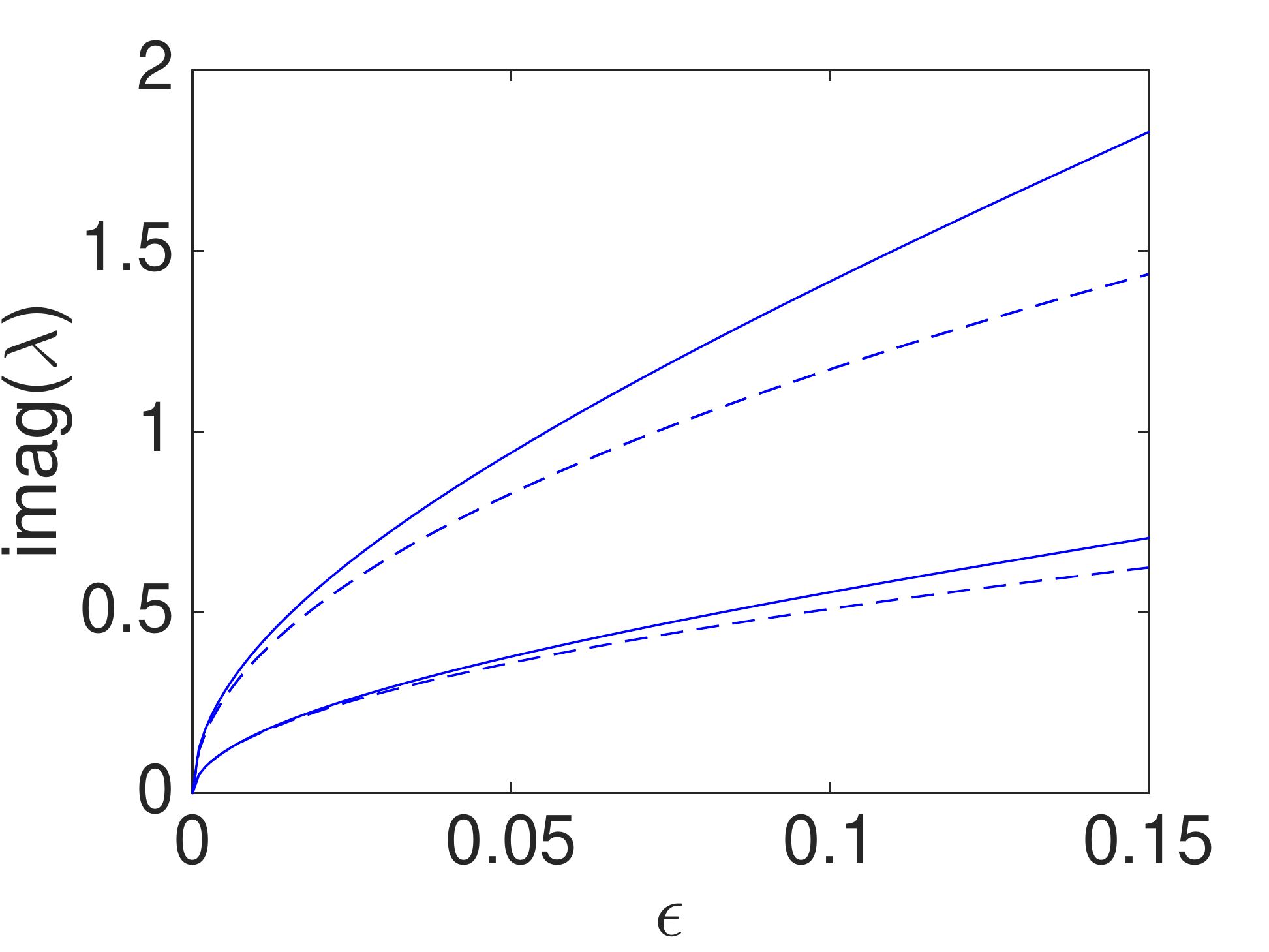}
 \includegraphics[width = 0.35\textwidth]{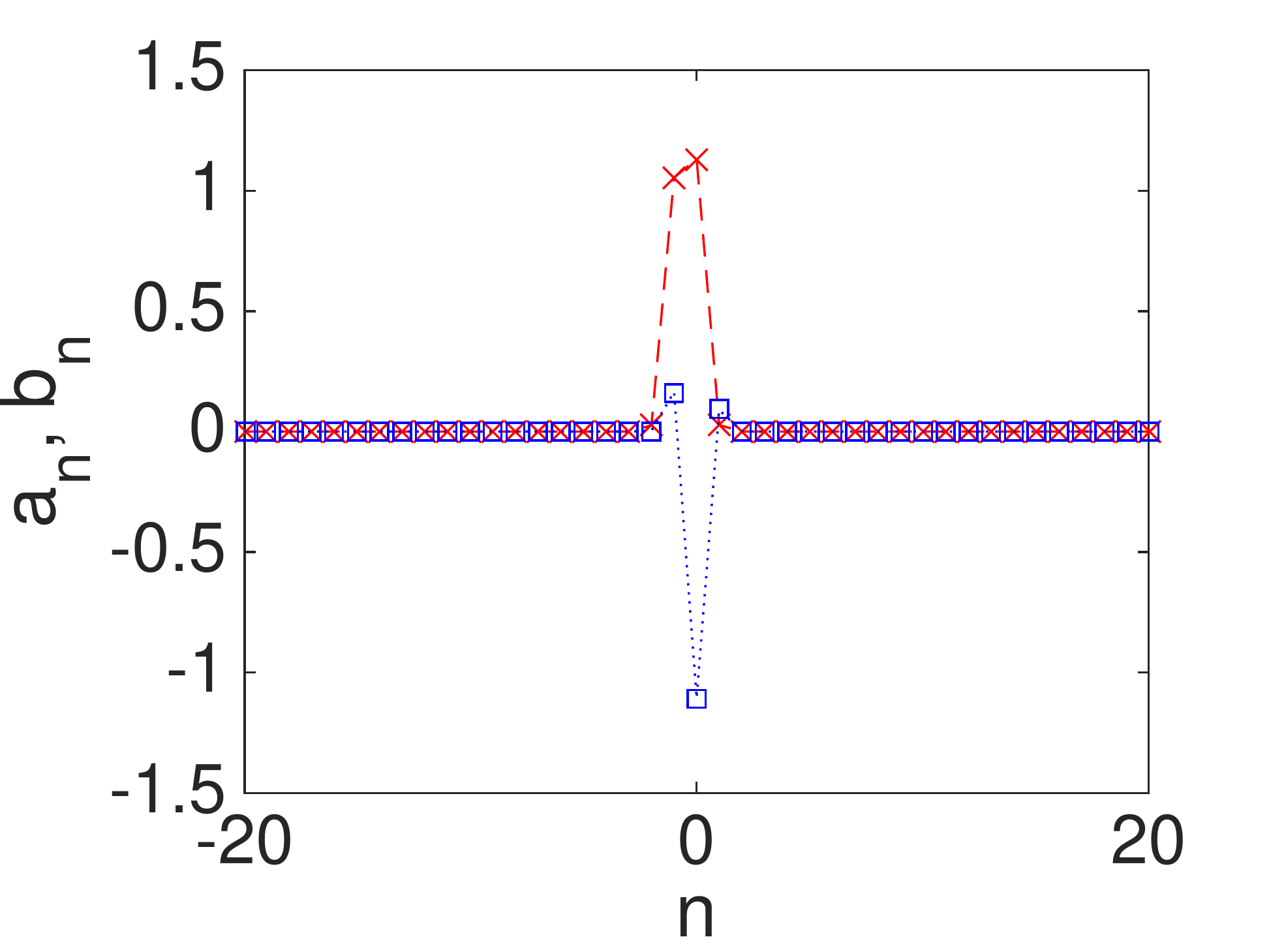}
 \includegraphics[width = 0.35\textwidth]{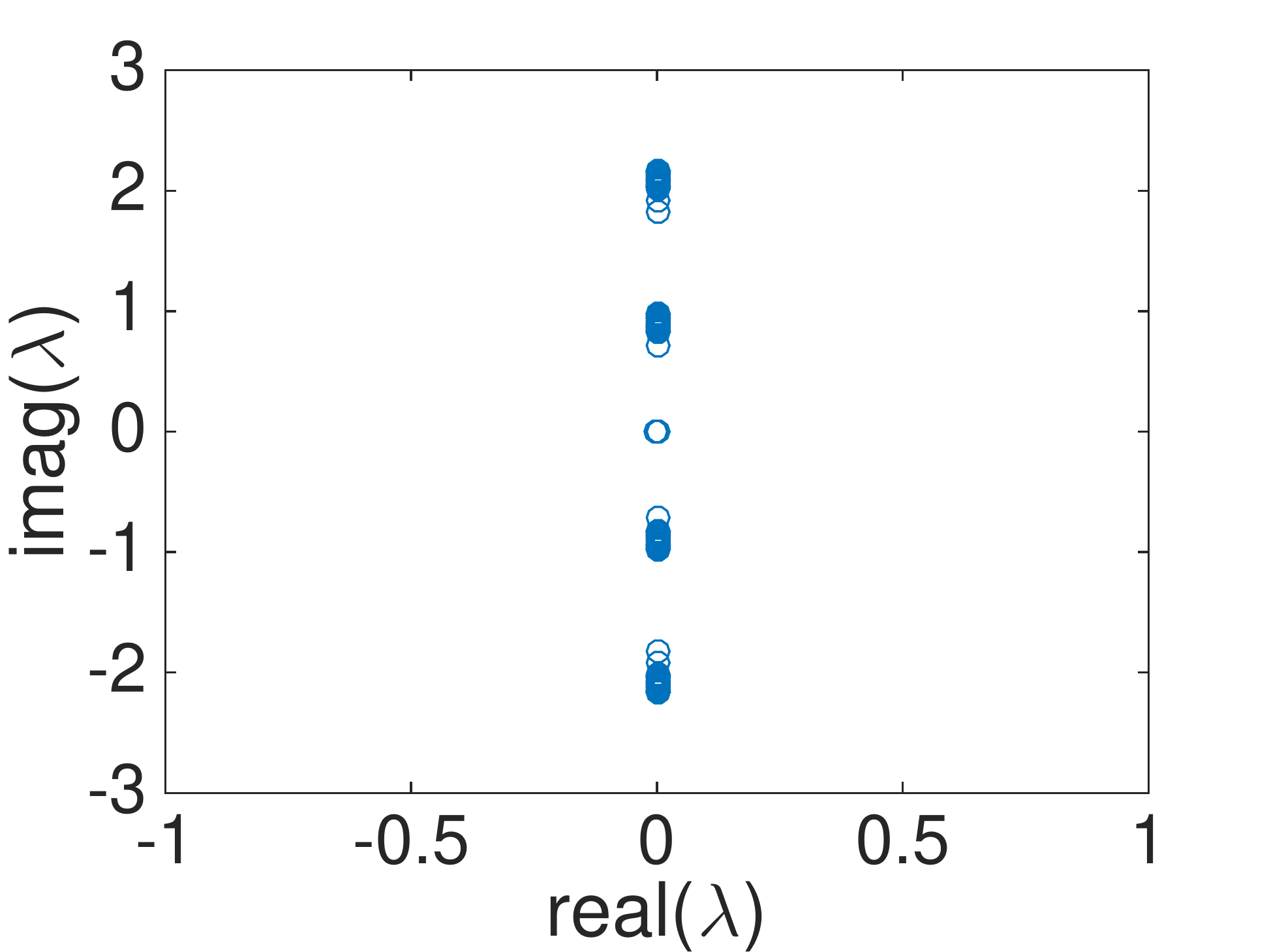}
 \caption{Profile of the $(++,0-)$ state for $\eps=0$ (top left)
and $\eps=0.15$ (bottom left). The spectral plane of the latter
is shown in the bottom right, while the imaginary eigenvalues
(predicted by theory through the dashed lines, and computed numerically
in the solid lines) emanating from $0$ are shown in the top right.
%Absolute value of statoinary solutions of $a_n$ in red crosses and $b_n$ in blue squares.  Starting from $\eps = 0$ on the top left panel,  continuously increasing  to $\eps=0.15$. We get the stationary solution for $\eps =0.15$ in bottom left panel and its spectrum in bottom right panel. The eigenvalue bifurcate from zero (solid line) and their lead order approximation (dash line) are shown as a function of $\eps$ in the top left panel.
}
 \label{34}
 \end{figure} 
\end{itemize}

\subsection{Four excited sites}
Finally, we consider some prototypical families of
configurations where the nonzero entries span $4$
sites. In that case, we have
\beq
(a_n^{(0)}, b_n^{(0)}) = \left\{
\begin{array}{lll}
(ae^{ic_{0}}, & be^{id_0}),& n=0, \\ 
(ae^{ic_1}, & be^{id_1}),& n=1, 
\end{array}
\right.
\eeq
then
\beq
(a_n^{(1)}, b_n^{(1)}) = \left\{
\begin{array}{lll}
(\frac{be^{id_0}}{\gamma_a a^2},& 0), & n=-1\\
(\frac{C_1be^{id_0}+be^{id_1}}{-2\gamma_a a^2}, &\frac{C_1a e^{ic_0}}{-2\gamma_b b^2}),& n=0, \\
(\frac{C_1be^{id_1}}{-2\gamma_a a^2}, &\frac{ae^{ic_{0}}+C_1a e^{ic_1}}{-2\gamma_b b^2}),& n=1, \\
(0,&\frac{ae^{ic_1}}{\gamma_bb^2})& n=2.
\end{array}
\right.
\eeq
Letting $s_1 = e^{i(c_0-d_0)} $, $s_2 = e^{i(d_0-d_1)}$, $s_3 = e^{i(d_1-c_1)}$,
the matrix determining the stability of the configuration now reads:
\bes
{\bf M} = s_1
\begin{pmatrix}
 (C_1+s_2)\frac{b}{a}&-C_1 & 0 &-s_2\\
-C_1&\frac{C_1a}{b}&0&0\\
0 &0& \frac{s_3C_1b}{a}&-s_3C_1\\
-s_2&0&-s_3C_1& (s_3C_1+s_2)\frac{a}{b}
\end{pmatrix}.
\ees

We only consider, for illustration purposes, the following two cases
\begin{itemize}
\item{$(++,++)$}: 

As seen in Fig.~\ref{41}, there will be three unstable pairs of 
eigenvalues bifurcating from zero, all of which are well predicted by their 
leading order theoretical 
approximations in the dashed lines in top right panel. 
%The stationary 
%solution and its linear stability are show in bottom panels
Hence, this configuration is highly unstable, following also
the general prediction of the theory of~\cite{Kevrekidis}
for $C_1=1$, suggesting that a configuration with $n$
in-phase excited sites will lead to $n-1$ real eigenvalue pairs.

\begin{figure}[!htbp]
 \centering
 \includegraphics[width = 0.35\textwidth]{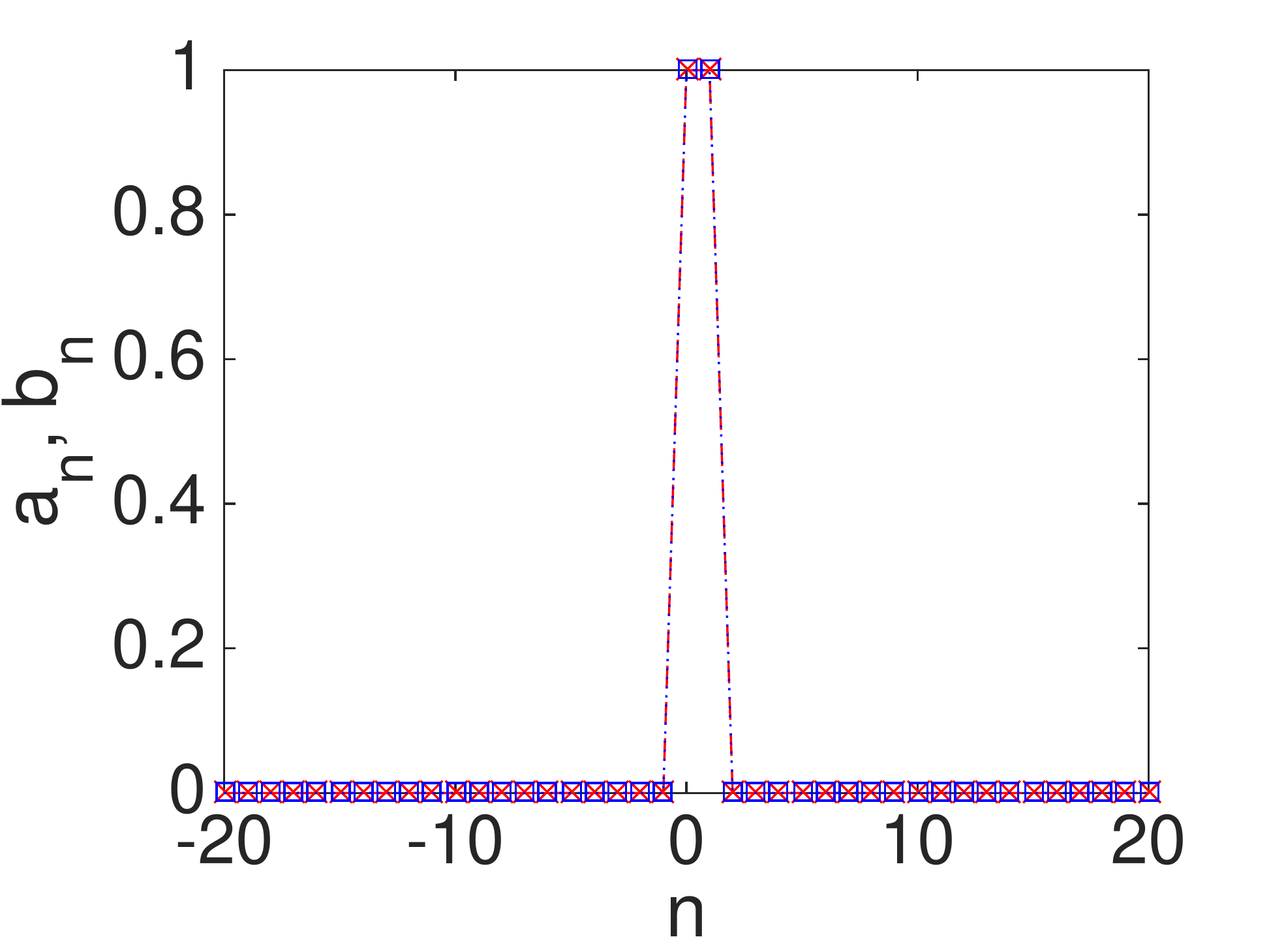}
 \includegraphics[width = 0.35\textwidth]{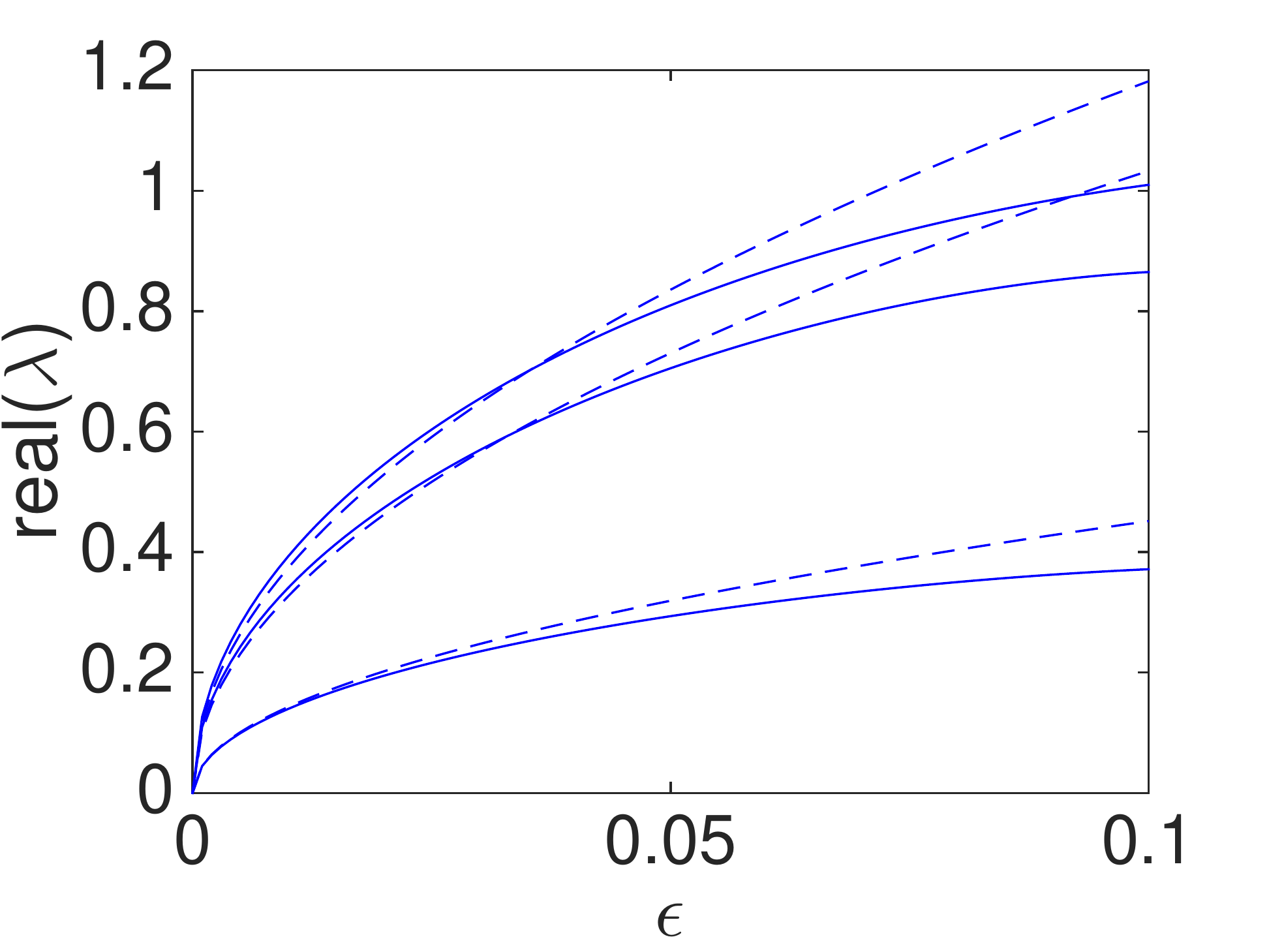}
 \includegraphics[width = 0.3\textwidth]{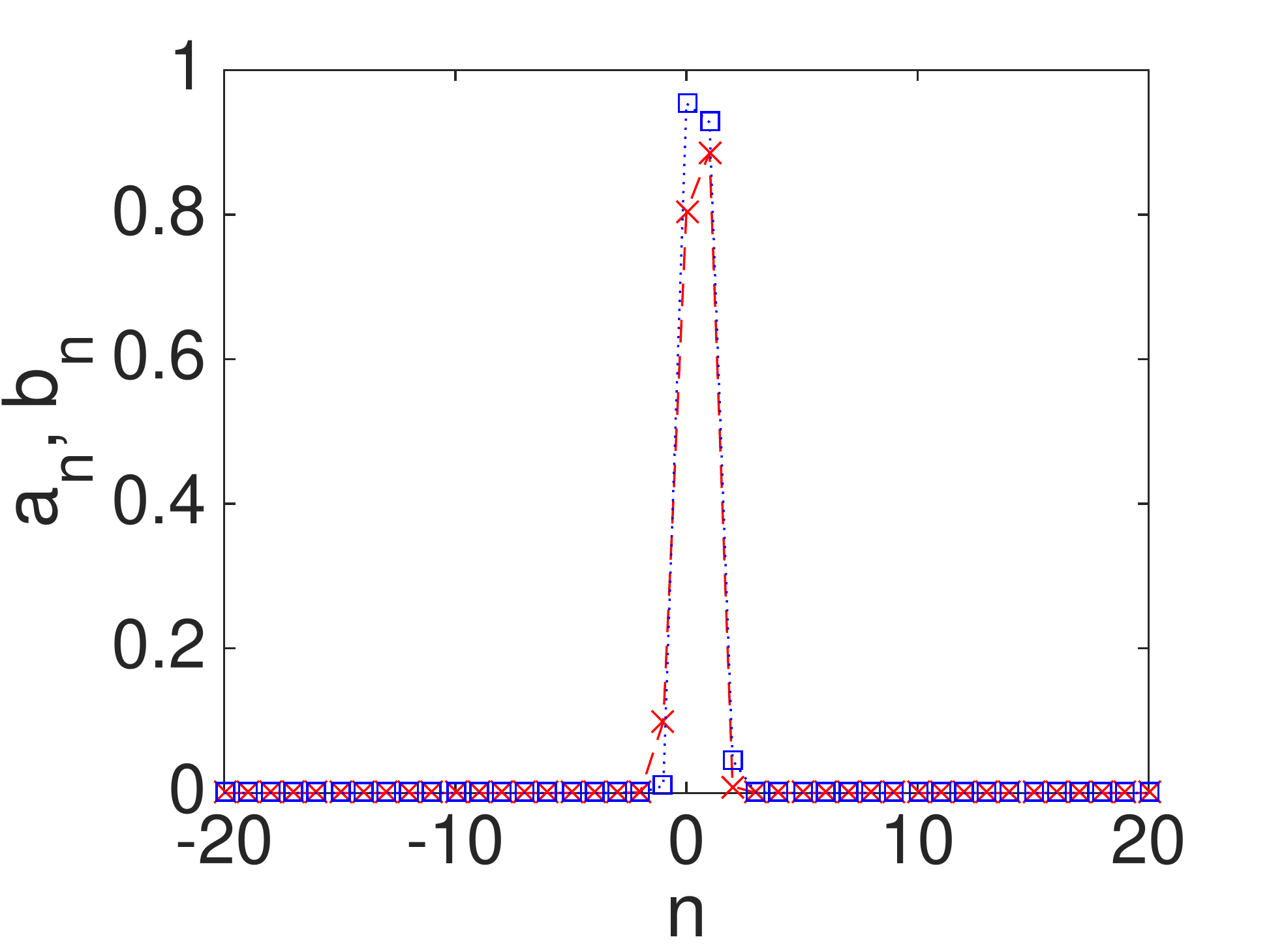}
 \includegraphics[width = 0.3\textwidth]{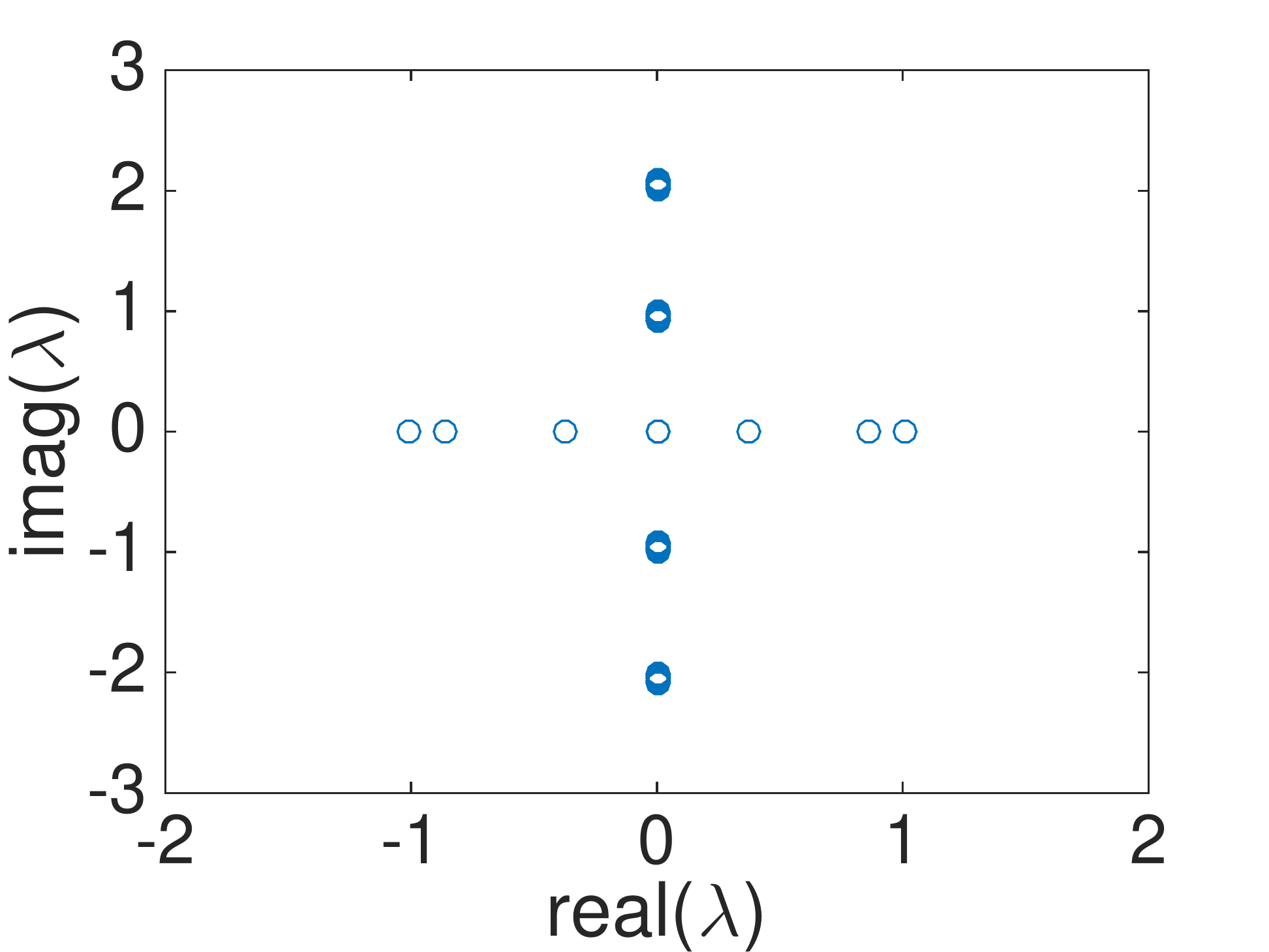}
 \includegraphics[width = 0.3\textwidth]{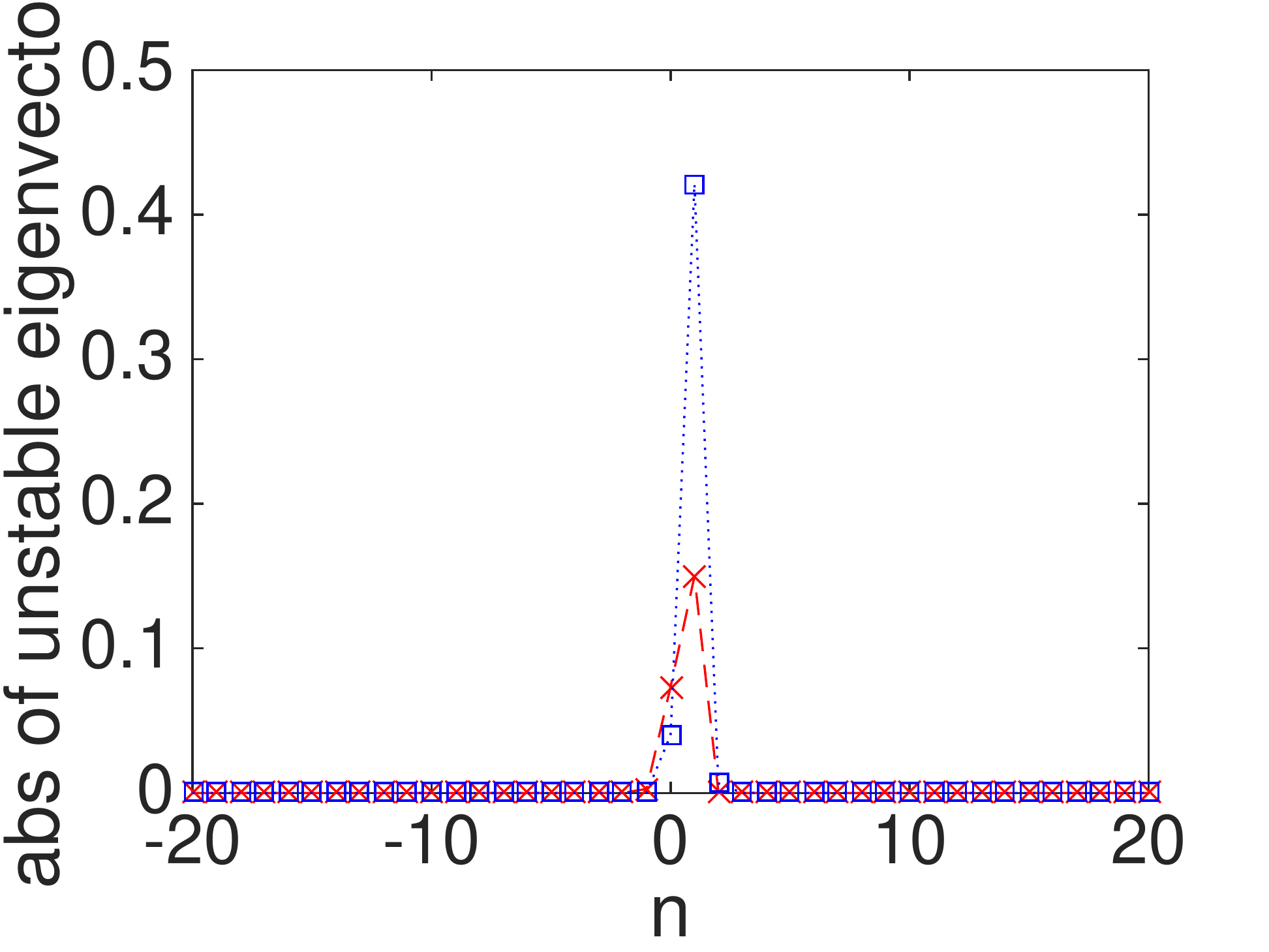}
 \includegraphics[width = 0.3\textwidth]{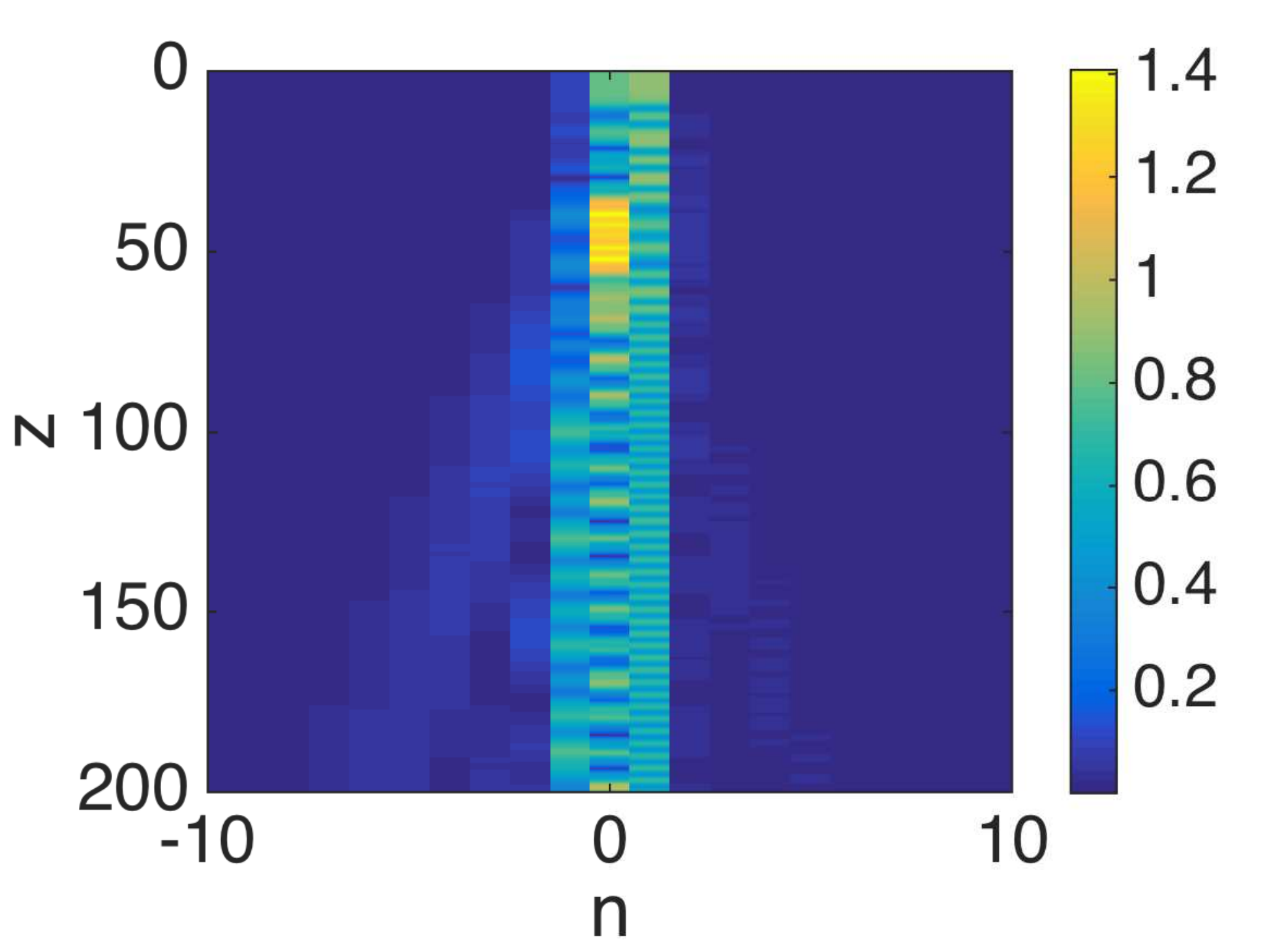}
 \includegraphics[width = 0.3\textwidth]{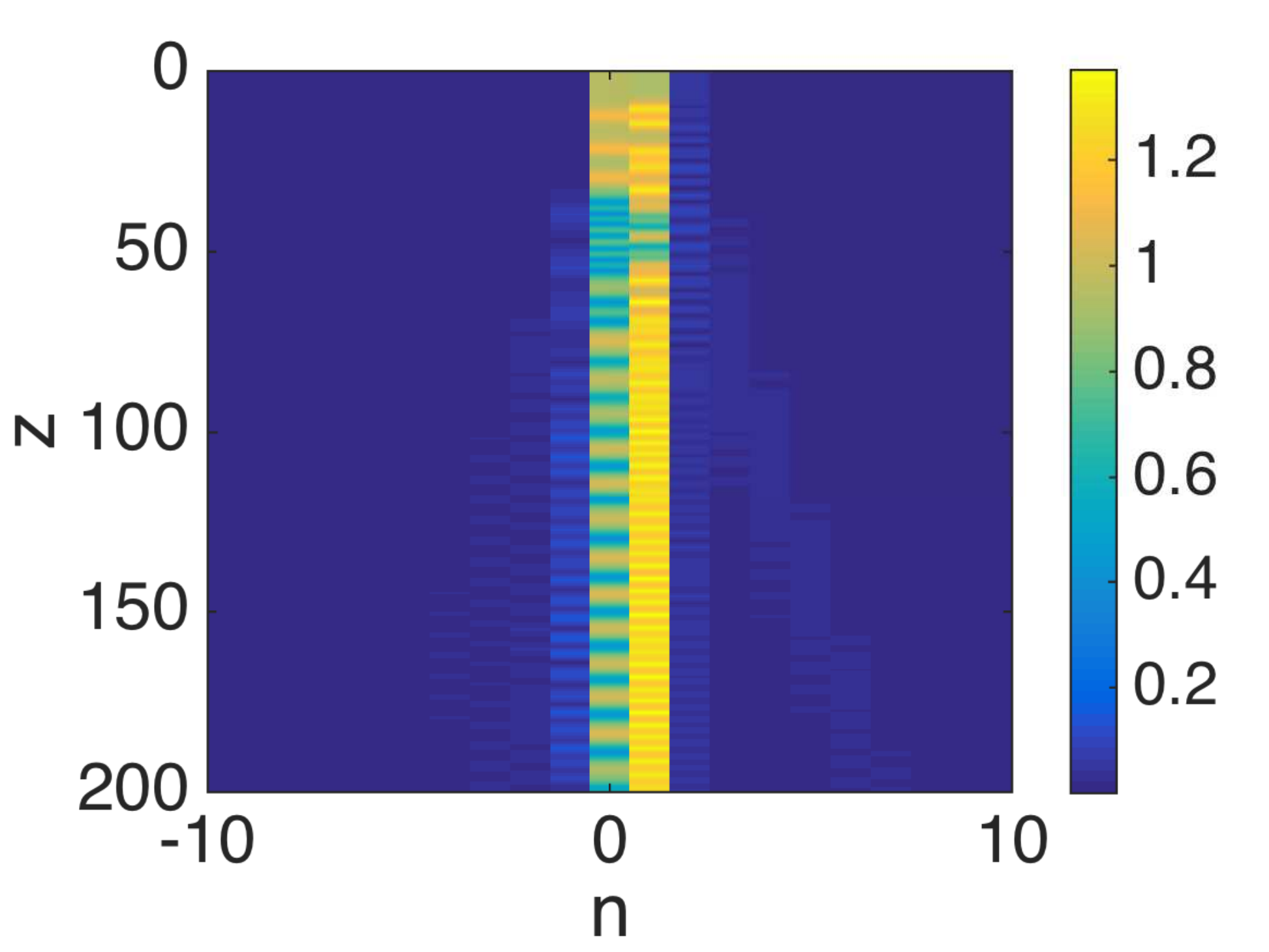}
 \includegraphics[width = 0.3\textwidth]{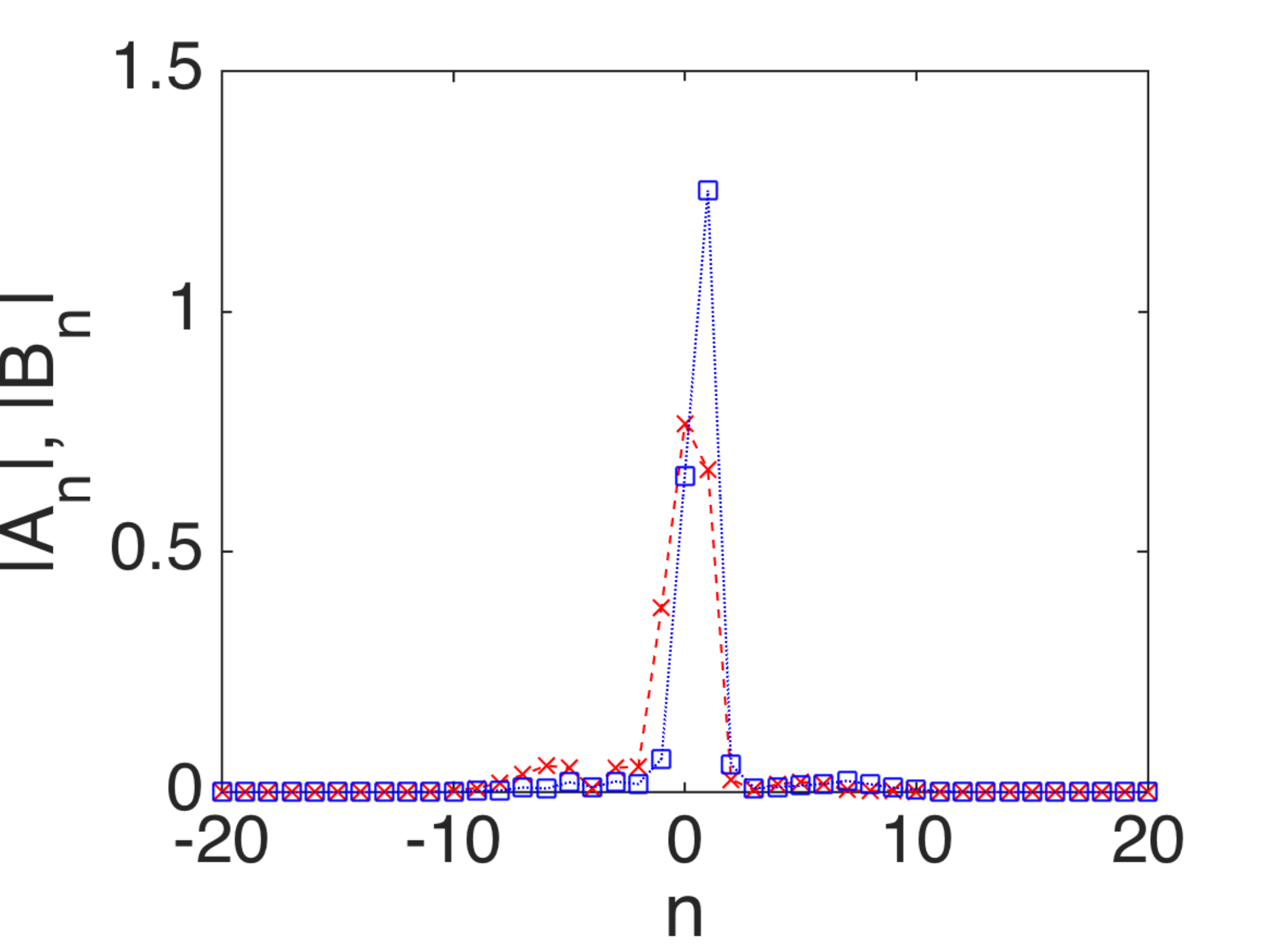}
 \caption{Similar to Fig.~\ref{34}, but now for the configuration
$(++,++)$ bearing three real eigenvalue pairs (whose real part
is shown in the top right). The configuration and the associated
spectral plane are shown in the middle panels of the figure for
$\eps=0.1$. The unstable dynamics are in the bottom panels with the modulus profile at $z=200$ at bottom right.
While the dynamics is not genuinely periodic in the modulus, it is
fairly proximal to that for sufficiently long times.
}
%Absolute value of statoinary solutions of $a_n$ in red crosses and $b_n$ in blue squares.  Starting from $\eps = 0$ on the top left panel,  continuously increasing  to $\eps=0.1$, we get the stationary solution and its spectrum in bottom panels.}%From second row to bottom are for  $\eps =0.1,\  0.15,\ 0.2$}
 \label{41}
 \end{figure} 

\item{$(++,--)$}:

In this configuration, again bearing alternating phases, 
we have three imaginary eigenvalue pairs bifurcate from zero, as shown in 
Fig.~\ref{42}. As $\eps$ increases, the largest one will collide with 
the edge of the continuous spectrum first and become unstable. As $\eps$ 
increases further, the second one will collide for larger values 
of $\eps$, hence in the top right panel, there are two humps of the real 
part of the eigenvalues associated with these two intervals
of oscillatory instabilities. 
%As $\eps$ further increase there can be a third collision. 
The presence of three imaginary eigenvalues in this case can yield
up to three distinct sets of oscillatory instabilities
and corresponding eigenvalue quartets.

\begin{figure}[!htbp]
 \centering
 \includegraphics[width = 0.3\textwidth]{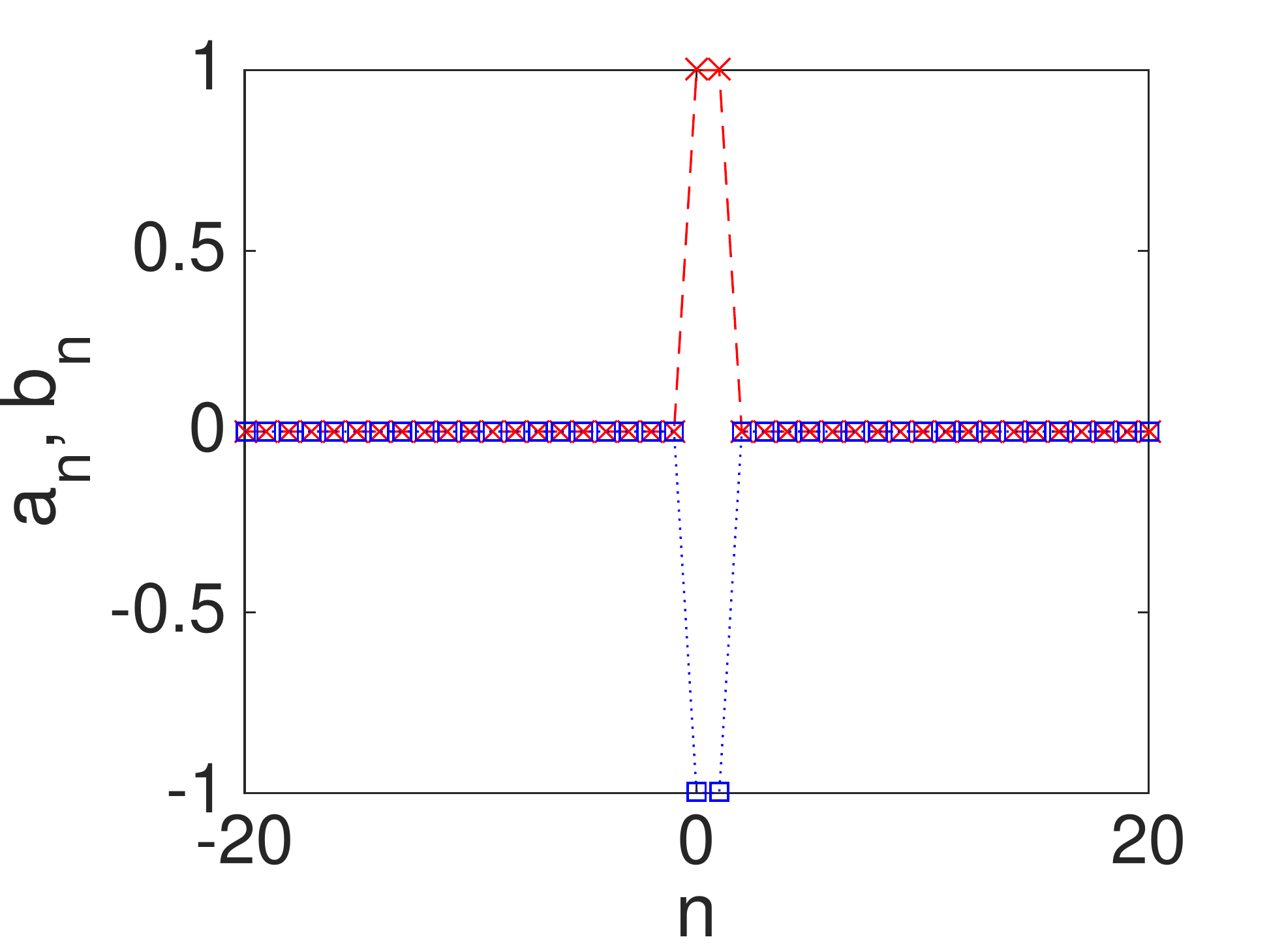}
 \includegraphics[width = 0.3\textwidth]{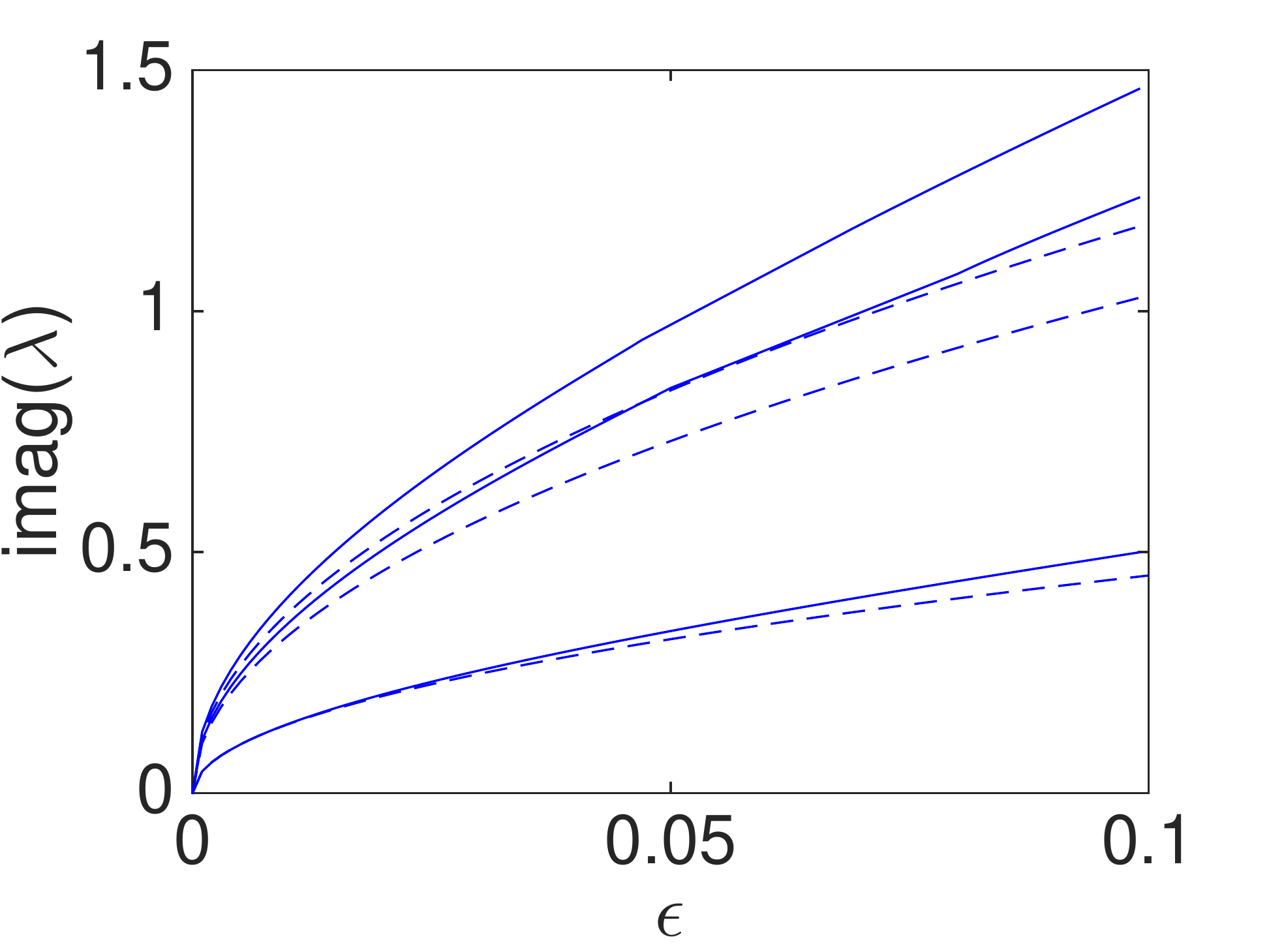}
 \includegraphics[width = 0.3\textwidth]{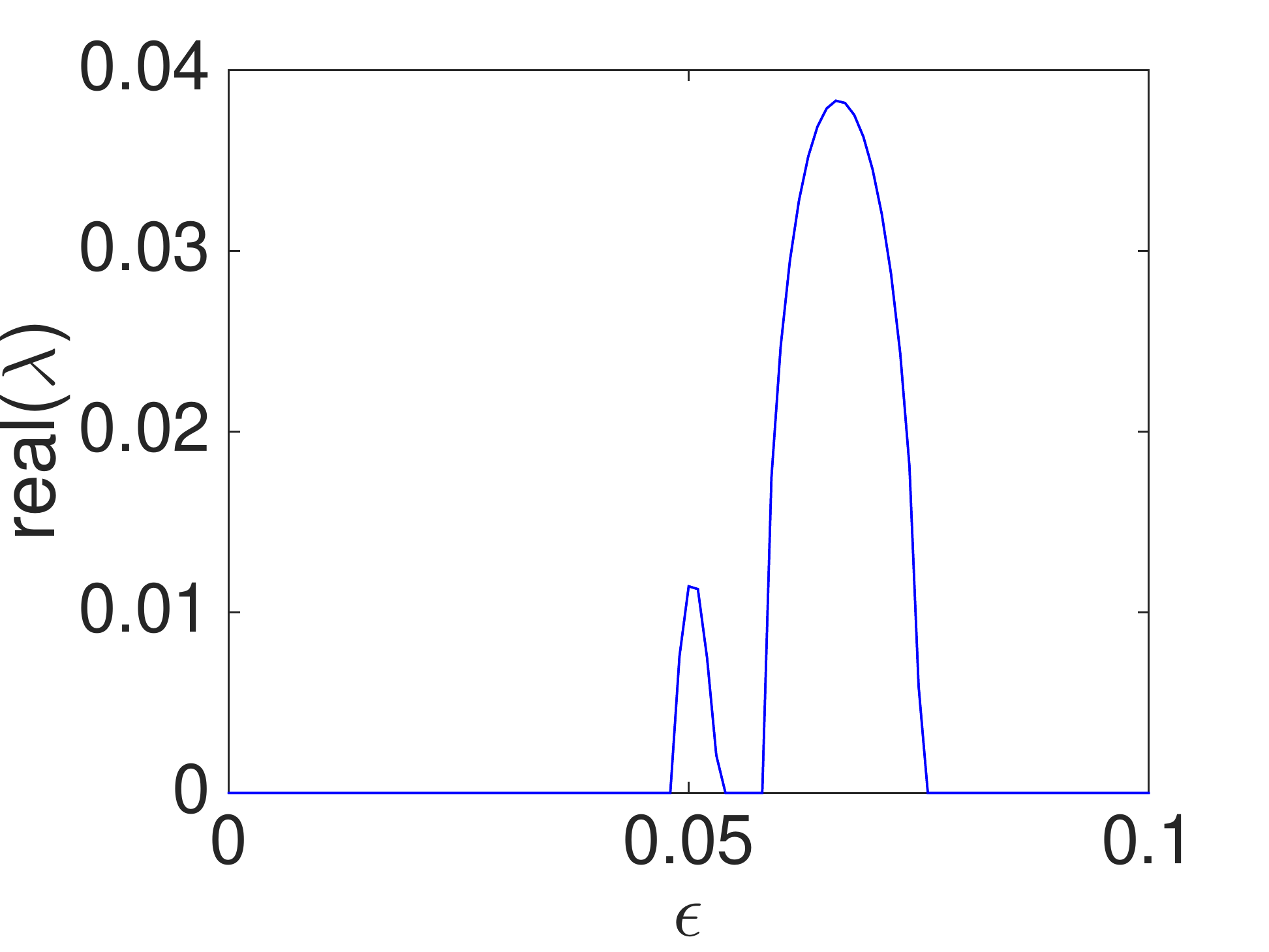}
 \includegraphics[width = 0.3\textwidth]{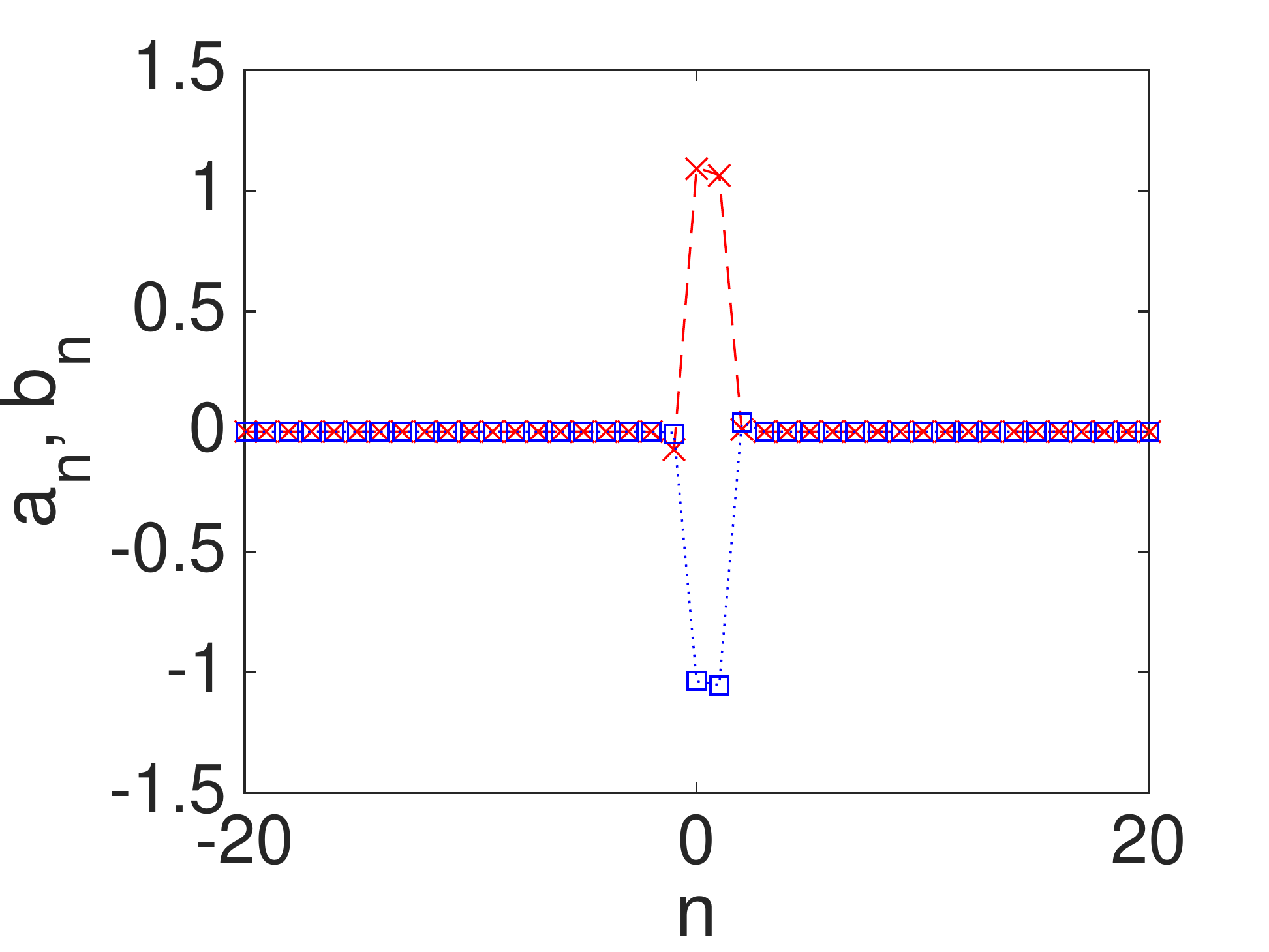}
 \includegraphics[width = 0.3\textwidth]{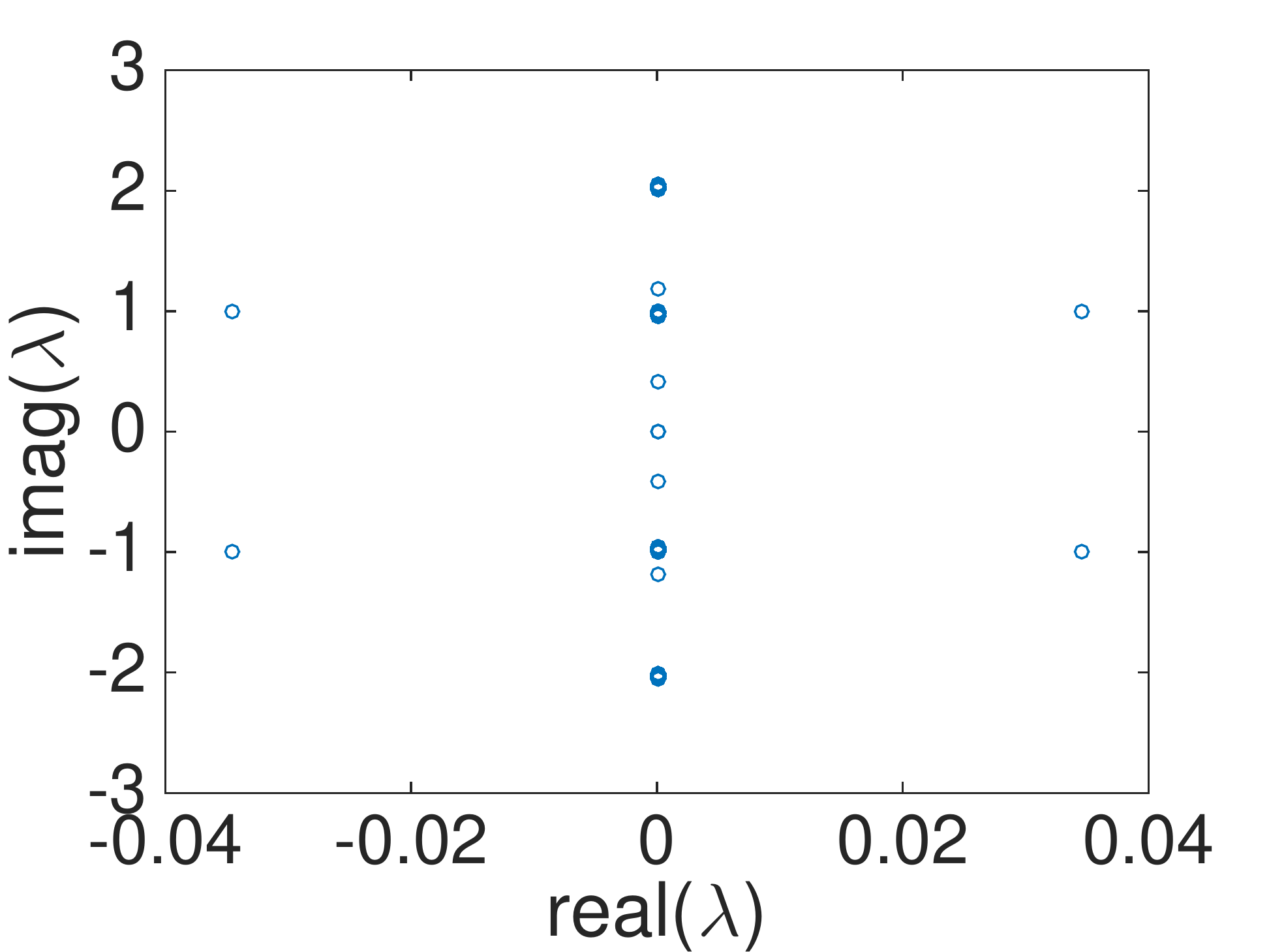}
 \includegraphics[width = 0.3\textwidth]{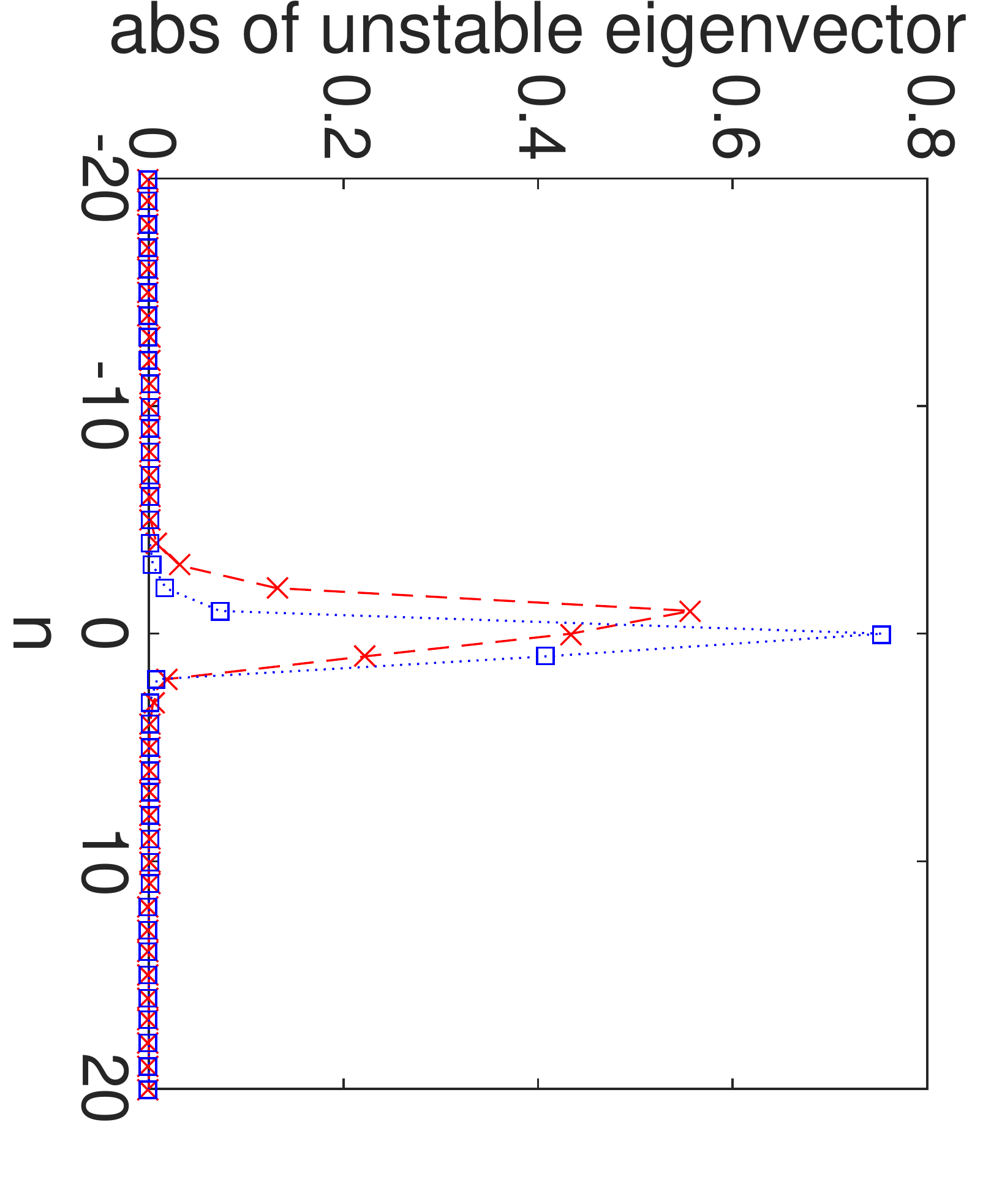}
 \includegraphics[width = 0.3\textwidth]{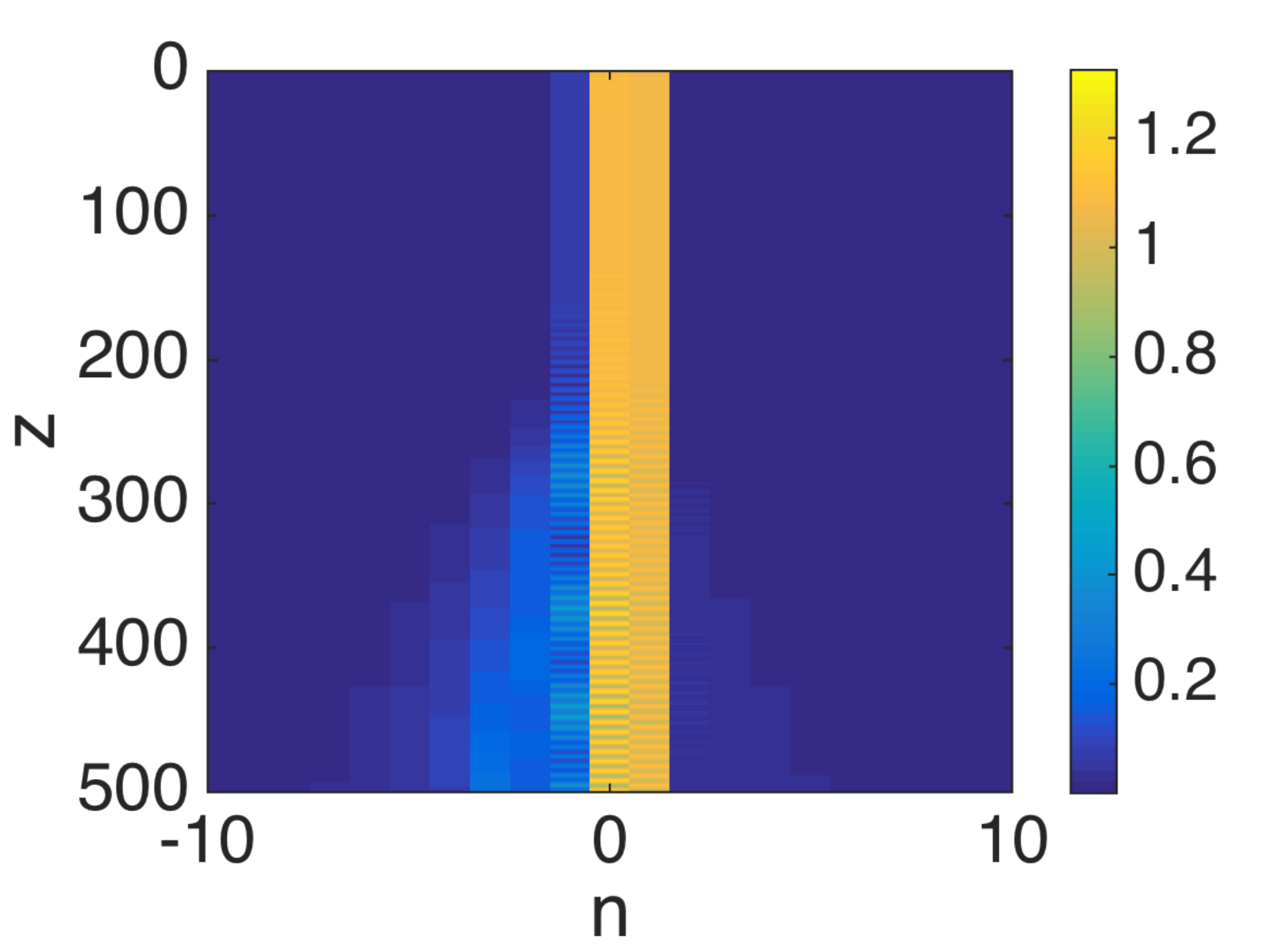}
 \includegraphics[width = 0.3\textwidth]{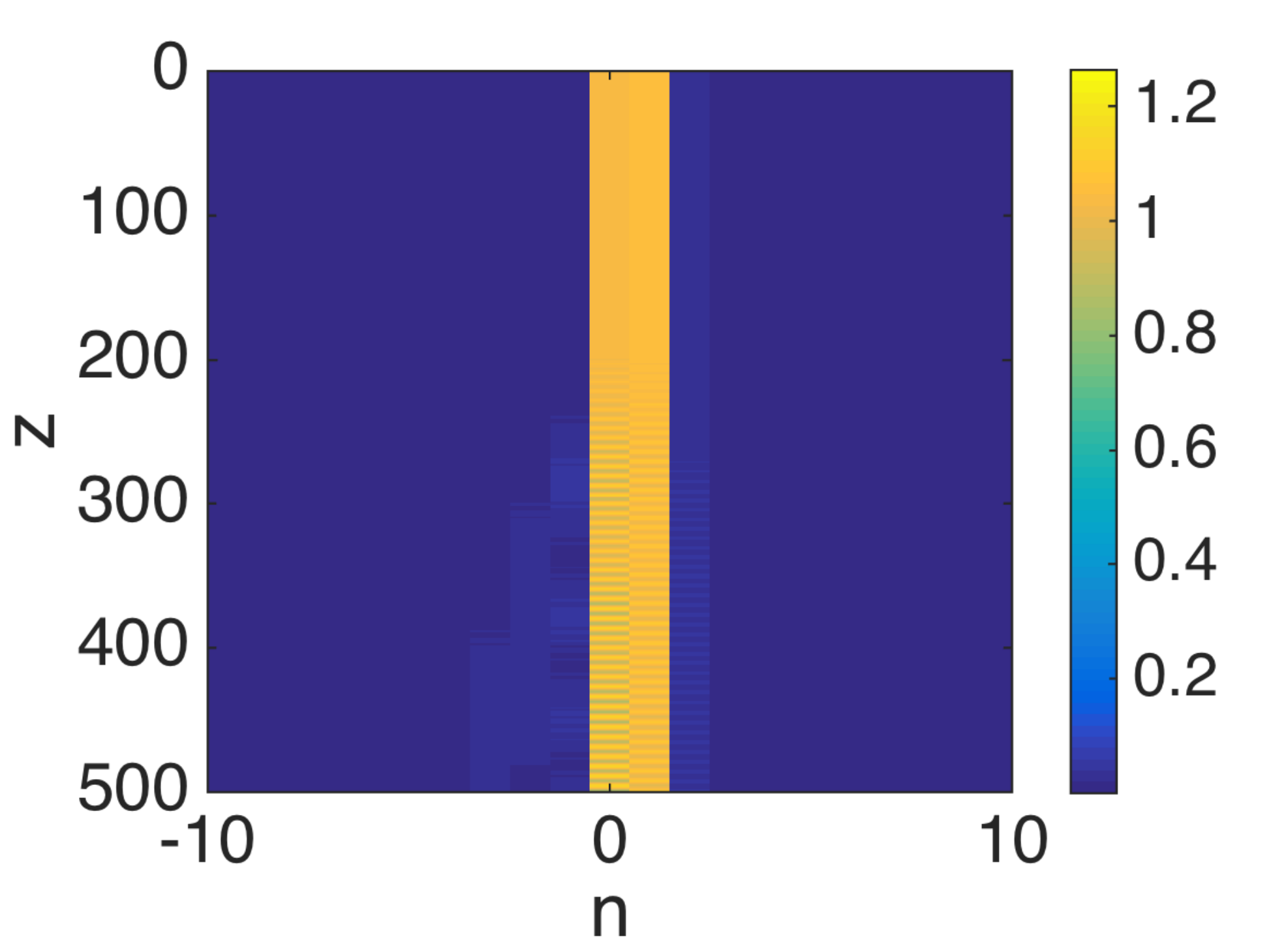}
 \includegraphics[width = 0.3\textwidth]{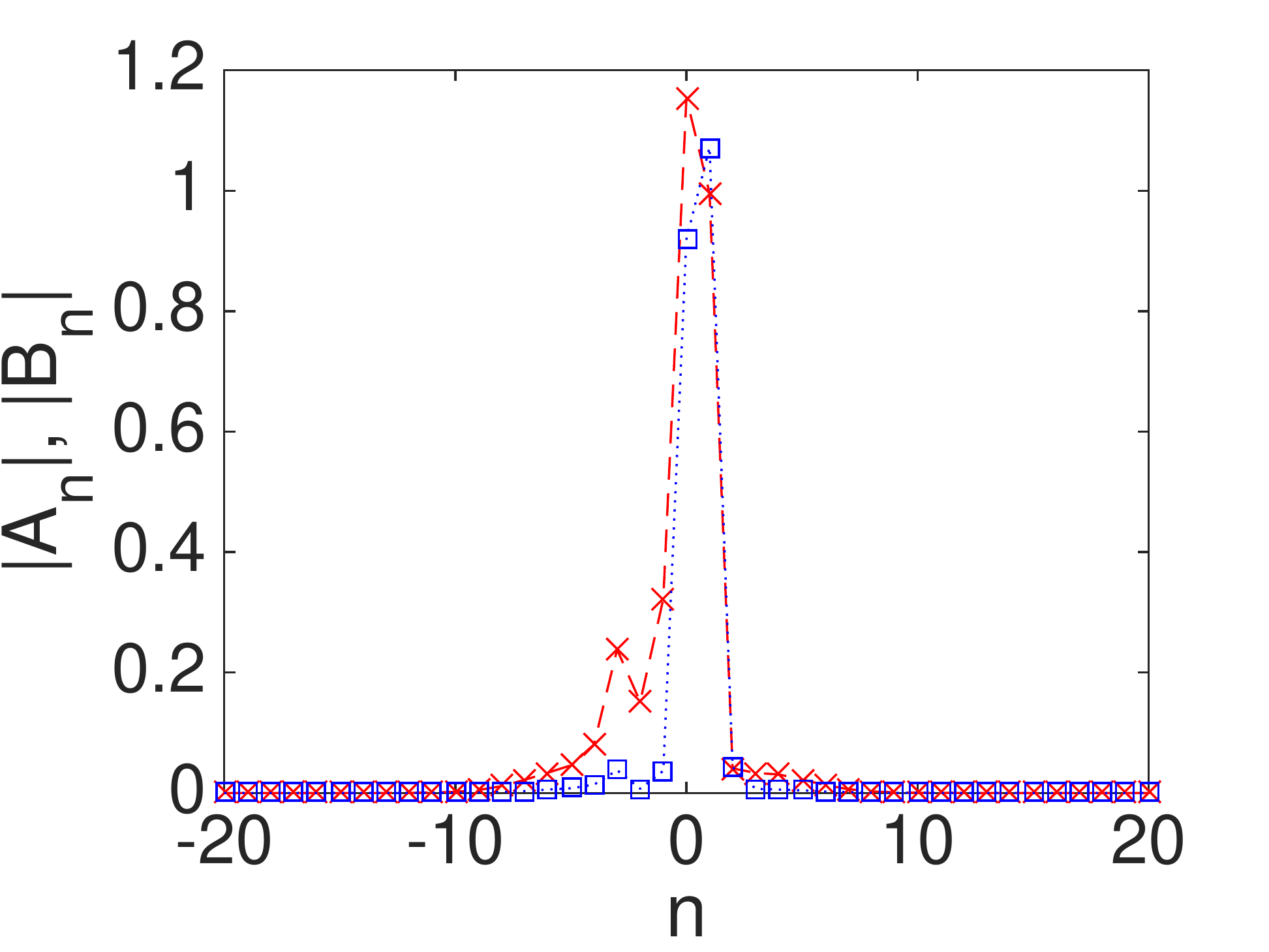}
 \caption{Similar to the previous figure but now for the configuration
of the form $(++,--)$. Here there are three imaginary eigenvalues
shown in the top middle, but their collisions with the continuous spectrum
give rise to quartets that possess real parts shown in the top right panel.
From left to right, the middle row shows the stationary solution profile and spectral plane
 and the unstable eigenmode for $\eps=0.07$. The bottom panels are the unstable dynamics and the modulus profile at $z= 500$.}
 \label{42}
 \end{figure} 

\end{itemize}

\section{Conclusions and Future Challenges}
 
In the present work, we have considered a binary waveguide system
in the vicinity of the anti-continuous limit. We have developed our
theoretical analysis of the existence and stability of few site
configurations in as general a manner as possible. We were able to
parametrically characterize the perturbed configurations 
(at the level of existence) and their corresponding linearization
eigenvalues (at the level of spectral stability). While the qualitative
characteristics of the principal cases we considered were reminiscent
of regular waveguide chains, we illustrated that this is
strongly dependent (as is even the conclusion of stability/instability 
itself) on the sign and magnitude of the binary coupling parameter
$C_1$. We illustrated that variations of this parameter can even
switch specific configurations from stable to unstable or vice-versa.
Whenever our examined configurations were found to be unstable,
we also used direct numerical simulation in order to study
their dynamical evolution. { For instance, for two excited nodes the relevant states result in robust breathing evolution which persists for long propagation intervals. As the number of excited nodes increases, the more complicated interactions of the excited nodes make the unstable dynamics less regular 
in the resulting oscillation amplitudes.}
%on the oscillation of amplitude.}

%Here, it was typically found that
%the relevant states result in robust breathing states which
%persist for long propagation intervals.

There are several directions in which it would be relevant to
extend the present considerations in the future. On the one hand,
it would be relevant to generalize such binary states to 
``checkerboard'' waveguide lattices in two-dimensional settings
and to seek both the near continuum (as in~\cite{akyl,Aceves})
and the highly discrete limits of these and what can be said about
the resulting nonlinear wave states. On the other hand, one could
try to connect the states identified herein for small $\eps$ with
the ones found in the above works for high $\eps$. Naturally, only
a few of the relevant configurations will persist all the way to
the continuum limit of $\eps \rightarrow \infty$, hence it would
be useful to identify the bifurcations thereof and how they
may be similar or different (also depending on the specifics of
$C_1$ and other parameters) from the standard DNLS case, explored
e.g. in~\cite{Konotop}. These topics are presently under study,
and will be reported in future publications.

\vspace{5mm}

{\it Acknowledgements.} PGK gratefully acknowledges support from
NSF-DMS-1312856, as well as from BSF-2010239 and from
the ERC under FP7, Marie Curie Actions, People, 
International Research Staff Exchange
Scheme (IRSES-605096). He also acknowledges the hospitality of
the Center for Nonlinear Studies and the Los Alamos National
Laboratory during the preparation of this work. Research
at Los Alamos is supported in part by the US DoE. Work by ABA was supported 
by the National Science Foundation through the NSF-ECCS-1128593 IDR grant.

\end{document}